\documentclass[12pt]{article}
\usepackage{amsthm,amsmath,amssymb}
\usepackage[hidelinks,hypertexnames=false]{hyperref}
\usepackage{graphicx}
\usepackage{enumerate}
\usepackage{natbib}
\usepackage{url} 
\usepackage{algorithmicx}
\usepackage{algpseudocode}
\usepackage{algorithm}
\usepackage{multirow}
\usepackage{nameref}
\usepackage{float}
\usepackage{xcolor}
\usepackage{authblk}
\newcommand\numberthis{\addtocounter{equation}{1}\tag{\theequation}}
\usepackage[title]{appendix}
\usepackage{enumitem}

\newcommand{\blind}{1}

\usepackage{soul}

\begin{document}

\def\spacingset#1{\renewcommand{\baselinestretch}%
{#1}\small\normalsize} \spacingset{1}


\if1\blind
{
  \title{\bf Compartmental Disease Models with Time-Varying Transmission Rates: A Bayesian
Generalized Hamiltonian Monte Carlo framework with Applications to COVID-19}
  \author{Hristo Inouzhe Valdes\\
    Basque Center for Applied Mathematics (BCAM)\\
    Universidad Autonoma de Madrid\\
    hristo.inouze@uam.es\\
    and \\
    Maria Xosé Rodríguez-Álvarez\\
    Universidade de Vigo and Galician Center for Mathematical Research and Technology (CITMAga)\\
    mxrodriguez@uvigo.gal\\
    and \\
    Lorenzo Nagar\\
    BCAM and Euskal Herriko Unibertsitatea (EHU)\\
    lnagar@bcamath.org \\
    and \\
    Elena Akhmatskaya\\
    BCAM and Basque Foundation for Science (IKERBASQUE)\\
    eakhmatskaya@bcamath.org }
    \date{}
  \maketitle
} \fi

\if0\blind
{
  \bigskip
  \bigskip
  \bigskip
  \begin{center}
    {\LARGE\bf Compartmental Disease Models with Time-Varying Transmission Rates: Bayesian Inference with Hamiltonian Monte Carlo and Applications to COVID-19}
\end{center}
  \medskip
} \fi

\bigskip
\begin{abstract} 

{Estimating how disease transmission changes over time from incidence data is challenging because transmission may be shaped by evolving behaviour, interventions, and other unobserved factors, whereas the available observations are noisy and incomplete. 
This paper develops a flexible Bayesian statistical-mechanistic framework for inferring time-varying transmission rates from incidence data using compartmental epidemic models. Rather than introducing the general idea of combining mechanistic epidemic models with statistical inference, our objective is to provide a unified framework that accommodates a broad class of staged compartmental models, flexible representations of temporal changes in transmission, and efficient Bayesian computation.
 The mechanistic component includes the classical susceptible-infectious-removed and susceptible-exposed-infectious-removed models as special cases, while allowing more flexible assumptions about the time spent in the infectious and exposed states. 
 The transmission rate is modelled using Bayesian P-splines, providing a smooth time-varying representation that avoids imposing a fixed parametric form. 
 The resulting compartmental dynamics are linked to observed incidence through a Negative Binomial model that can incorporate external information on incomplete case detection. 
 Posterior inference is performed using adaptively tuned Generalised Hamiltonian Monte Carlo, supported by gradient calculations, adaptive numerical integration, and structured chain initialisation. 
 The proposed framework is evaluated using synthetic data and compared with established alternatives before being applied to COVID-19 incidence data from the Basque Country, Spain, covering the period from February 2020 to January 2021. 
 The results show that the framework can recover plausible temporal patterns in transmission, while revealing that structurally different models may provide similarly good fits to the same incidence data yet imply substantially different transmission dynamics. These findings underline the importance of sensitivity analyses when using flexible compartmental models with incidence data alone. A documented, open-source implementation is available on GitHub.
}
\end{abstract}

\noindent%

\spacingset{1.8} 

\section*{Introduction}

The COVID-19 pandemic had a profound and devastating impact on many aspects of human life, with several million reported deaths worldwide. Given this emergency context, it is unsurprising that, from 2020 onward, the scientific community directed substantial attention toward various aspects of SARS-CoV-2 infection and transmission. After the identification of the novel coronavirus and the description of the early clinical and epidemiological characteristics of the disease \citep{Zhu2020, Huang2020, Li2020}, mathematical and statistical modelling became central to studying its spread and consequences in terms of incidence, hospitalisation and mortality \citep{Flaxman2020, Kucharski2020, Zelner2021}. This effort built on a long tradition of modelling for infectious diseases \citep{Heesterbeek2015}.

Our focus in this work is on dynamic models for infectious-disease transmission, particularly on statistical-mechanistic formulations that combine epidemiologically informed epidemic dynamics with a probabilistic model for the available data \citep{Coelho2011, Chatzilena2019, Osthus2019, Grinsztajn2020, Zelner2021}. We aim to use such models to partially describe the complex reality of COVID-19 transmission, as well as other infectious diseases, and to utilise available data for estimating some of the unknown quantities governing the transmission process.

Many such dynamic models are based on compartmental models, which split the population into disjoint compartments, specify a flow diagram between compartments (see Figure \ref{erlang_diagrams}), and, finally, encode this flow diagram into an Ordinary Differential Equation (ODE) system that governs the transmission process of the disease \citep{Li2018}. The best-known compartmental models include the Susceptible-Infectious-Removed (SIR) model, originating in the work of \citet{Kermack1927}, and the Susceptible-Exposed-Infectious-Removed (SEIR) model. These classic compartmental models can be seen as approximations of corresponding Markov processes in which the infectious period, and the exposed period when present, follow exponential distributions \citep{Anderson1980, Li2018}. Equivalently, the same conclusion can be reached from the dynamics of the occupation of a compartment \citep[see Section 1.4.1 in][]{Li2018}. Generalisations of such classical models, obtained by subdividing the exposed and/or infectious compartments into consecutive sub-compartments, were discussed, for example, by \citet{Anderson1980}, \citet{Bailey1964} and \citet{Hurtado2019}. This sub-compartmental construction, commonly known as the linear or gamma chain trick, induces Erlang distributions for the time spent in these states while preserving an ODE formulation. We denote the resulting models as SI$_K$R and SE$_M$I$_K$R, with $K$ ($M$) standing for the split of the infectious (exposed) compartment into $K$ ($M$) sub-compartments. The classical SIR and SEIR models are recovered when the relevant numbers of sub-compartments are equal to one, so this formulation defines a general compartmental family rather than a distinct modelling framework. The exponential assumption underlying the classical SIR/SEIR formulation can be restrictive when external knowledge points to more concentrated or otherwise more realistic residence-time distributions \citep{Wearing2005}, and, if such structure is ignored, part of the resulting misspecification may instead be absorbed by the estimated time-varying transmission rate $\beta(t)$, making it less interpretable as a transmission signal. The effects of these more flexible time-in-state distributions were studied, among others, by \citet{Lloyd2001} and \citet{Krylova2013}. In these SIR- and SEIR-like models, the driving force of disease transmission is the transmission rate $\beta$, i.e., the average number of contacts per person per time unit multiplied by the probability of infection in a contact. Thus, the transmission rate guides the transition from the susceptible compartment to the infectious or exposed compartments, respectively. A related quantity is the basic reproduction number ($\mathsf{R_0}$), commonly interpreted as the number of secondary infections generated by a single infectious individual in a fully susceptible population.

In textbook epidemiological modelling, the transmission rate is usually assumed to remain constant over the period of study \citep{Kermack1927, Li2018}. However, this approach can be overly restrictive and unrealistic, especially in the context of diseases like COVID-19, where transmission dynamics are influenced by multiple factors. Changes in transmission can occur due to non-pharmacological interventions, such as mask mandates, lockdowns, curfews, and capacity restrictions \citep{Flaxman2020, Haug2020, Brauner2021}, as well as shifts in group behaviour driven by factors like increased awareness, behavioural fatigue or risk perception. Other mechanisms, such as changes in pathogen characteristics or seasonal effects, may also affect transmission \citep{Alleman2023}. These factors may slow down or accelerate disease transmission at the population level, and in SIR- and SEIR-like models this can be represented by decreases or increases in the transmission rate. Furthermore, while pharmacological measures such as vaccines can be explicitly incorporated through additional compartments or transition mechanisms \citep{Alleman2023}, their effects can also be effectively captured by using a time-dependent transmission rate. In the case of a rapidly spreading pandemic caused by a new disease, like COVID-19, group behaviour changes frequently, and a variety of non-pharmacological and pharmacological measures are adopted at different times. Therefore, it is reasonable and justifiable to adopt a flexible time-dependent transmission rate instead of a constant one when employing SIR- and SEIR-like models. This approach allows for a more accurate representation of the dynamic nature of disease transmission and the diverse factors that influence it. Moreover, the specification of the time-dependent transmission rate should be sufficiently flexible to capture these changes.

In this work, we consider the SI$_K$R/SE$_M$I$_K$R model family with a time-dependent transmission rate parameterised using P-splines \citep{Eilers1996}. Spline-based representations of time-varying transmission or reproduction dynamics have previously been used in statistical-mechanistic epidemic models \citep{Frasso2016, Hong2020, Girardi2020}. Various alternative techniques for modelling time-dependent transmission rates have also been proposed, including logistic functions \citep{Hauser2020}, Legendre polynomials \citep{Smirnova2019}, change points \citep{dehning2020inferring}, and diffusion processes \citep{Dureau2013}. We focus on P-splines for two primary reasons. Firstly, P-splines offer a high degree of flexibility in capturing transmission dynamics without imposing a fixed parametric form or a prespecified change-point structure for the transmission rate. Secondly, P-splines are particularly well-suited for our purpose of employing Hamiltonian Monte Carlo (HMC) methods {\citep{duane1987hmc, Brooks2011}}.


As pointed out by \citet{Zelner2021}, a crucial task in modelling disease transmission is accounting for the diverse sources of uncertainty. In our setting, the observed daily incidence provides a noisy and partial view of the underlying transmission process. Hence, to account for uncertainty in the data, we use a probabilistic/generative compartment-based model, and resort to the Bayesian paradigm for parameter estimation. This allows us to naturally incorporate prior knowledge on the parameters of the mechanistic model and to quantify the uncertainty of the results in a consistent and coherent way. Bayesian methods have shown to be well suited for the epidemiological context \citep[see, for example,][]{Coelho2011, Frasso2016, Chatzilena2019, Osthus2019, Hauser2020, Zelner2021}.


To obtain a sample of the posterior distribution of the parameters we employ the HMC method {\citep{duane1987hmc, Brooks2011}}. HMC makes use of Hamiltonian dynamics and the gradient of the log-posterior to propose candidates in a Markov chain. Its ability to avoid a random walk behaviour (when properly tuned) helps to achieve a better convergence to the target distribution as well as superior performance for high-dimensional problems compared to conventional Markov Chain Monte Carlo (MCMC) methods  \cite[see section 5.3.3 in][]{Brooks2011}. In particular, we choose to use a generalised formulation of HMC \citep{Kennedy2001} that, in contrast to a conventional HMC approach, results in an irreversible sampler, GHMC \citep{Fang2014}. Our choice is supported by multiple theoretical and numerical results demonstrating the advantage of irreversible samplers over reversible algorithms in terms of asymptotic variance of an estimator and convergence rates \cite[see, for example,][]{Duncan_2016, Song_2022}. As far as we are aware, HMC methods have not been used in dynamic models when the transmission rate or the basic reproduction number is time-dependent and belongs to a wide space of functions (for a time-dependent logistic shape see \citealp{Hauser2020}; for a time-independent case see, for example, \citealp{Chatzilena2019}). As for GHMC, to the best of our knowledge, the sampler has never been employed in the context of epidemiological modelling.

Our main contributions in this work are the following. 1) Inspired by \cite{Frasso2016}, \cite{Girardi2020} and \cite{Hong2020} we enrich SI$_K$R and SE$_M$I$_K$R models with a time-dependent transmission rate parameterised using Bayesian P-splines \citep{Lang2004}. Hence, we provide well known and widely successful compartmental models with a time-dependent transmission rate in a functional space capable of capturing a wide range of transmission patterns. 2) We propose a dynamic model based on these compartmental models in order to handle uncertainty in the data. 3) We develop a pragmatic MAP-centred initialisation workflow designed to place chains in epidemiologically plausible high-posterior-density regions, thereby improving numerical stability and reducing the cost of warm-up in this challenging partially observed setting. 4) We make GHMC sampling feasible for the proposed models through explicit and semi-analytical calculation of the gradients required by the sampler. 5) We provide a means for realisation of our approach which relies on the tools s-AIA for adaptive integration and ATune for adaptive HMC tuning developed in \cite{nagar2023adaptive} and \cite{AKHMATSKAYA2026116892}, respectively. 6) We supply a synthetic dataset and use it to validate the proposed methodology, while comparing its behaviour with corresponding diffusion-based models fitted with the well-established SMC$^2$ sampler \citep{Chopin2013} and with the popular EpiEstim method \citep{Cori2013}. 7) We apply our methods to estimate the transmission behaviour of the COVID-19 pandemic in the Basque Country in Spain. 8) Finally, we release a documented, MIT-licensed reference implementation on GitHub, together with the datasets and scripts needed to reproduce the main numerical experiments.

The rest of the paper is organised as follows. The proposed modelling framework is detailed in \nameref{sec_Methods}. First, we introduce the mechanistic models SE$_M$I$_K$R with a spline-based transmission rate. Next, we discuss a Negative Binomial generative probabilistic model based on the solutions of the mechanistic dynamics. The main elements of the suggested Bayesian approach are developed in \nameref{bayesian_set-up}. The summary of the Hamiltonian-based sampling strategy concludes the method section. The synthetic dataset and the COVID-19 daily incidence data of the Basque Country are presented in \nameref{sec_Data}, whereas two case studies built on them are discussed in \nameref{section_results_and_discussion}. We summarise our contributions and future plans in \nameref{sec_Conclusions}. Some technical details and extra results for the two case studies are available as Supplementary Materials.

\section*{Methods}
\label{sec_Methods}
In this section, we describe in-depth all components of the proposed methodology. This includes a formulation and implementation of (i) a mechanistic dynamic model for the disease transmission and (ii) a probabilistic model incorporating the solutions of the mechanistic dynamics, the data to fit and a HMC sampling.
\subsection*{Mechanistic dynamics}
\begin{figure}
\centering
\includegraphics[scale=0.2]{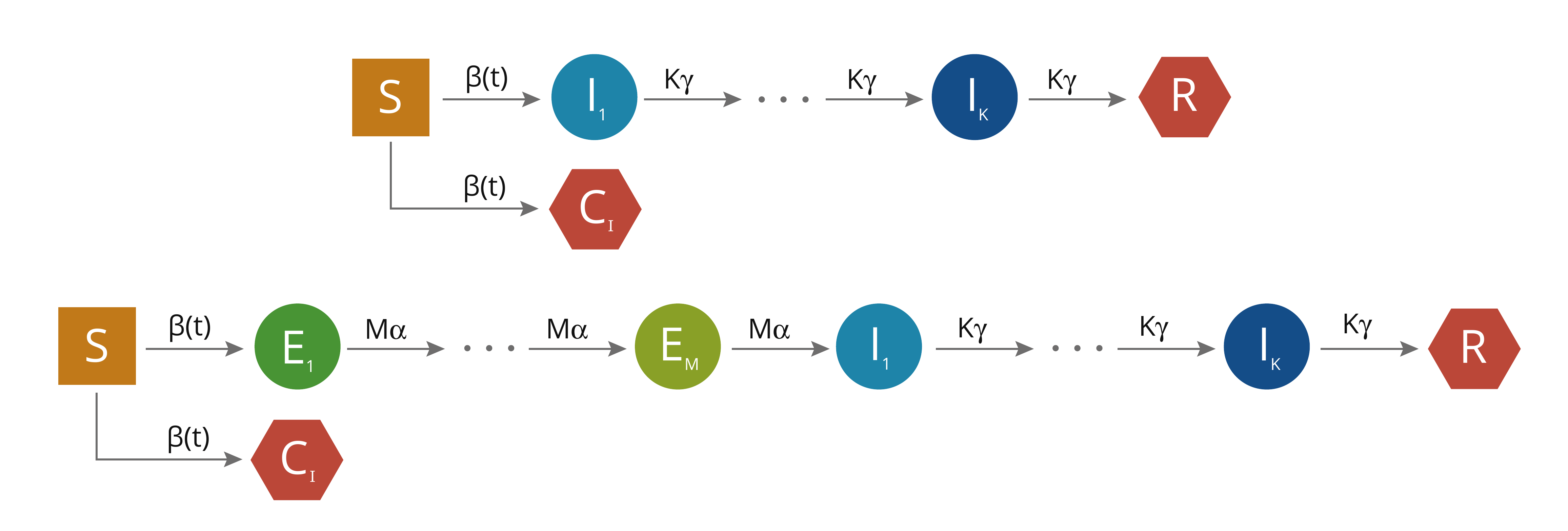}
\caption{
Flow Diagrams for SI$_K$R (top) and SE$_M$I$_K$R (bottom) models with time-dependent transmission rate $\beta(t)$. $S$: number of individuals in the population susceptible to be infected. $E_1,\dots, E_M$: number of individuals at the different stages of exposure (infected but not infectious). $I_1,\dots, I_K$: number of individuals at the different stages of infectiousness (infected and infectious). Here $M\alpha$  and $K\gamma$ are constant rates associated with compartments $E_i$ and $I_j$, respectively. The average time spent in an exposed compartment ($E_i$) is given by $\frac{1}{M\alpha}$, and the average time being exposed is given by $\frac{1}{\alpha}$. Similarly, the average time spent in an infectious compartment ($I_j$) is given by  $\frac{1}{K\gamma}$, and the average time being infectious is given by $\frac{1}{\gamma}$. The $R$ compartment is the number of individuals removed from the pool of susceptible (dead or with long lasting immunity). The time-dependent transmission rate $\beta(t)$ governs the transition between being susceptible and being infected/exposed. The $C_I$ compartment is out of the transmission process and just counts the total number of individuals that have been infected.
}
\label{erlang_diagrams}	
\end{figure}

The compartmental models that we choose as a foundation of our framework are SI$_K$R and SE$_M$I$_K$R, introduced in \cite{Anderson1980} (with a constant transmission rate $\beta$), and their flow diagrams are depicted and explained in Figure~\ref{erlang_diagrams}. These models are extensions of the well-known SIR and SEIR models obtained by splitting the infectious state, and when present the exposed state, into multiple sequential sub-compartments. For brevity, we omit subscripts when the values of $K$ and $M$ are equal to 1; in particular, SI$_1$R and SE$_1$I$_1$R are written as SIR and SEIR, respectively. We use this formulation as a general compartmental family: the classical SIR and SEIR models are recovered as special cases, whereas larger values of $M$ and $K$ allow the exponential assumption on the time spent in the corresponding states to be relaxed when supported by epidemiological knowledge or by the aims of the analysis. Specifically, the sub-compartment structure induces Erlang distributions, i.e., Gamma distributions with integer shape parameters, for the corresponding exposed and infectious periods, while retaining the compartmental ODE structure. If the residence-time distribution is misspecified, part of this misspecification may instead be absorbed by the estimated time-varying transmission rate $\beta(t)$, making $\beta(t)$ less interpretable as a transmission signal; the SI$_K$R/SE$_M$I$_K$R formulation gives practitioners a direct way to encode disease-stage knowledge in the compartmental structure instead. This construction can also be interpreted as a modelling strategy to implement delays in the transitions between compartments, e.g., from $I$ to $R$, or from $E$ to $I$. We remark that, for the models displayed in Figure~\ref{erlang_diagrams}, the average time spent in the exposed state is $1/\alpha$, and the average time being infectious is $1/\gamma$, where $\alpha$ and $\gamma$ are the parameters governing the overall average duration of the exposed and infectious periods, respectively. The per-compartment transition rates are $M\alpha$ and $K\gamma$, so $M$ and $K$ control the shape of the corresponding Erlang distributions while preserving these mean durations. An informative study of the effects of the distribution for the infectious period in these Erlang-based models is presented in \citet{Krylova2013}. It is important to note that the split of the compartments does not have to reflect a real distinction in the development of the disease (although it could); rather, it is a mathematical artefact used to impose a desired temporal behaviour in the model.

To model the time-dependent transmission rate $\beta(t)$, we follow \cite{Frasso2016, Girardi2020, Hong2020} and use splines. In particular, we approximate (the logarithm of) $\beta(t)$ by a linear combination of $m$ B-spline basis functions
\begin{equation}
\label{log_beta}
\log\beta(t)=\sum_{i=1}^m\beta_iB_i(t),
\end{equation}
where $B_i(t)$ denotes the $i$th B-spline basis function evaluated at time $t$, and $\boldsymbol{\beta} := (\beta_1, \dots, \beta_m)'$ is the vector of coefficients. The B-spline basis is defined over the time interval $[t_0, t_1]$ and determined by the degree $d$ and the number of internal knots $Q$, with $m = Q + d - 1$. For an in-depth discussion on splines, we refer to \cite{Dierckx1993}. We emphasise that splines are particularly well suited to our setting, as they yield a differentiable representation of $\beta(t)$ that is compatible with HMC-based inference.

With the previous considerations, the ODE equations corresponding to the SE$_M$I$_K$R model, which is represented by the bottom flow diagram in Figure \ref{erlang_diagrams}, are the following: 
\begin{small}
\begin{align*}
\frac{dS(t)}{dt}&=-\beta(t)S(t)\frac{I(t)}{N},\\
\frac{dE_1(t)}{dt}&=\beta(t)S(t)\frac{I(t)}{N}-M\alpha E_1(t),\\
\frac{dE_2(t)}{dt}&=M\alpha E_1(t)-M\alpha E_2(t),\quad \dots,\quad
\frac{dE_M(t)}{dt}=M\alpha E_{M - 1}(t)-M\alpha E_M(t),\\
\frac{dI_1(t)}{dt}&=M\alpha E_M(t)-K\gamma I_1(t),\numberthis \label{SEMIKR_ODES}\\
\frac{dI_2(t)}{dt}&=K\gamma I_1(t)-K\gamma I_2(t),\quad
\dots,\quad
\frac{dI_K(t)}{dt}=K\gamma I_{K-1}(t)-K\gamma I_K(t),\\
\frac{dR(t)}{dt}& = \frac{d}{dt}\left(N - S(t) - E(t) - I(t)\right) =  K\gamma I_K(t),\\
\frac{dC_I(t)}{dt}&=\beta(t)S(t)\frac{I(t)}{N},
\end{align*}
\end{small}
with initial conditions:
\begin{equation}\label{init_cond}
\begin{split}
S(t_0)& = S_0,\;E_1(t_0) = E_0,\;E_2(t_0)=\dots=E_M(t_0)=0,\; I_1(t_0) = I_0,\\
I_2(t_0) &=\dots = I_K(t_0) = 0,\;R(t_0) = R_0 = N - (S_0 +E_0+I_0),\;
C_I(t_0)=N-S_0,
\end{split}
\end{equation}
where $\beta(t) = \exp\left(\sum_{i=1}^{m}\beta_iB_{i}(t)\right)$ (see Eqn. (\ref{log_beta})), $E(t) = \sum_{i = 1}^{M}E_i(t)$, $I(t) = \sum_{j= 1}^{K}I_j(t)$, and $N = S(t)+E(t)+I(t)+R(t)$ is the fixed size of the total population. The initial conditions in \eqref{init_cond} are specified to ensure a unique solution to the system under standard regularity assumptions. Specifically, they indicate that, at time $t_0$ (typically corresponding to the beginning of the transmission process), there were $S_0$, $E_0$, $I_0$, and $R_0$ susceptible, exposed, infectious, and recovered individuals, respectively.

A solution of the system~\eqref{SEMIKR_ODES} provides $S(t), E_1(t), \dots, E_M(t), I_1(t), \dots, I_K(t), R(t)$ for $t$ in a certain time period $[t_0, t_1]$. That is, the solution yields the number of individuals at each stage (compartment) of the transmission process at a given time $t$, i.e., the number of prevalent cases. To facilitate the computation of cumulative and incident cases, we introduce a counting compartment $C_I$ in the model dynamics. This auxiliary compartment accumulates all new infections over time, with no outflow. Specifically, individuals who become infected at time $t$, given by $\beta(t) S(t) I(t)/N$, are added to $C_I(t)$, so that $C_I(t)$ denotes the cumulative number of infected individuals up to time $t$. Let time $t$ be measured in days, and let $0 < j < t_1 - t_0$ with $j \in \mathbb{N}$. Then, $C(t_0 + j) = C_I(t_0 + j) - C_I(t_0 + j - 1)$ gives the number of new infected individuals on day $j$ of the disease transmission, i.e., the daily incidence at day $j$. Although $C_I(t)$ is redundant from a dynamical perspective (since cumulative cases can also be computed as $N - S(t)$), we include it for clarity and to make explicit the link between the modelled transmission dynamics and the observed daily incidence data (see next section). We note that the SI$_K$R model is a direct simplification of the SE$_M$I$_K$R model, in which the exposed compartment is removed and susceptible individuals transition directly to the first infectious stage upon infection. For completeness, the corresponding equations are provided in \ref{SI_K_R_model}.

\subsection*{Probabilistic model}
Once the decision on a mechanistic model for disease transmission is made and its parameters specified, the next step consists in defining a suitable probabilistic model that links the compartmental dynamics to the observed incidence data, and in setting up the full Bayesian inference scheme.

In a pandemic, daily incidences, $\{\widetilde{C}_{t_0+j}\}_{j=1}^n$, that account for the new daily positive cases over $n$ days, are typically noisy due to multiple factors, such as lack of information, changes in measurement criteria, human behaviour, and inherent variability. A standard approach to handle such noise is to use a probabilistic model centred on the output of a mechanistic one. Following this idea, we treat $\widetilde{C}_{t_0+j}$ as a realisation of a random variable $\widetilde{C}(t_0+j)$, whose mean is the daily incidence at day $j$, 
\[ 
C(t_0+j) = C_I(t_0 + j) - C_I(t_0 + j - 1), 
\]
where $C_I(t)$ is obtained from the numerical solution of~\eqref{SEMIKR_ODES}. Additionally, it is important for COVID-19 modelling to explicitly account for the underestimation of incidence in surveillance data, which may arise from under-ascertainment and under-reporting of infections \citep{Gibbons2014}, for instance due to limited testing and the presence of asymptomatic or mildly symptomatic individuals. Information on the extent of underestimation is typically obtained from seroprevalence studies \citep{ENE_COVID_2020, Pollan2020}. Following \citet{Frasso2016}, we incorporate incomplete detection into the probabilistic model via a time-varying function $0 < \eta(t) \leq 1$, which represents the fraction of new cases that are actually detected. This function modifies the expected value of the observed incidence by scaling the output of the compartmental model. In the absence of underestimation, $\eta(t) = 1$ for all $t$.

All in all, the probabilistic model that describes new infected cases at time $t$, $\widetilde{C}(t)$, and relies on the solution $C_I(t)$ of the SE$_M$I$_K$R equations (\ref{SEMIKR_ODES}), is as follows
\begin{align}
\label{NegBinModel}
\widetilde{C}(t)\sim \text{NegativeBinomial}(\eta(t)C(t), \phi),
\end{align}
where $C(t) = C_I(t)- C_I(t - 1)$ and $\phi > 0$. The Negative Binomial distribution is parametrised as
\[
\mathbb P\left(\widetilde{C}(t) = k\right) = \binom{k + \phi - 1}{k}
\left(\frac{\eta(t)C(t)}{\eta(t)C(t) + \phi}\right)^k
\left(\frac{\phi}{\eta(t)C(t) + \phi}\right)^\phi.
\]
This implies
\[
\mathbb{E}\left[\widetilde{C}(t)\right] = \eta(t)C(t), \quad 
\mathrm{Var}\left[\widetilde{C}(t)\right] = \eta(t)C(t) + \frac{\left[\eta(t)C(t)\right]^2}{\phi}.
\]
The parameter $\phi$ controls the amount of overdispersion around the mean. The use of a Negative Binomial distribution, which can be viewed as an overdispersed alternative to the Poisson distribution, is well supported in the literature and is a standard choice in epidemic modelling \cite[see, e.g.,][]{Frasso2016}.

In summary, the adopted probabilistic model links the observed data, $\{\widetilde{C}_{t_0+j}\}_{j=1}^n$, to the true daily incidence $C(t)$ derived from the compartmental model, with incomplete detection represented by $\eta(t)$, and incorporates overdispersion through a Negative Binomial distribution.

\subsection*{Bayesian set-up}\label{bayesian_set-up}
We now formulate a Bayesian framework to estimate the parameters of the mechanistic and probabilistic models using the available data.
\subsubsection*{Parameters and prior specification}
Our goal is to obtain a posterior distribution of the parameters
\begin{equation}
\label{parameters}
\mathbf{p} = (p_0,\dots,p_5, p_6, \dots, p_{m+5})' = (\alpha, S_0,E_0,I_0,\phi^{-1},\tau^2, \boldsymbol{\beta}')',
\end{equation}
given the observed data $\mathcal{D} = \{\widetilde{C}_{t_0 + j}\}_{j=1}^n$ and the prior knowledge on these parameters. All parameters in (\ref{parameters}) but $\phi^{-1}$ (it appears in the probabilistic model (\ref{NegBinModel})) are associated with the mechanistic model (\ref{SEMIKR_ODES}). The parameter $\tau^2$ is the variance of the Gaussian increments in the random walk prior for $\boldsymbol{\beta}$, which we discuss in detail below. Before proceeding, we note that a key modelling challenge in our context is that incidence data alone do not contain enough information to jointly identify all parameters. In particular, $\gamma$ (the inverse of the recovery time) and the time-varying transmission rate $\beta(t)$ are not simultaneously identifiable: different combinations of these parameters may give rise to similar observed incidence data. For example, a large number of new infections may result from either a high $\beta(t)$ or a small $\gamma$, and vice versa. Without external constraints, posterior inference may therefore be dominated by the prior distributions rather than by the likelihood. To address this, we fix $\gamma$ to an epidemiologically plausible value based on prior knowledge, which allows us to focus inference on the transmission dynamics, our main target of interest. The implications of this modelling choice are revisited in ~\nameref{section_results_and_discussion}, where we examine the robustness of our conclusions under alternative plausible values of $\gamma$. Although $\alpha$, which influences the rate of progression from exposed to infectious, also enters the model dynamics, its interaction with $\beta(t)$ is less critical from an identifiability perspective, and we retain it as an unknown parameter to be estimated. 

The choices of prior distributions for $\alpha, S_0, E_0, I_0$ and $\phi^{-1}$ will be analysed in the context of the studied datasets. Here, we focus on the prior specification for the vector of coefficients $\boldsymbol{\beta} :=(\beta_1,\dots,\beta_m)'$ in (\ref{log_beta}), i.e., the coefficients associated with the time-dependent transmission rate $\beta(t)$. In particular, we propose using Bayesian P-splines for that purpose. In brief, in (Bayesian) P-splines, a moderately large number $m$ of B-spline basis functions defined over a sequence of equidistant knots is used to ensure sufficient flexibility to capture complex temporal patterns in $\beta(t)$. However, using many basis functions without further constraints can lead to overfitting and overly wiggly estimates. To prevent this, smoothness is enforced through a penalty on the coefficients $\boldsymbol{\beta}$. In the Bayesian framework, this penalty is introduced via a prior distribution: specifically, a random walk of order $q$ is placed on $\boldsymbol{\beta}$. For example, a second-order random walk --the most common in the literature and our choice here-- assumes that each coefficient satisfies
\begin{equation}
\beta_k = 2\beta_{k-1} -\beta_{k-2} + u_k, \quad u_k \sim \mathcal{N}(0, \tau^2), \quad k=3,\dots,m,
\label{eq:random_walk}
\end{equation}
with $\beta_1$ and $\beta_2$ given diffuse (flat) priors. This formulation penalises deviations from local linear trends across neighbouring coefficients, thereby promoting smoothness in the resulting function $\beta(t)$. The random walk prior distribution variance, $\tau^{2}$, controls the amount of smoothing: smaller values yield smoother functions, while larger values allow greater flexibility. This parameter can either be fixed or assigned a prior distribution. Here, we adopt the latter approach and, following \cite{Lang2004}, assume $\tau^2\sim\text{InvGamma}(a_{\tau^2}, b_{\tau^2})$.

The second-order random walk prior in \eqref{eq:random_walk} induces the following partially improper multivariate Gaussian prior for the vector of coefficients $\boldsymbol{\beta}$
\begin{align*}
\mathbb P(\boldsymbol{\beta}:=(\beta_1,\dots,\beta_m)'|\tau^2)\propto \exp\left(-\frac{1}{2\tau^2}\boldsymbol{\beta}' \mathsf{K}\boldsymbol{\beta}\right),\numberthis\label{prior_improper}
\end{align*}
where the precision matrix $\mathsf{K} = \mathsf{D}'\mathsf{D}$ with $\mathsf{D}$ being a second-order difference matrix. For the explicit form of $\mathsf{K}$ and $\mathsf{D}$ and further technical details, see Section 4.2.2.1 in \cite{Kneib2005}. A comprehensive overview of the Bayesian P-spline methodology is provided in \cite{Lang2004}.

\subsubsection*{Implementation}\label{implementation}
In the proposed setting, the log-posterior distribution takes the following form:
\begin{align*}
\ell(\mathbf{p} \mid \mathcal{D})& \propto \ell_{like}(\mathcal{D} \mid \mathbf{p}) + \ell_{prior}(\mathbf{p})\\
&= \sum_{j=1}^n\ell_{like}^j(\mathcal{D} \mid \mathbf{p}) + \ell_{prior}(\mathbf{p})\numberthis\label{log_post}\\
&=\sum_{j=1}^n\log\Bigg(\frac{\Gamma\left(\widetilde{C}_{t_0 + j} + \phi\right)}{\Gamma(\phi)\Gamma\left(\widetilde{C}_{t_0 + j} + 1\right)}\left(\frac{\eta(t_0 + j)C(t_0 + j)}{\eta(t_0 + j)C(t_0 + j)+\phi}\right)^{\widetilde{C}_{t_0 + j}}\\& \quad\left(\frac{\phi}{\eta(t_0 + j)C(t_0 + j)+\phi}\right)^{\phi}\Bigg) +\ell_{prior}(\mathbf{p}),
\end{align*}
\noindent where $\Gamma$ denotes the gamma function.

To proceed with Bayesian estimation of parameters (\ref{parameters}), we use GHMC to sample from the posterior distribution (\ref{log_post}). A practical challenge in this setting is that incidence-only data, latent compartmental states, and a flexible time-dependent transmission rate can induce weakly identified directions and multiple high-posterior-density regions. Therefore, the optimisation step described below should not be understood as a substitute for posterior exploration, but rather as a device for finding stable and epidemiologically plausible regions from which Hamiltonian-based sampling can be initialised.

Accordingly, the posterior summaries reported in this work should be interpreted conditionally on the selected model structure, the externally fixed quantities such as $\gamma$ and $\eta(t)$, and the high-posterior-density region explored by the retained chains. We do not claim exhaustive exploration of all possible posterior modes. Instead, our goal is to obtain stable and interpretable inference for the time-varying transmission dynamics in the region of the parameter space that is both well supported by the data and epidemiologically plausible. For this purpose, we adopt a four-stage workflow that separates model specification, posterior mode localisation, and sampling.

\paragraph{\textbf{Stage 0: Model specification.}}  
Before sampling begins, we specify the structure of both the mechanistic and spline components of the model. This involves selecting the disease-phase orders $M$ and $K$ in the $\text{SE}_M\text{I}_K\text{R}$ formulation and the number $m$ of B-spline basis functions used to represent the time-varying transmission rate $\beta(t)$.
\begin{itemize}
  \item  \emph{Disease-phase orders $(M,K)$.} When empirical estimates for the distributions of the exposed and infectious periods are available, we choose the smallest values of $(M,K)$ such that the corresponding Erlang distributions approximate those targets. Increasing $(M,K)$ improves biological realism but increases computational cost.
  
  \item  \emph{Basis dimension $(m)$.} Fixing the spline degree (we use $d = 3$), we increase the number of internal knots $Q$ until a preliminary P-spline fit captures the main features of the observed incidence curve, including all peaks and troughs. We then set $m = Q + d - 1$. This ensures sufficient flexibility while keeping the dimension of the parameter space manageable.
\end{itemize}

\paragraph{\textbf{Stage 1: Maximum A Posteriori (MAP) search.}}  
We use a quasi–Newton optimiser (L–BFGS–B) to maximise the posterior distribution \eqref{log_post}, starting from $S$ randomly generated initial points. The initial values for $\alpha$, $\phi^{-1}$, ($S_0$, $E_0$, $I_0$, $R_0$), and $\tau^2$ are drawn from their prior distributions. The spline coefficients $\beta_k$ are all initialised to a common value $\beta_0 \sim \operatorname{Uniform}(a,b)$, corresponding to a constant transmission rate. This allows the optimiser to introduce temporal variation in $\beta(t)$ only when supported by the data.

\paragraph{\textbf{Stage 2: Localised initial values.}} Around the MAP $\hat{\mathbf{p}}$ we first generate a pool of $S_{\text{cand}}$ candidate points:
\begin{itemize}
    \item \emph{Unconstrained parameters} $(\boldsymbol\beta)$: sample $\beta^{*}_k \sim \mathcal{N}\bigl(\hat{\beta}_k,\, \sigma_{\mathrm{prop}}^{2}\, \hat{\beta}^{2}_k\bigr)$, $1\leq k\leq m$.
    
    \item \emph{Strictly positive parameters} $(\alpha, \phi^{-1})$: add Gaussian noise on the log scale and exponentiate to preserve positivity.
    
    \item \emph{Compartments} $(\mathbf{p}^{\mathrm{prop}} = (S_0,E_0,I_0,R_0))$: draw $\mathbf{p}^{\mathrm{prop},*}$ such that the total variation distance from $\hat{\mathbf{p}}^{\mathrm{prop}}$ satisfies $\mathrm{TV}\bigl(\mathbf{p}^{\mathrm{prop},*}, \hat{\mathbf{p}}^{\mathrm{prop}}\bigr) \le \mathrm{TV}_0$, where $\mathrm{TV}(\mathbf{q}, \mathbf{q}') = \tfrac{1}{2} \sum_i |q_i - q'_i|$. For a hierarchical approach to sampling initial occupancies, see \ref{sup_sec:Initial_Ocuppancy}.
\end{itemize}
We then evaluate the log-posterior $\ell$ (see \eqref{log_post}) at each of the $S_{\text{cand}}$ candidates, cluster them into $n_{\text{chains}}$ groups in $\ell$-space (e.g., using $k$-means on the log-posterior values), and select one seed at random from each cluster. This procedure ensures that the final initial values are not only close to $\hat{\mathbf{p}}$ but also well dispersed in terms of posterior density, promoting independent exploration by the $n_{\text{chains}}$ parallel chains.

\paragraph{\textbf{Stage 3: GHMC sampling and diagnostics.}}
Each selected initial point from Stage~2 is used to launch an independent chain evolved using GHMC.
\begin{itemize}
\item \emph{Sampling algorithm and tuning.}
We employ the irreversible variant of Hamiltonian Monte Carlo (GHMC), which enhances mixing and reduces sensitivity to trajectory length by incorporating partial momentum refreshment (see~\ref{Hamiltonian_dynamics} for details). Since GHMC requires gradient evaluations of the log-posterior, we derive them analytically wherever feasible. In our setting, the log-posterior $\ell$ and most components of its gradient can be computed analytically, except for terms involving the solution of the ODE system and its partial derivatives with respect to model parameters. These derivatives, known as sensitivities, quantify how the compartment trajectories respond to changes in parameter values. We compute them by numerically solving an extended ODE system augmented with sensitivity equations, using CVODES from the SUNDIALS suite \cite[see Section 2.6 in][]{CVODES_2020}. Full details are provided in~\ref{SEIR_Sensitivity_dynamic}. To optimise the performance of the sampler, we follow the adaptive strategy proposed by \citet{AKHMATSKAYA2026116892}, which nominates a model-specific numerical integrator with a complete set of reliable parameters; the full tuning procedure is described in~\ref{Tuning_HMC}.

\item \emph{Convergence monitoring.}
We report split-$\hat R$ for all epidemiologically interpretable scalar parameters and use $\hat R<1.05$ as a pragmatic threshold in this computationally demanding setting. This value lies between the classical guideline $\hat R<1.1$ \citep{geyer1992practical} and the more stringent recommendation $\hat R<1.01$ \citep{vehtari2021rank}, and is sufficient for our main target: stable functional summaries of $\beta(t)$ and $\mathsf{R_0}(t)$. Values above this threshold are treated as evidence that the corresponding fit has not fully converged under the fixed computational budget, which throughout the paper refers to the prescribed maximum number of GHMC warm-up and production steps unless stated otherwise. Such chains or model specifications are therefore not used for the main pooled posterior summaries. We also inspect posterior predictive checks for daily incidence, chain-wise summaries of $\beta(t)$ and $\mathsf{R_0}(t)$, and numerical stability of the Hamiltonian trajectories.
For the high-dimensional spline coefficients $\boldsymbol{\beta}$, we do not rely exclusively on component-wise $\hat R$, since individual coefficients are nuisance parameters and may be strongly correlated. Instead, we assess agreement at the functional level by comparing chain-wise posterior medians and credible bands of $\beta(t)$ and the derived $\mathsf{R_0}(t)$. Persistent chain-specific deviations in these functions are treated as evidence of non-convergence or multimodality.

\end{itemize}
The dispersion scale $(\sigma_{\mathrm{prop}})$ and compartment tolerance $(\mathrm{TV}_0)$ in Stage 2 are key hyperparameters that jointly control how far the initial values used to initialise sampling chains can deviate from the MAP estimate. Larger values encourage broader exploration but increase the risk of placing chains in regions of low posterior density. To manage this trade-off, we adopt a cautious annealing strategy: we begin with conservative values of $(\sigma_{\mathrm{prop}}, \mathrm{TV}_0)$ that restrict initialisation to a high-posterior-density neighbourhood around the MAP, assess within-region mixing and convergence, and gradually increase them only if no convergence issues are detected. This approach enables robust inference by avoiding unnecessary computation in implausible regions while still allowing the sampler to uncover alternative, yet plausible, epidemic scenarios. The result is a set of chains that (i) converge more reliably, (ii) explore a salient enough region of the posterior, and (iii) provide uncertainty quantification for the time-varying transmission rate conditional on the selected model, the fixed epidemiological inputs, and the high-density region explored by the retained chains.
 
\section*{Data}\label{sec_Data}
In this section we present a brief overview of the synthetic data that we use to calibrate and explore the behaviour of the proposed methodology, and the real data to which we apply our approach.
\subsection*{Synthetic data}
To obtain a meaningful synthetic dataset, we aim to emulate a plausible real situation and proceed as follows. We fix the population size $N$ to 2189138, i.e. the population of the Basque Country as given by the Spanish National Statistical Institute (INE in Spanish) for the year 2020 (available at the following \href{https://www.ine.es/prensa/cp_e2021_p.pdf}{link}) and choose a SEI$_3$R model to go beyond a standard case where the ground truth is a SEIR model. The SEI$_3$R model parameters $\mathbf{p}^{syn}$ are then obtained by solving numerically (\ref{SEMIKR_ODES}) at varying parameters values until two waves, i.e., two peaks of daily incidence separated by a region of much lower incidence, in a period of 100 days, are obtained. Specifically, we set $\gamma$, $\alpha$, $S_0$, $E_0$, $I_0$, and $\phi^{-1}$ (see \eqref{synt_params}) to values that are epidemiologically reasonable for a SARS-like virus and consistent with the early phase of the transmission process. Then, the parameters $a$ and $b$ of a transmission function $$\beta(t;a,b) = \frac{e^{\sin(2\pi t/a) - t/a}}{b},$$
are varied until two waves are obtained (we finally consider $a = 50$ and $b = 4$). Notice that $\log\beta(t; a,b)$ is clearly oscillatory and that is why we chose this functional shape. Finally, a cubic B-spline approximation of $\log\beta(t;50,4)$ is used, with $m = 12$ functions, which allows to obtain the B-spline coefficients $\boldsymbol{\beta}$ in \eqref{log_beta}.

The resulting parameters are: 
\begin{align*}
\mathbf{p}^{syn} = (\gamma = 0.1, \alpha = 0.5,S_0 &=2189128, E_0 = 10, I_0 =0, \phi^{-1} = 0.1, \beta_1 =-1.8699,\beta_2 =-1.3014,\\
\beta_3 =-0.2422, \beta_4 &=-1.5110, \beta_5 =-3.3045, \beta_6 =-3.0917, \beta_7 =-1.5683,\numberthis\label{synt_params}\\
\beta_8 =-1.5705, \beta_9  &=-3.4479, \beta_{10} =-4.5214,
\beta_{11} =-3.3348, \beta_{12} =-2.8091).
\end{align*}
The numerical solution of (\ref{SEMIKR_ODES}) for a SEI$_3$R model with parameters $\mathbf{p}^{syn}$, $\{C^{syn}(j)\}_{j=1}^{100}$, is used to yield
$$\widetilde{\mathbf{C}}^{syn}=\{\widetilde{C}^{syn}_j:\quad\text{draw from}\quad \text{NegativeBinomial}(C^{syn}(j), \phi^{-1} = 0.1)\}_{j=1}^{100}.$$
In this synthetic-data experiment, we set $\eta(t)=1$ for all $t$, so that no underestimation of incidence is introduced. Our purpose is to study the ability of the proposed workflow to recover a known time-dependent transmission rate from incidence data. The dataset $\widetilde{\mathbf{C}}^{syn}$ used in the synthetic experiments is shown as black dots in Web Figure \ref{synt_inc_fig}.

\subsection*{COVID-19 daily incidence data}
The COVID-19 daily incidence data used in this work were obtained from the Spanish National Epidemiological Center (CNE in Spanish). The data is available at the tab `Documentación y Datos' in this \href{{https://cnecovid.isciii.es/covid19/#documentación-y-datos}}{link}. Data for all Spanish Autonomous Communities is available on the accompanying \href{https://github.com/HristoInouzhe/BayesTemporalEpidemicDynamics}{GitHub repository}.

The data collection started on the 1st of January 2020 when the first positive case was detected in Madrid. As a starting point for the spread of the disease we consider the first day when a positive case is followed by at least one more positive case in the next 7 days. As we are interested in the Basque Country data, we selected the 10th of February 2020 as an initial moment of the pandemic and the 31st of January 2021 as an ending point of the data series in order to avoid the effect of vaccination. Vaccination in Spain started in early January 2021, and was relatively slow at the beginning, hence until the end of January it was almost negligible for our purposes. 
 
\section*{Case studies}\label{section_results_and_discussion}
The results of numerical experiments on the synthetic and real data, as well as the discussion of the behaviour of the developed methods, are provided in this section.

\subsection*{Case study 1: Synthetic data}\label{sec_cs1_synth_data}
Here we use the synthetic data introduced above in order to validate the proposed spline-based modelling approach and get a deeper insight into its behaviour. We not only examine the spline-based SEI$_3$R model, which closely matches the data-generating model but also investigate the behaviours of the SIR, SI$_3$R, and SEIR models. Furthermore, comparisons with two alternative methods are provided: one is a diffusion-based approach, retaining an epidemiologically informed dynamics, and the other is a purely stochastic process aimed at estimating the time-dependent reproduction number. We note that this synthetic experiment is intentionally favourable to the proposed modelling family, since the data-generating mechanism belongs to the same broad class of spline-based compartmental models. Its purpose is therefore not to prove general superiority over competing approaches, but to verify that the inference workflow can recover the main features of a known time-varying transmission process under controlled conditions. 

Following \cite{Dureau2013} and with the help of the \texttt{R}-package \texttt{rbi} \citep{rbi}, we implemented SIR/SEIR-like dynamic models with a time-dependent transmission rate modelled by a diffusion process (i.e., given by Stochastic ODE's (SODEs); see \ref{SODE_section} for details). The parameters estimation for these models is performed using a type of Particle Filter Monte Carlo, known as SMC$^2$ \citep{Chopin2013}. This provided us with a good baseline for investigating the behaviour of the proposed spline-based dynamic models SE$_M$I$_K$R (\ref{SEMIKR_ODES}) and SI$_K$R (\eqref{SEMIKR_ODES_2} in \ref{SI_K_R_mechanistic_model}) combined with (\ref{NegBinModel}) against the diffusion-based SE$_M$I$_K$R and SI$_K$R (see, respectively, \eqref{SODES} and \eqref{SODES_2} in \ref{SODE_section}) also combined with (\ref{NegBinModel}). The second comparison method is the popular EpiEstim \citep{Cori2013, thompson2019improved} which assumes that transmission follows a Poisson process such that an infection at time step $t-s$ generates a new infection at time step $t$ with a rate $\mathsf{R_t}w_s$, where $\mathsf{R_t}$ is the time-dependent reproduction number, constant over a time period $[t-\xi, t]$, and $w$ is a discrete probability distribution describing the average infectiousness profile after infection. Additional details on EpiEstim can be found in \ref{comparison_methods}.

To use GHMC along with the spline-based dynamic models introduced in this work we selected the following prior distributions:
\begin{align}
\label{priors_HMC}
\alpha &\sim \mathcal{N}(0.5,0.05^2), \quad \phi^{-1} \sim \operatorname{Exp}(20), \nonumber\\
(S_0, E_0, I_0, R_0)/N&\sim \operatorname{Dirichlet}(999993.424608, 4.575392, 1, 1),\\
\tau^2&\sim \operatorname{InvGamma}(1, 0.005),\nonumber
\end{align}
and fix $\gamma=0.1$, an epidemiologically plausible value for SARS-like viruses (see \citealp{cevik2021sars}; sensitivity to $\gamma$ is explored in Section \nameref{sec:sensitivity_study}). In addition, a prior distribution for the coefficients, $\boldsymbol{\beta}$, associated with the transmission rate was chosen as discussed in \nameref{bayesian_set-up} section (see Eqn. (\ref{prior_improper})). We chose cubic B-splines ($d = 3$), and set the number of basis functions to the same value as the true generator, i.e., $m = 12$ (the influence of the number of spline basis functions is studied in the sensitivity study provided below). All in all, the proposed priors either codify confidence in values, as for $\alpha$ and $(S_0, E_0, I_0, R_0)$ (for which a near-deterministic Dirichlet centred at $\mathbb E[E_0]=10$ is considered; see \ref{sup_sec:Initial_Ocuppancy}), force dispersion $\phi^{-1}$ to be close to 0, or provide  almost no prior knowledge on the transmission rate $\boldsymbol{\beta}$, except the smoothness conditions imposed by the B-spline basis and $\tau^2$. Notice that the prior on $\tau^2$ is a standard choice from the P-splines literature. For the comparison methods, we selected priors that allow for fair comparisons and also performed some manual tuning to get the best possible results. Details are provided in \ref{comparison_methods}. 

Regarding the four-stage workflow described in \nameref{implementation}, we proceeded as follows. For Stage 1 (MAP estimation) we used Stan's \texttt{optimize} routine with default settings \citep{carpenter2017stan} while adjusting the convergence tolerance for our setting (\texttt{tol\_rel\_obj} $= 0.5\times10^{-4}$, \texttt{tol\_obj} $= 0.1$). We ran 100 random initialisations with the spline coefficients initialised as $\beta_0 \sim \operatorname{Uniform}(-4,4)$ and retained the point with the highest log-posterior as the MAP estimate $\hat{\mathbf{p}}$. In Stage 2, we drew 100 perturbed candidates around $\hat{\mathbf{p}}$ with  $(\sigma_{\mathrm{prop}}, \mathrm{TV}_0) = (0.25, 10^{-4})$;  
the former placed about 65\% of draws within $\pm25\%$ of $\hat{\mathbf{p}}$, while the latter allowed at most $N\times\mathrm{TV}_{0}=219$ individuals to shift initial compartments. We then clustered these 100 candidates by their log-posterior values into $n_{\mathrm{chains}}=10$ groups and sampled one seed per cluster. In Stage 3, GHMC sampling was performed with initial integrator settings: step size $h=0.002$, number of integration steps per iteration 1, and momentum-refreshment parameter $\psi=0.5$. These served as inputs to the adaptive tuning algorithm of \ref{Tuning_HMC}, which refined all numerical parameters before production sampling. We ran 10 chains (20,000 warm-up, 100,000 draws) for each compartmental specification—SIR, SI$_3$R, SEIR, and SEI$_3$R. Models whose latent structure aligned with the data achieved split-$\hat R \le 1.05$ for the required parameters $\alpha,\ \phi^{-1},S_0,E_0,I_0$. In misspecified structures (e.g., omitting the exposed phase or coarsening infectious staging), the same fixed sampling budget was occasionally insufficient to meet this threshold, making the impact of structural misspecification on convergence explicit. We kept the budget constant across models to allow a like-for-like comparison.

\begin{figure*}[h!]
\begin{center}
\includegraphics[scale=0.27]{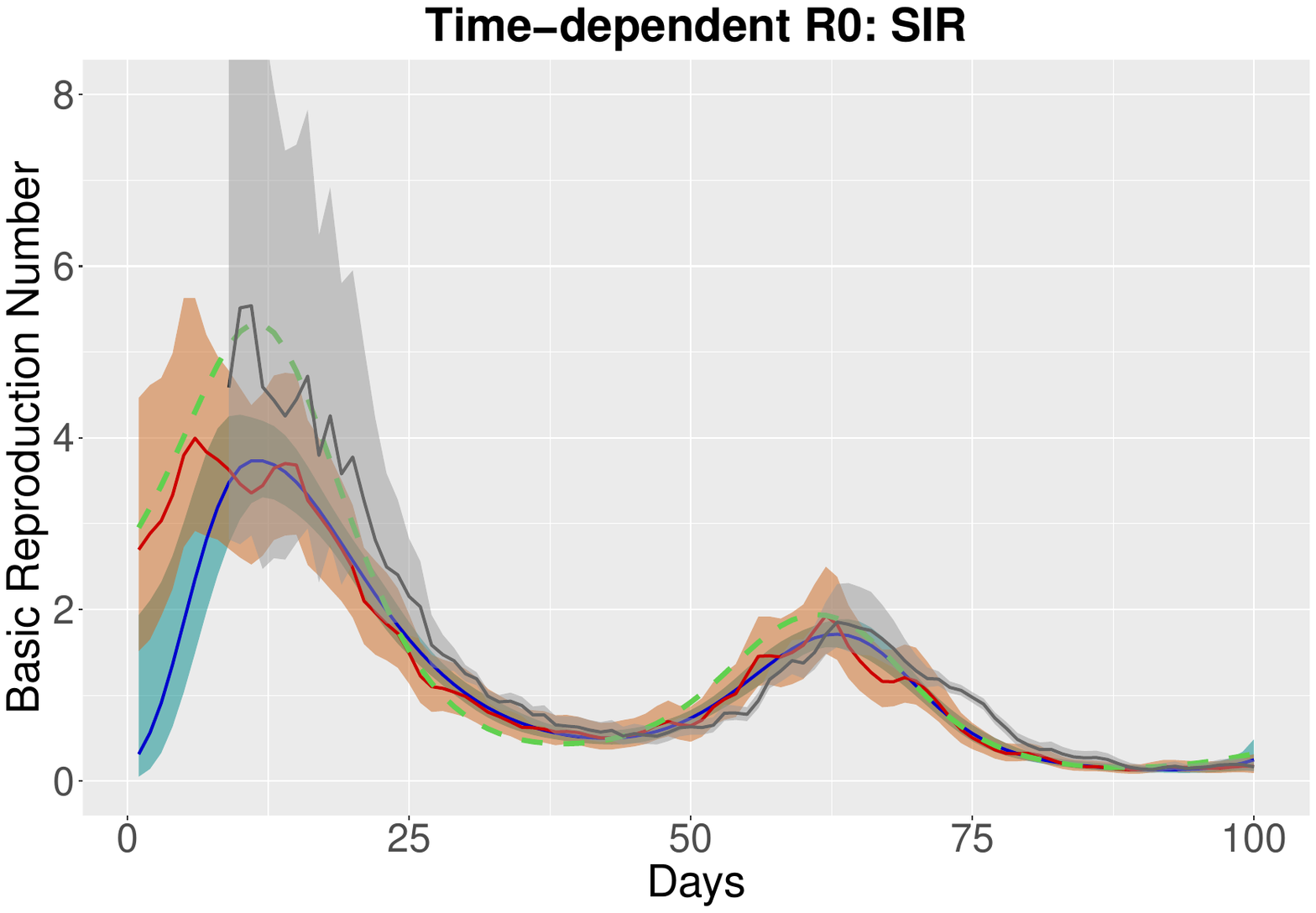}\includegraphics[scale=0.27]{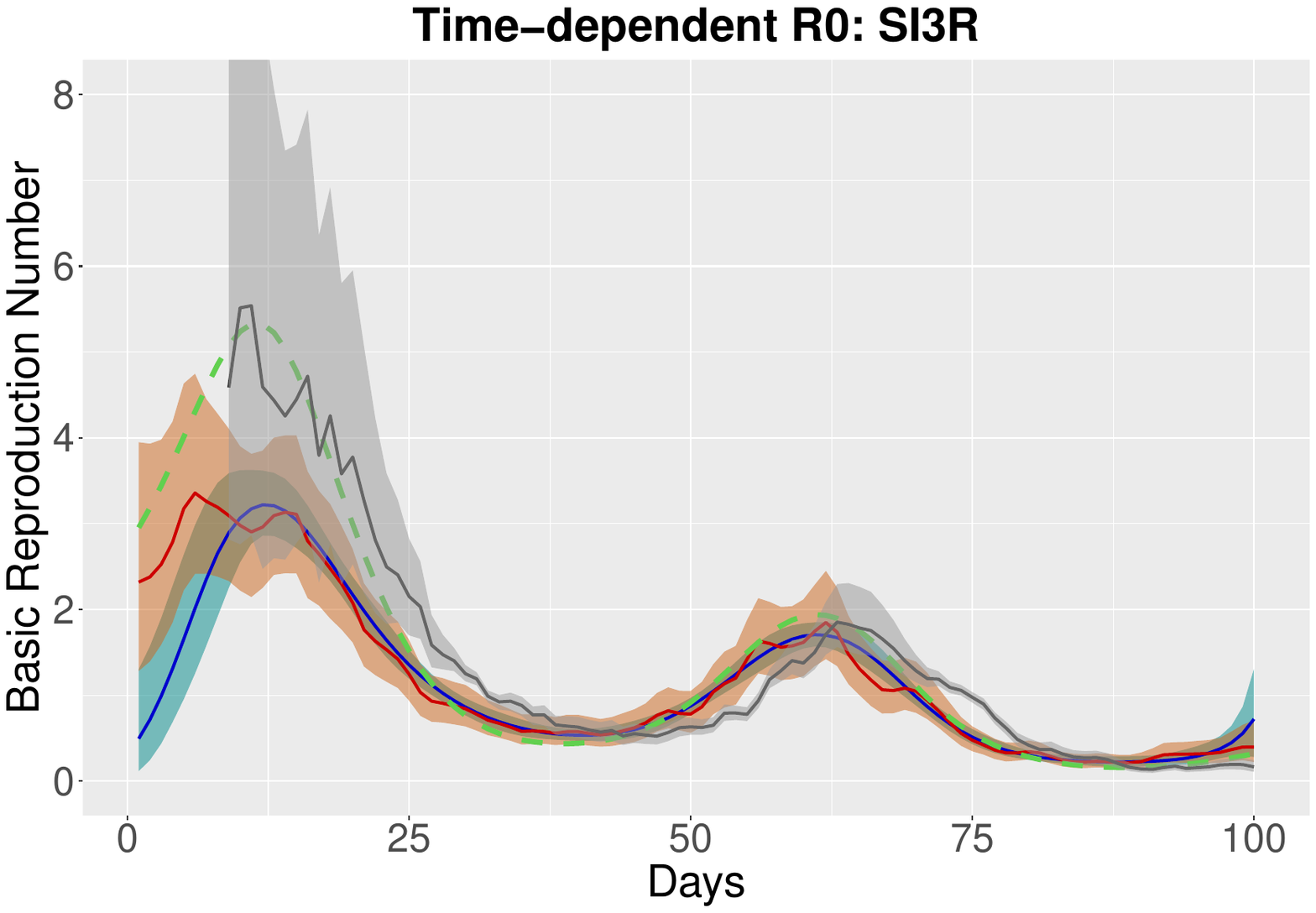}
\includegraphics[scale=0.27]{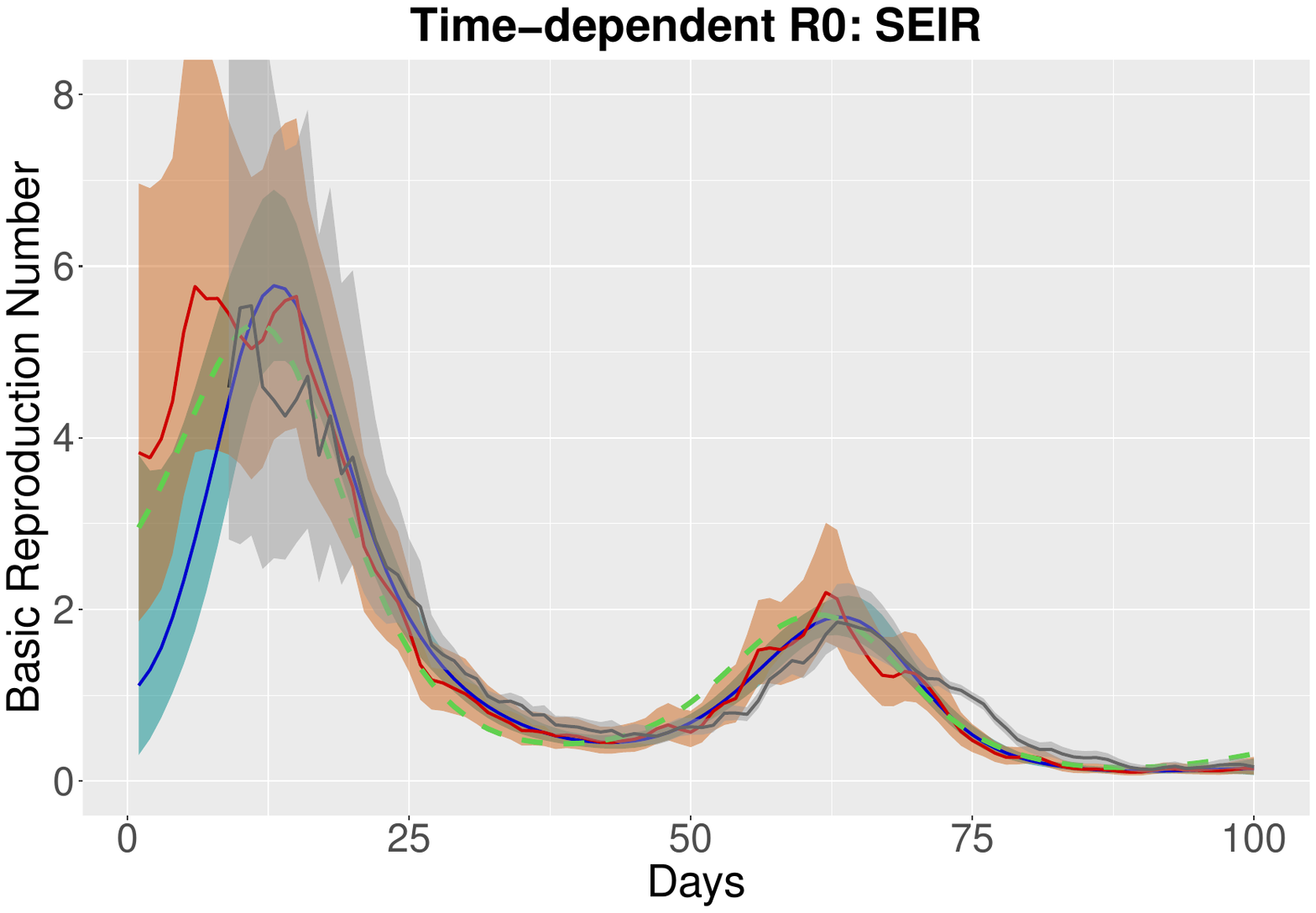}\includegraphics[scale=0.27]{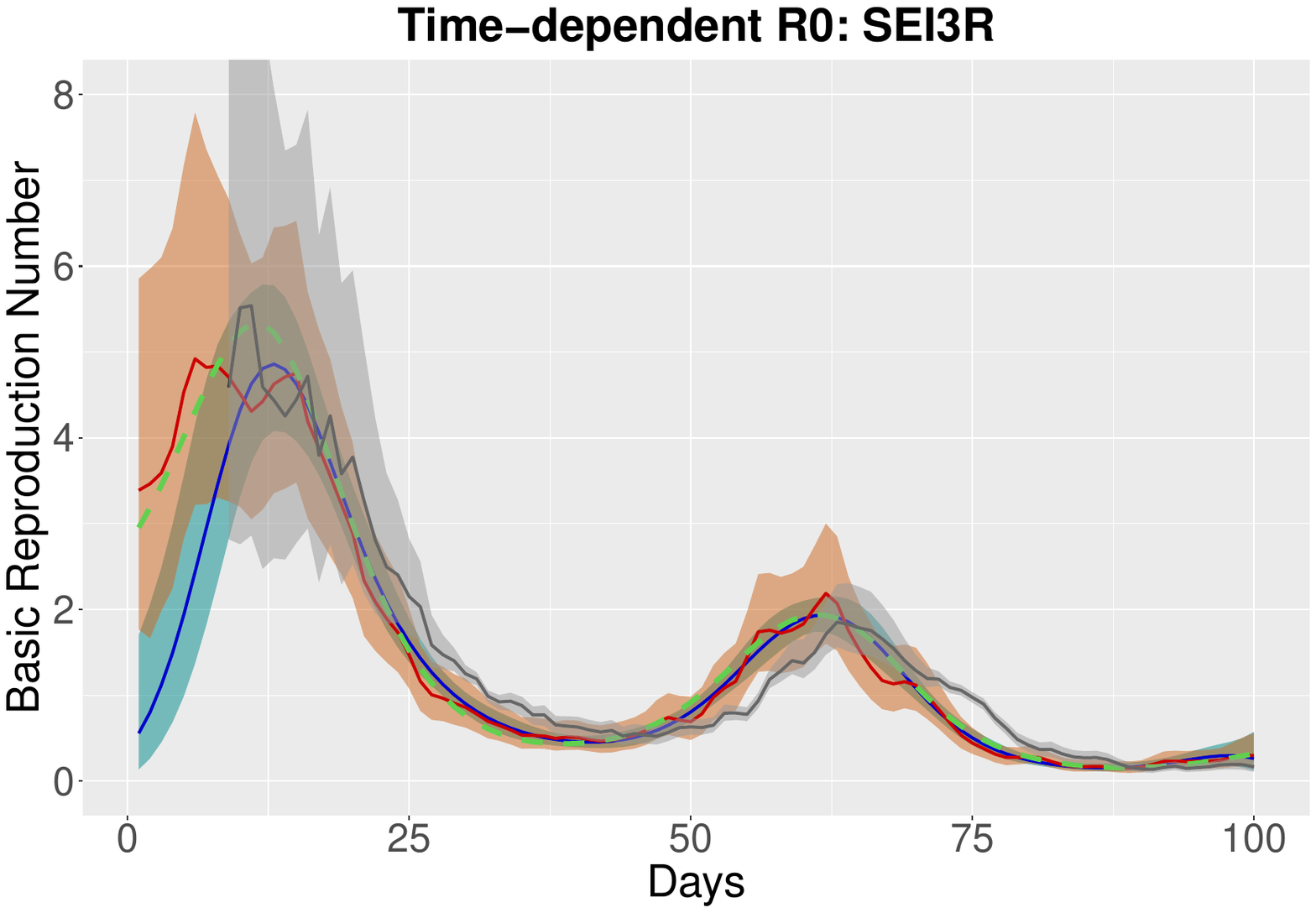}
\includegraphics[scale=0.61]{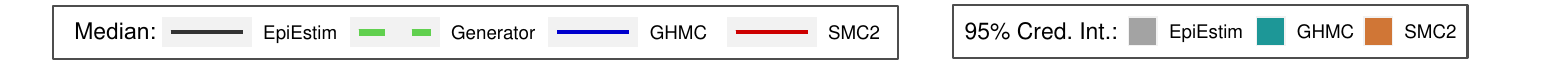}
\end{center}
\caption{For the synthetic data: posterior medians (solid lines) and 95\% credible intervals (shaded areas) of the time-dependent basic reproduction number, $\mathsf{R_0}(t) = \beta(t)/\gamma$, for a spline-based dynamics sampled with GHMC (combination of 10 chains with 100000 production steps), a diffusion-based dynamics sampled with SMC$^2$ (combination of 5 chains with 1000 particles and 1000 production steps) and EpiEstim. Dashed purple lines represent the starting values of the GHMC chains corresponding to the dispersed initialization procedure (result of Stage 2). Dashed green lines show the true generator $\mathbf{p}^{syn}$ (see \eqref{synt_params}).
}
\label{synt_R0_fig}	
\end{figure*}
In Figure \ref{synt_R0_fig}, the estimated time-dependent reproduction numbers for the three different methods are provided (more figures and details are provided in \ref{sec_additional_plots}). For the spline and diffusion-based models, we supply $\mathsf{R_0}(t)=\beta(t)/\gamma$ with the different models (SIR, SI$_3$R, SEIR, SEI$_3$R) and samplers. For EpiEstim, we provide $\mathsf{R_t}$ accounting not only for data noise but also for uncertainty in the infectiousness distribution $w$. Recall that the same result for EpiEstim is shown for all four plots. Regarding the spline and diffusion-based models, despite the clear similarities between SIR and SI$_3$R and also between SEIR and SEI$_3$R, significant differences between models with and without an exposed compartment are observed. For the 
models with an exposed compartment, SEIR and SE$_3$IR, the 95\% credible intervals (CIs) are very close to containing the true values (in dashed green), while this is not the case for the models without an exposed compartment. For some of the less conservative dispersion radii in our sensitivity study (see below) the SEI$_3$R's CIs fully contain the true generator. It is worth noticing that the posterior median for the time-dependent reproduction number (for all three methods), in 
solid lines, captures well the wave-like behaviour of the true generator. We see that the GHMC produces results comparable to those of the SMC$^2$ procedure. Posterior distributions of the epidemiologically relevant parameters and further transmission trajectories for the spline-based methods are available at \href{https://hristoinouzhe.github.io/BayesTemporalEpidemicDynamics/#synthetic}{the following link}; the corresponding plots for the diffusion-based methods can be found \ref{sec_additional_plots}. 

Among all specifications, the correctly specified SEI$_3$R model yields posterior medians that most closely match the true dynamics under both the spline-based and diffusion-based methods. Additionally, methods with an exposed compartment seem to converge better with our fixed budget and initialization procedure (see Web Table \ref{sup-table:r-hat_synthetic}). EpiEstim is also very competitive but produces a slight lag on the peak of the second wave. Notice that the first 7 days are not estimated since the time window $\xi$ is set to that amount. These results highlight the importance of the choice of model for recovering unobserved quantities such as the basic reproduction number. The appropriate compartmental structure, and hence, appropriate infectious time distribution, allows to recover the underlying transmission dynamics.

Noticeable differences between the compared approaches are the following. The trajectories obtained with GHMC  are smoother as a result of the choice in the mechanistic model of the transmission rate (\ref{log_beta}) which belongs to a space spanned by a B-spline basis. This regularity is not required when $\beta(t)$ is governed by a diffusion process or in $\mathsf{R_t}$ for EpiEstim. In our experiments, the Hamiltonian-based implementation was less sensitive to some prior choices than the SMC$^2$ implementation. For example, trial and error was needed to specify the prior on $\phi^{-1}$ for SMC$^2$, whereas the spline-based GHMC implementation remained stable under the exponential prior used in \eqref{priors_HMC}. This observation is empirical and specific to the present implementation, but it illustrates a practical advantage of having access to gradient information in this model class.

Finally, while all three approaches can encode a flexible time-varying reproduction number, the spline-based dynamic models have a smooth deterministic structure that is naturally compatible with gradient-based inference. This makes them a promising candidate for future extensions involving automatic differentiation, variational inference, amortised calibration, or neural components.

\subsubsection*{A sensitivity study}\label{sec:sensitivity_study}
Because the SE$_M$I$_K$R model is highly flexible, and its inference depends on a small number of tuning and prior choices, we conducted a sensitivity analysis with four targeted perturbations:
\begin{itemize}
    \item \textbf{Stage-2 dispersion radii.}
    We refitted the model with
    $$
    (\sigma_{\text{prop}},\mathrm{TV}_0)\in
    \bigl\{(0.50,10^{-4}),\;(0.75,10^{-4}),\;
           (0.50,10^{-3}),\;(0.75,10^{-3})\bigr\},
    $$
    probing how much the width of the MAP-centred seed cloud affects convergence and mixing. (For convenience we refer to these experiments as Sets 01-04)
    \item \textbf{Recovery-rate prior.}
    Keeping all other settings fixed, we lowered and raised the fixed value of $\gamma$ to $0.05$ and $0.20$, respectively, to quantify how strongly the posterior for $\beta(t)$ must compensate when the mean infectious period is mis-specified. (Set 05-06.)
    \item \textbf{Size of the B-spline basis.} To assess sensitivity to basis dimension, we held all other settings fixed and varied $m$ to $8$ (under-parameterised) and $16$ (over-parameterised) relative to the data-generating basis. (Set  07-08.)
    \item \textbf{Initial-state Dirichlet precision.} Keeping $a_3=a_4=1$, we reduced the total concentration to $a_0=10^{5}$ and set the prior mean to $\mathbb{E}[E_0]=100$ (i.e., $\mathbb{E}[E_0/N]=100/N$). This probed whether a weaker, higher-mean prior on the initially exposed population materially alters early-epidemic estimates. (Set 09.)
\end{itemize}
Sensitivity plots are available in the web gallery at \href{https://hristoinouzhe.github.io/BayesTemporalEpidemicDynamics/#sensitivity}{the following link}; the corresponding split–$\hat R$ values are reported in \ref{monte_carlo_details} (Web Table \ref{sup-table:r-hat_sensitivity}). In the following we describe the main takeaways. Increasing the Stage-2 dispersion radii produced seeds farther from the MAP and hence greater heterogeneity across chains. 
Across all but the largest setting, chains met our criterion \(\hat R<1.05\) for \(\alpha,\ \phi^{-1},S_0,E_0,I_0\); at \((\sigma_{\mathrm{prop}},\mathrm{TV}_0)=(0.75,10^{-3})\) two chains failed to converge (persistent divergences/poor mixing observable in the daily incidence posterior predictive checks), indicating that dispersion is too wide for the sampling setup. 
Misspecifying the recovery rate (\(\gamma=0.05\) or \(0.20\)) shifted both the height and timing of the transmission peak, underscoring the need for epidemiologically informed specification or strong prior information on \(\gamma\). 
Varying the B-spline basis size (e.g., \(m=8\) vs.\ \(m=16\)) yielded very similar \(\beta(t)\) and \(\mathsf{R_0}(t)\) trajectories, suggesting that once the basis is moderately rich, smoothness is governed primarily by \(\tau^2\) rather than \(m\). It is worth mentioning that with moderate increase of the dispersion radii or with the change in spline basis, the confidence intervals of the SEI$_3$R model did contain the true generator. Finally, increasing the prior mean for \(E_0\) while reducing the Dirichlet precision produced noticeable discrepancies in the earliest posterior predictive incidence, but these dissipated within about five days as the trajectories align with the data.

\subsubsection*{Identifiability and model-structure sensitivity}
We remark that four different model families, with transmission rates modelled in different ways, and sampled with different techniques, can generate almost identical data (see Web Figure \ref{synt_inc_fig}). This is a consequence of the fact that not only some parameters but also the model structure are not easily identifiable from incidence data \citep{souto_maior2019}. This should be kept in mind when drawing conclusions from a model and when selecting prior distributions.

Despite these difficulties, the proposed models and methodology were enough to retrieve plausible generators for the data and to capture the wave-like behaviour of the basic reproduction number and transmission-rate, which are not directly observed. Therefore, with the proposed methods one can recover meaningful information about the transmission evolution of a disease. Our synthetic experiments show that, not surprisingly, the choice of models and samplers has an influence in the obtained transmission evolution, and practitioners should be aware of that.

The previous results reinforce a central limitation of incidence-only epidemic inference: a good posterior predictive fit to daily incidence does not imply that the latent compartmental structure, the transmission rate, or the reproduction number have been uniquely identified. Consequently, posterior predictive checks must be complemented with sensitivity analyses, convergence diagnostics, and epidemiological judgement about plausible model structures and fixed parameters.

\subsection*{Case study 2: Basque Country data}
The methods proposed in this work are applied to daily incidence data for the Basque Country. A crucial task is to provide adequate prior distributions for the parameters of interest. The synthetic example of the previous section supports fixing $\gamma$, while still leaving a lot of freedom through the time-dependent transmission rate $\beta(t)$ (for more details see \nameref{bayesian_set-up} section and Equation (\ref{prior_improper}) therein). Given we model a coronavirus, there is strong prior evidence on the mean infectious period. We therefore fix the recovery rate at $\gamma=1/5\ \text{day}^{-1}$ (mean infectious duration $\approx 5$ days), in line with current estimates \citep{CDC_YellowBook_COVID19_2025, puhach2023sars}. Following our early-epidemic initialization (\ref{subsec:dirichlet}), we assume a very small initial exposed fraction, setting $\mathbb{E}[E_0/N]=100/N$, and fix the Dirichlet parameters to $a_3=a_4=1$ and $a_0=10^5$. Hence, our choice for priors is the following
$$\alpha \sim \mathcal{N}(0.5,0.05^{2}),\quad
    \phi^{-1}\sim\operatorname{Exp}(20),$$
\begin{equation}
    \frac{(S_0,E_0,I_0,R_0)}{N}\sim\operatorname{Dirichlet}(99993.424608,4.575392,1,1),
\end{equation}
$$\tau^{2}\sim\operatorname{Inv\!Gamma}(1,0.005).$$

The next step is to choose a proper value for $m=Q+d-1$, the number of B-spline basis functions. As for the synthetic data, we used cubic B-splines. To determine a suitable number of internal knots $Q$, we applied Bayesian P-spline regression to the Basque Country Incidence data for $Q\in\{10,11,\dots,24,25\}$ (see Supplementary Figure \ref{regression_fig}), and selected the one that provided the lowest widely applicable information criterion \citep[WAIC;][]{Gelman2014} (see Web Table \ref{sup-table:waic}).  In our case, the criterion chose $Q = 21$, and hence we set $m=21+3-1=23$. This step is relatively inexpensive in terms of computational power.

We treat the Basque Country incidence series as a demanding testbed with partially observed states. To ensure comparability across compartmental specifications, the computational budget was fixed to 10 parallel chains per model, with 20{,}000 warm-up iterations followed by 100{,}000 draws, and a conservative MAP-centred initialisation with dispersion radii \((\sigma_{\mathrm{prop}}, \mathrm{TV}_0)=(0.15,10^{-4})\). Generalised Hamiltonian Monte Carlo (GHMC) tuning followed the same procedure across all fits (\ref{Tuning_HMC}). The correction for incomplete detection $\eta(t)$ in (\ref{NegBinModel}) was set to: $\eta(t)=0.15$ for $0<t\leq 92$, $\eta(t)=0.15 + (0.54-0.15)(t-92)/(281-92)$ for $92<t<281$, $\eta(t)=0.54$ for $281\leq t$. It is a linear interpolation of the seroprevalence found in \cite{ENE_COVID_2020, Pollan2020}, which reflects that at the beginning of May (day 92) only 15\% and at mid November (day 281) only 54\% of positive cases were reported in the Basque Country. In our current implementation, end-to-end wall time for a single model at this budget is well above one day, which motivates reporting rules that prioritise converged behaviour in the main text and defer full diagnostics to the appendix.

We fit four compartmental structures to the same data: SIR, SI\(_3\)R, SEIR, and SEI\(_3\)R. Convergence is assessed as previously via split–\(\hat R\) with the requirement \(\hat R<1.05\) for the epidemiologically interpretable parameters \(\alpha\), \(\phi^{-1}\), and \((S_0,E_0,I_0)\). For the high-dimensional spline coefficients \(\boldsymbol\beta\) we check functional agreement of \(\beta(t)\) and derived \(\mathsf{R_0}(t)\) across chains. To avoid mixing converged and non-converged trajectories in summaries, the main text pools only the chains that meet the convergence criterion; more details are provided in~\ref{sensitivity_analysis}.

Under the fixed computational budget and conservative initialisation, only the SEIR specification achieved the convergence criterion for all ten chains (Web Table \ref{sup-table:r-hat_basque}). Therefore, the main quantitative interpretation of the Basque Country data in the manuscript is based on the SEIR fit. The remaining specifications are still informative as diagnostic and sensitivity analyses: they show that several compartmental structures can reproduce the observed incidence, but they also reveal that richer or misspecified latent structures may lead to weakly identified directions and incomplete convergence under the same computational budget.
For example, for SEI$_3$R, the evidence points to a shallow multimodality/identifiability ridge: several chains agree on both the median $\mathsf{R_0}(t)$ and its credible bands; others share a similar median but display wider uncertainty; and a single chain shows a noticeably shifted median while retaining credible intervals comparable to some of the others. All of these trajectories fit the incidence well, suggesting that alternative $(\alpha,\beta(t))$ trade-offs can achieve similar likelihoods yet imply subtly different transmission dynamics. Posterior plots for both parameters and transmission dynamics for all models are available at \href{https://hristoinouzhe.github.io/BayesTemporalEpidemicDynamics/#basque}{the following web gallery}.

\begin{figure}[h!]
    \centering
    \includegraphics[width=1\linewidth]{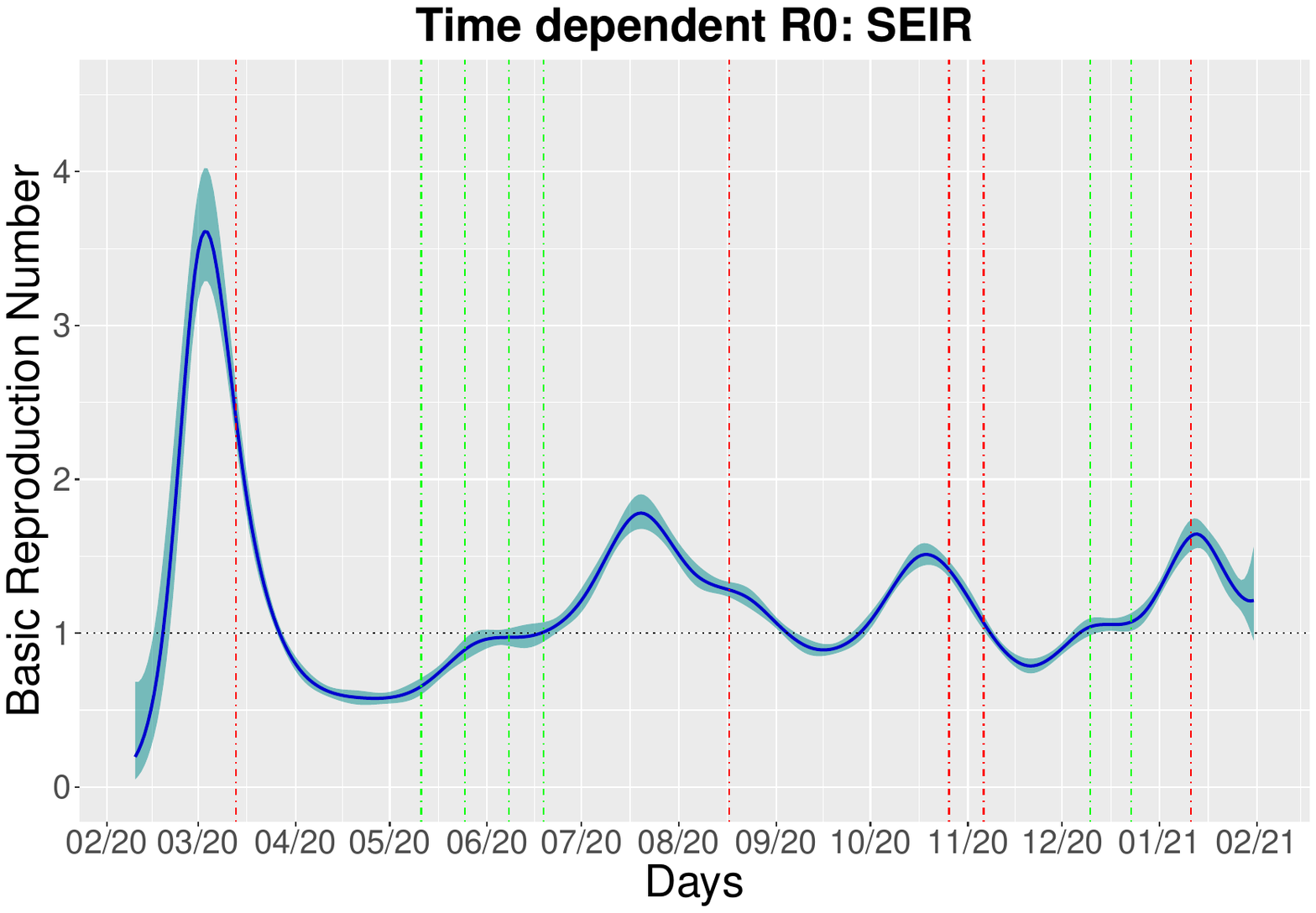}
    \caption{Posterior median (solid line) and 95\% credible intervals (shaded area) of the time-dependent basic reproduction number, $\mathsf{R_0}(t)=\beta(t)/\gamma$, for the SEIR spline-based dynamic model  (\ref{SEMIKR_ODES}) and (\ref{NegBinModel}) sampled with GHMC. Results are shown for a combination of 10 chains with 100000 production steps.}
    \label{fig:R0_SEIR_basque}
\end{figure}

Pooling the ten converged SEIR chains yields stable estimates of $\mathsf{R_0}(t)$ (Figure \ref{fig:R0_SEIR_basque}). Posterior predictive checks closely track the observed incidence (see “Posterior Predictive Daily Incidence” in the \href{https://hristoinouzhe.github.io/BayesTemporalEpidemicDynamics/#basque}{web gallery}).
Although the other specifications did not achieve full convergence under this budget, most retained chains still produce accurate posterior-predictive incidence. We therefore use them only as exploratory evidence that the incidence data alone do not uniquely determine the latent compartmental structure or the associated transmission dynamics.

A crucial practical concern is how non-pharmacological measures correlate with the transmission dynamics of a disease. In the time-dependent setting of this work this information is codified in the observed changes in the basic reproduction number.  In Figure \ref{fig:R0_SEIR_basque} the posterior median and 95\% credible intervals for the basic reproduction number ($\mathsf{R_0}$) for the SEIR model is plotted, alongside the different measures taken in the Basque Country, such as lockdowns and emergency protocols (in dot-dashed red), as well as the lifting and easing of these extreme measures (in dot-dashed green). Recall that $\mathsf{R_0}$ is a good indicator of the state of the disease transmission, with values below 1 indicating the dying out of the transmission and values above 1 implying the further propagation. The higher the value of $\mathsf{R_0}$ the more individuals are infected on average by a single infectious individual. What we see in Figure \ref{fig:R0_SEIR_basque} is that there were three time intervals where median $\mathsf{R_0}$ went consistently below 1 (a dotted grey line). One was during the state wise lockdown that started in the middle of March 2020, another some time after a sanitary emergency was declared in the middle of August 2020, and the final one during the lockdown of the Basque Country commenced at the end of October 2020. We also see that during restrictive measures there is a decrease in transmission, whereas the ease of restriction comes together with an increase in transmission. Additionally, after applying the correction for incomplete detection, one sees that the highest transmission occurred in the first wave, and the next waves had lower transmission rates, though high enough to keep the pandemic going on. Finally, restrictions in the fourth wave seem to precede or coincide with the peak of the transmission rate, while the restrictive measures in the first three waves were likely taken after the peaks in transmission. We think that one should be extremely cautious when trying to interpret this last observation as a measure of the efficacy of non-pharmacological actions, and we abstain from doing so. Thus, these plots should be read as descriptive temporal associations between inferred transmission dynamics and public-health measures, not as estimates of the causal effect of those measures.

\section*{Discussion and conclusions} \label{sec_Conclusions}
In this work, we implemented a flexible Bayesian framework for inferring time-dependent transmission rates in compartmental epidemic models. The framework combines established ingredients such as mechanistic SI$_K$R/SE$_M$I$_K$R structures, Bayesian P-splines, Negative Binomial observation models, and Hamiltonian-based sampling into a single computational workflow. The contribution is therefore not the invention of these components in isolation, but their integration into a reproducible framework for studying time-varying epidemic transmission under partial observation.
This Bayesian formulation allows us to quantify uncertainty in both the model parameters and the unobserved transmission process, and was implemented in a user-friendly tool based on \texttt{C} and \texttt{R} \citep{R23}, available as \href{https://github.com/HristoInouzhe/BayesTemporalEpidemicDynamics}{a GitHub repository}.
The synthetic case study and the comparison with established alternative methods show that our procedure can produce results comparable to those of diffusion-based approaches and EpiEstim in a controlled setting, while providing a smooth mechanistic representation of the transmission process.

Experiments on COVID-19 daily incidence data from the Basque Country show that our methods can capture important temporal variation in the transmission dynamics. We emphasise that we are not promoting a single best model that captures the information in a dataset. On the contrary we are offering a broad family of models that can be used for that purpose. Our findings indicate that while structurally different models can yield similarly good fits to incidence data, they may lead to distinct inferences for key epidemiological quantities, such as the basic reproduction number. This underscores the importance of considering multiple plausible model structures and prior choices in parallel, as focusing solely on fit quality may obscure model misspecification. The final model choice should be guided by expert judgement and sensitivity analyses.

Finally, we note that reliable inference in this setting required a careful parameter initialisation strategy. This was observed both in our GHMC implementation and in the Stan implementation used for comparison, which relied on the No-U-Turn Sampler (NUTS), a widely used adaptive variant within the HMC family \citep{Hoffman2014,carpenter2017stan}. This suggests that these difficulties are not specific to GHMC, but are instead related to the complex posterior geometry resulting from flexible transmission rates combined with partially observed dynamics. Our use of GHMC was motivated by the fact that the proposed workflow was developed around the semi-analytical gradient information obtained from the ODE sensitivity equations, together with the sampling and tuning strategy described in the Methods. In both implementations, convergence and efficiency improved markedly when the sampling chains were initialised near high-probability regions, identified through optimisation. Thus, our use of GHMC should not be read as a claim that Stan/NUTS is unsuitable for this class of models. While a full comparison lies beyond the scope of this work, we provide the Stan code used in our experiments in the same \href{https://github.com/HristoInouzhe/BayesTemporalEpidemicDynamics}{GitHub repository} and plan to report further comparisons in future work. Some information on comparative  performance of Stan/NUTS and adaptively tuned GHMC can be found in \cite{AKHMATSKAYA2026116892} 


The methods presented can be generalised to situations that have not been discussed in this work. To begin with, it is conceptually straightforward, although technically involved, to produce predictions for the future transmission of a disease of interest. Models more suitable for a particular modelling purpose can easily replace the Negative Binomial model presented in (\ref{NegBinModel}). Another possible generalisation is to use SEIR-like models with mortality and hospitalisation where the time-dependent mortality and hospitalisation rates are modelled using (Bayesian) B-splines. In that situation, computing the semi-analytical form of the gradients may be more involved, but it is completely feasible. Furthermore, while our current work assumes homogeneous mixing across the entire population under investigation, there is potential to expand our methods to incorporate age stratification with inhomogeneous mixing, thereby addressing age assortativity within the population.

Finally, a notable extension arises from the good differentiability properties inherent in the considered models, facilitating the computation of the Hessian of the log-posterior (see (\ref{extended_ODE_2}) in \ref{Sensitivity_dynamic}) in a manner akin to gradient calculation. This feature holds promise for extending HMC and applying the models in other Machine Learning procedures \citep{Radivojevic2020}. Although our current approach relies on semi-analytical derivatives, which may complicate certain generalisations, the aforementioned favourable differentiability properties and the current Stan implementation pave the way for leveraging automatic differentiation to achieve greater flexibility. This prospect represents future work that is poised to significantly alleviate the burden on the user.

\clearpage
\section*{Supporting Information}
\addcontentsline{toc}{section}{Supporting Information}
\setcounter{section}{0}
\setcounter{subsection}{0}
\setcounter{equation}{0}
\setcounter{figure}{0}
\setcounter{table}{0}
\renewcommand{\figurename}{Web Figure}
\renewcommand{\thefigure}{S\arabic{figure}}
\renewcommand{\tablename}{Web Table}
\renewcommand{\thesection}{Web Appendix \Alph{section}}
\renewcommand{\theequation}{\Alph{section}\arabic{equation}}
\renewcommand{\theHsection}{supp.\arabic{section}}
\renewcommand{\theHsubsection}{supp.\arabic{section}.\arabic{subsection}}
\renewcommand{\theHequation}{supp.\arabic{section}.\arabic{equation}}
\renewcommand{\theHfigure}{supp.\arabic{figure}}
\renewcommand{\theHtable}{supp.\arabic{table}}

\section{SE$_M$I$_K$R model gradients and sensitivity dynamics}\label{Sensitivity_dynamic}\label{SEIR_Sensitivity_dynamic}

For ease of notation and brevity, we make the following definition
\begin{align*}
\mathbf{y}(t)&=(y_1(t), \dots y_{M+K+3}(t))'=(S(t), E_1(t),\dots E_M, I_1(t),\dots, I_{K}(t),\allowbreak R(t), C_I(t))'.
\end{align*}
where the prime notation $'$ denotes the transpose operation. With this, a condensed form of Eqn. \eqref{SEMIKR_ODES} in the main text is given by
\begin{align}
\label{SEMIKR_ODES_condensed}
\dot{\mathbf{y}}(t)=\frac{d\mathbf{y}(t)}{dt}=\mathbf{f}(t,\mathbf{y},\mathbf{p}), \quad	\mathbf{y}(t_0)=(S_0, E_0,0, \dots,0,I_0,0,\dots,0, R_0,  N - S_0)',
\end{align}
where $\mathbf{f}(t)$ is the vector formed by the right hand side of the equalities in the ODE system (see Eqn. \eqref{SEMIKR_ODES} in the main text) and $\mathbf{p}$ are the parameters of the model (see Eqn. \eqref{parameters} in the main text)
\subsection{Log-likelihood gradient}\label{sec:loglike-grad}
For HMC, the ability to effectively compute the gradient of (\ref{log_post}) (in the main text) is vital. This gradient is supplied here, where a key notion is sensitivity, $\mathbf{s}_i(t) = \frac{\partial\mathbf{y}(t)}{\partial p_i}$, which broadly speaking measures the dependence of a compartment on the parameters of the model. More specifically,
\begin{align*}
\mathbf{s}_i(t)=\frac{\partial\mathbf{y}(t)}{\partial p_i}&=(\partial_{p_i}y_1(t),\dots, \partial_{p_i}y_{M+K+3}(t))'=\left(s_{i,1}(t), \dots, s_{i, M + K+3}(t)\right),
\end{align*}
which obeys the following equations
\begin{align}
\label{Sens_condensed}
\dot{\mathbf{s}}_i(t) = \frac{d\mathbf{s}_i(t)}{dt} = \frac{\partial \mathbf{f}(t)}{\partial \mathbf{y}}\mathbf{s}_i(t) + \frac{\partial \mathbf{f}(t)}{\partial p_i},\quad\text{with initial condition} \quad \mathbf{s}_i(t_0),
\end{align}
that come from the interchange property of partial and ordinary differentiation and the chain rule applied to (\ref{SEMIKR_ODES_condensed}).

 In our setting, all non-analytically computable elements (i.e., the solution of the ODE system and its sensitivities) can be obtained by numerically solving the extended ODE system 
\begin{align}
\label{extended_ODE}
\dot{\mathbf{y}}(t)&=\mathbf{f}(t,\mathbf{y},\mathbf{p}),\nonumber\\
\mathbf{y}(t_0)&=(p_1,p_2,0, \dots,0, p_3,0\dots, 0, N - (p_1 + p_2 + p_3), N - p_1)',\nonumber\\
\dot{\mathbf{s}}_i(t) &=  \frac{\partial \mathbf{f}(t)}{\partial \mathbf{y}}\mathbf{s}_i(t) + \frac{\partial \mathbf{f}(t)}{\partial p_i},\\
\mathbf{s}_i(t_0)&=(\delta_{i1},\delta_{i2},0,\dots,0, \delta_{i3}, 0,\dots,0, -(\delta_{i1}+\delta_{i2}+\delta_{i3}),-\delta_{i1})',\nonumber
\end{align}
where $\delta_{ij}$ is the usual notation for the Kronecker delta, i.e., $\delta_{ij} = 1$ if $i=j$ and 0 otherwise. Analytic expressions for the Jacobian matrices $\frac{\partial \mathbf{f}(t)}{\partial \mathbf{y}}$ and $\frac{\partial \mathbf{f}(t)}{\partial \mathbf{p}}$ are provided below. Note that (\ref{extended_ODE}) is a system of $(M + K + 3) + (M + K +3)\times (m + 4)$ ODEs, where $M + K + 3$ is the number of compartments, and $m+4$ is the number of parameters involved in the ODEs.
We stress that $\mathbf{s}_4(t)=(0,\dots,0)'$ and $\mathbf{s}_5(t)=(0,\dots,0)'$ for any $t\geq 0$ since $p_4 =\phi^{-1}$ and $p_5 = \tau^2$ do not appear in the ODEs or the initial conditions and, hence, there is no need to incorporate the sensitivity for $p_4$ and $p_5$ in the system of ODEs. This results in only $m+4$ parameters required to solve the ODEs which are solely related to the dynamic system. To solve (\ref{extended_ODE}), an ODE solver with forward sensitivity capabilities is needed. In our implementation we use CVODES from the SUNDIALS suite \cite[see section 2.6 in ][]{CVODES_2020}.

The gradient of the log-posterior, using the chain rule, takes the following form
\begin{align*}
\frac{d\ell}{dp_i} &= \sum_{j=1}^n\frac{\partial\ell^j_{like}}{\partial C(t_0 + j)}\frac{\partial C(t_0 + j)}{\partial p_i} + \frac{\partial \ell^{j}_{like}}{\partial p_i} +\frac{\partial \ell_{prior}}{\partial p_i}\\
&= \sum_{j=1}^n\left(\frac{\partial\ell^j_{like}}{\partial C(t_0 + j)}\left(\frac{\partial C_I(t_0 + j)}{\partial p_i} - \frac{\partial C_I(t_0 + j-1)}{\partial p_i}\right) + \frac{\partial \ell^j_{like}}{\partial p_i}\right) +\frac{\partial \ell_{prior}}{\partial p_i}.
\end{align*}
The term inside the inner parentheses is related with the notion of sensitivity.
With a bit of algebra and using the notion of sensitivities one obtains
\begin{align}
\label{log_gradient}
\frac{d\ell}{dp_i}=\begin{cases}
\sum_{j=1}^n\frac{\partial\ell_{like}^j}{\partial C(t_0 + j)}\left(s_{i,M + K + 3}(t_0 + j) - s_{i,M + K + 3}(t_0 + j - 1)\right) + \frac{\partial \ell_{prior}}{\partial p_i} & i=0,1,2,3,6,\dots,m + 5\\
\sum_{j=1}^n\frac{\partial \ell^j_{like}}{\partial p_i} + \frac{\partial \ell_{prior}}{\partial p_i} & i = 4\\
\frac{\partial \ell_{prior}}{\partial p_i} & i = 5
\end{cases}
\end{align}
and
\begin{align*}
\frac{\partial \ell_{like}^j}{\partial C(t_0 + j)} &= \frac{\eta(t_0 + j) \tilde{C}_{t_0 + j}}{\eta(t_0 + j) C(t_0 + j)} - \frac{\left(\eta(t_0 + j)\tilde{C}_{t_0 + j} + \phi\eta(t_0 + j)\right)}{\eta(t_0 + j) C(t_0 + j) + \phi}\\
&=\frac{\phi\left( \tilde{C}_{t_0 + j} -\eta(t_0 + j) C(t_0 + j)\right)}{C(t_0 + j)\left(\eta(t_0 + j) C(t_0 + j) + \phi\right)},\\
\frac{\partial \ell_{like}^j}{\partial p_i} &=-\delta_{i4}\Bigg(\psi\left(\tilde{C}_{t_0 + j} + \phi\right) - \psi(\phi) + \frac{\eta(t_0 + j)C(t_0 + j) - \tilde{C}_{t_0 + j}}{\eta(t_0 + j)C(t_0 + j) + \phi}+\log(\phi) - \\
&\quad \log(\eta(t_0 + j)C(t_0 + j)+\phi)\Bigg)\phi^2.
\end{align*}
Here $\psi(x)=\frac{d}{dx}\log(\Gamma(x))=\frac{\Gamma'(x)}{\Gamma(x)}$ denotes the digamma function. Once priors are chosen, computing $\frac{\partial \ell_{prior}}{\partial p_i}$ is also analytical.

\subsection{Jacobian matrices}\label{sec:jacobian}
Differentiating the right hand sides of system (\ref{SEMIKR_ODES}) in the main text, one obtains the following Jacobian matrix:
\begin{footnotesize}
\begin{align*}
\left[\frac{\partial \mathbf{f}(t)}{\partial \mathbf{y}}\right]_{1,1} &= -\exp\left(\sum_{i=1}^{m}\beta_iB_{i}(t)\right) \frac{1}{N}\sum_{j=M +2}^{M +K + 1}y_j(t),\left[\frac{\partial \mathbf{f}(t)}{\partial \mathbf{y}}\right]_{2,1} =- \left[\frac{\partial \mathbf{f}(t)}{\partial \mathbf{y}}\right]_{1,1}, \left[\frac{\partial \mathbf{f}(t)}{\partial \mathbf{y}}\right]_{M+K+3,1} =- \left[\frac{\partial \mathbf{f}(t)}{\partial \mathbf{y}}\right]_{1,1},\\
\left[\frac{\partial \mathbf{f}(t)}{\partial \mathbf{y}}\right]_{j,1} &= 0,\quad 2<j<M + K + 3,\\
\left[\frac{\partial \mathbf{f}(t)}{\partial \mathbf{y}}\right]_{i,i} &= -M\alpha, \left[\frac{\partial \mathbf{f}(t)}{\partial \mathbf{y}}\right]_{i+1,i} = M\alpha, \left[\frac{\partial \mathbf{f}(t)}{\partial \mathbf{y}}\right]_{j,i} = 0,\quad j\neq i,i+1,\quad i = 2,\dots, M+1,\\
\left[\frac{\partial \mathbf{f}(t)}{\partial \mathbf{y}}\right]_{1,i} &= -\exp\left(\sum_{i=1}^{m}\beta_iB_{i}(t)\right) \frac{y_1(t)}{N}, \left[\frac{\partial \mathbf{f}(t)}{\partial \mathbf{y}}\right]_{2,i} = -\left[\frac{\partial \mathbf{f}(t)}{\partial \mathbf{y}}\right]_{1,i}, \left[\frac{\partial \mathbf{f}(t)}{\partial \mathbf{y}}\right]_{M +K+3,i} = -\left[\frac{\partial \mathbf{f}(t)}{\partial \mathbf{y}}\right]_{1,i},\\
\left[\frac{\partial \mathbf{f}(t)}{\partial \mathbf{y}}\right]_{i,i} &= -K\gamma,  \left[\frac{\partial \mathbf{f}(t)}{\partial \mathbf{y}}\right]_{i+1,i} = K\gamma, \left[\frac{\partial \mathbf{f}(t)}{\partial \mathbf{y}}\right]_{j,i} = 0,\quad j\neq 1,2,i,i+1, M+K+3\quad i = M+2, \dots, M+K+1,\\
\left[\frac{\partial\mathbf{f}(t)}{\partial \mathbf{y}}\right]_{j,i} &= 0,\quad j = 1, \dots,  M+K+3,\quad i = M+K+2, M+K+3.
\end{align*}
\end{footnotesize}
Additionally, differentiating with respect to the parameters yields
\begin{align*}
\frac{\partial \mathbf{f}(t)}{\partial p_0} &= \left(0, -My_2(t), My_2(t) - My_3(t),\dots, My_{M}(t)-My_{M+1}(t),My_{M+1}(t),0,\dots,0\right),\\
\frac{\partial \mathbf{f}(t)}{\partial p_1} &=\cdots =\frac{\partial \mathbf{f}(t)}{\partial p_5} = 0,\\
\frac{\partial \mathbf{f}(t)}{\partial p_i} &= \Bigg(-\exp\left(\sum_{i=1}^{m}\beta_iB_{i}(t)\right) B_{i-5}(t)\frac{y_1(t)\sum_{j=3}^{K + 2}y_j(t)}{N},\\
&\exp\left(\sum_{i=1}^{m}\beta_iB_{i}(t)\right) B_{i-5}(t)\frac{y_1(t)\sum_{j=3}^{K + 2}y_j(t)}{N},0,\dots,0,\\
&\exp\left(\sum_{i=1}^{m}\beta_iB_{i}(t)\right) B_{i-5}(t)\frac{y_1(t)\sum_{j=3}^{K + 2}y_j(t)}{N}\Bigg)'\quad i=6,\dots, m+5.\\
\end{align*}
To compute the Hessian, a second differentiation leads to
\begin{align*}
\frac{d^2\ell}{dp_kdp_i}&= \sum_{j=1}^n\Bigg(\frac{\partial^2\ell^j_{like}}{\partial C^2(t_0 + j)}\frac{\partial C(t_0 + j)}{\partial p_k}\frac{\partial C(t_0 + j)}{\partial p_i}  + \frac{\partial\ell^j_{like}}{\partial C(t_0 + j)}\frac{\partial^2 C(t_0 + j)}{\partial p_k\partial p_i} + \frac{\partial^2 \ell_{like}^j}{\partial p_k\partial p_i}\Bigg) +\frac{\partial^2 \ell_{prior}}{\partial p_k\partial p_i}\\
&= \sum_{j=1}^n\Bigg(\frac{\partial^2\ell^j_{like}}{\partial C^2(t_0 + j)}\Bigg(\frac{\partial C_I(t_0 + j)}{\partial p_k} - \frac{\partial C_I(t_0 + j-1)}{\partial p_k}\Bigg)\Bigg(\frac{\partial C_I(t_0 + j)}{\partial p_i} - \frac{\partial C_I(t_0 + j-1)}{\partial p_i}\Bigg)\\
&+ \frac{\partial\ell^j_{like}}{\partial C(t_0 + j)}\Bigg(\frac{\partial^2 C_I(t_0 + j)}{\partial p_k\partial p_i} - \frac{\partial^2 C_I(t_0 + j-1)}{\partial p_k\partial p_i}\Bigg) + \frac{\partial^2 \ell^j_{like}}{\partial p_k\partial p_i}\Bigg) +\frac{\partial^2 \ell_{prior}}{\partial p_k\partial p_i}.
\end{align*}
Defining the second order sensibilities as
\begin{align*}
\mathbf{s}^k_i(t)=\frac{\partial^2\mathbf{y}(t)}{\partial p_k\partial p_i}&=\left(\partial^2_{p_k,p_i}y_1(t),\dots, \partial^2_{p_k,p_i}y_{M + K+1}(t)\right)'=\left(s^k_{i,1}(t), \dots, s^k_{i, M + K+1}(t)\right),\\
\end{align*}
one gets the following ODEs
\begin{align*}
\dot{\mathbf{s}}^k_{i,j}(t) &=  \frac{\partial^2 \mathbf{f}_j(t)}{\partial \mathbf{y}^2}\mathbf{s}_k(t)\mathbf{s}_i(t) + \frac{\partial \mathbf{f}_j(t)}{\partial \mathbf{y}}\mathbf{s}^k_i(t)+\frac{\partial^2 \mathbf{f}_j(t)}{\partial p_k\partial p_i},\quad\mathbf{s}^k_i(t_0)=\mathbf{0}.
\end{align*}
Hence the Hessian can be computed by solving the extended ODE system
\begin{align}
\label{extended_ODE_2}
\dot{\mathbf{y}}(t)&=\mathbf{f}(t,\mathbf{y},\mathbf{p}), \quad	\mathbf{y}(t_0)=(N-p_2,p_2,0, \dots,0, p_2)',\nonumber\\
\dot{\mathbf{s}}_i(t) &=  \frac{\partial \mathbf{f}(t)}{\partial \mathbf{y}}\mathbf{s}_i(t) + \frac{\partial \mathbf{f}(t)}{\partial p_i},\quad\mathbf{s}_i(t_0)=(-\delta_{i2},\delta_{i2},0,\dots,0,\delta_{i2})',\\
\dot{\mathbf{s}}^k_{i,j}(t) &=  \frac{\partial^2 \mathbf{f}_j(t)}{\partial \mathbf{y}^2}\mathbf{s}_k(t)\mathbf{s}_i(t) + \frac{\partial \mathbf{f}_j(t)}{\partial \mathbf{y}}\mathbf{s}^k_i(t)+\frac{\partial^2 \mathbf{f}_j(t)}{\partial p_k\partial p_i},\quad\mathbf{s}^k_{i,j}(t_0)=0, j = 1,\dots, M+K+1.\nonumber
\end{align}
\section{\texorpdfstring{SI$_K$R model}{SIKR model}}\label{SI_K_R_model}
\subsection{The mechanistic model}\label{SI_K_R_mechanistic_model}
The dynamics for the SI$_K$R model can be described by the following system of ODEs:
\begin{align*}
\label{SEMIKR_ODES_2}
\frac{dS(t)}{dt}&=-\beta(t) S(t)\frac{I(t)}{N},\\
\frac{dI_1(t)}{dt}&=\beta(t) S(t)\frac{I(t)}{N}-K\gamma I_1(t),\\
\frac{dI_2(t)}{dt}&=K\gamma I_1(t)-K\gamma I_2(t),\\
&\dots, \numberthis\\
\frac{dI_K(t)}{dt}&=K\gamma I_{K-1}(t)-K\gamma I_K(t),\\
\frac{dR(t)}{dt}& = \frac{d}{dt}\left(N - S(t) - I(t)\right) =  K\gamma I_K(t),\\
\frac{dC_I(t)}{dt}&=\beta(t) S(t)\frac{I(t)}{N},\\
\end{align*}
with initial conditions:
\begin{align*}
S(t_0)&= S_0, I(t_0)=I_1(t_0) = I_0, R(t_0)=R_0, C_I(t_0) = N - S_0,
\end{align*}
where $\beta(t) = \exp\left(\sum_{i=1}^{m}\beta_iB_{i}(t)\right)$, $I(t) = \sum_{j= 1}^{K}I_j(t)$, and $N = S(t)+I(t)+R(t)$ is the fixed size of the total population.

Using the parameterisation
\begin{align*}
\mathbf{y}(t)&=(y_1(t),y_2(t),\dots ,y_{K+1}(t),y_{K+2}(t), y_{K+3}(t))'\\
\\
&=(S(t), I_1(t),\dots, I_{K}(t),\allowbreak R(t), C_I(t))',
\end{align*}
we obtain the following condensed form of the equations
\begin{align}
\dot{\mathbf{y}}(t)=\frac{d\mathbf{y}(t)}{dt}=\mathbf{f}(t,\mathbf{y},\mathbf{p}), \quad	\mathbf{y}(t_0)=(S_0,I_0,0, \dots,0, R_0, N - S_0)'.
\end{align}
By combining Eqns. (\ref{SEMIKR_ODES_2}) and the probabilistic model (\ref{NegBinModel}) in the main text, the parameters to estimate can be expressed as
$$\mathbf{p} = (p_0,p_1,p_2,p_3,p_4, p_5, \dots, p_{m+3})' = (S_0, I_0,\phi^{-1},\tau^2, \beta_1,\beta_2,\dots, \beta_m)'.$$
Notice that for the SI$_K$R model one has the same probabilistic model (eqn. \eqref{NegBinModel} in the main text) and, therefore, the same log-posterior (Eqn. \eqref{log_post} in the main text).
\subsection{Log-likelihood gradient}\label{sec:loglike-grad_SIR}\label{SIR_Sensitivity_dynamic}
For the SI$_K$R model described above, the gradient of the log-posterior takes the following form
\begin{align}
\frac{d\ell}{dp_i}=\begin{cases}
\sum_{j=1}^n\frac{\partial\ell_{like}^j}{\partial C(t_0 + j)}\left(s_{i, K + 3}(t_0 + j) - s_{i,K + 3}(t_0 + j - 1)\right) + \frac{\partial \ell_{prior}}{\partial p_i} & i=0,1,4,\dots,m + 3,\\
\sum_{j=1}^n\frac{\partial \ell^j_{like}}{\partial p_i} + \frac{\partial \ell_{prior}}{\partial p_i} & i = 2,\\
\frac{\partial \ell_{prior}}{\partial p_i} & i = 3.
\end{cases}
\end{align}
Additionally, we have
\begin{align*}
\frac{\partial \ell_{like}^j}{\partial C(t_0 + j)} &= \frac{\phi \left( \tilde{C}_{t_0 + j} - \eta(t_0 + j) \, C(t_0 + j) \right)}
{C(t_0 + j) \left( \eta(t_0 + j) \, C(t_0 + j) + \phi \right)},\\
\frac{\partial \ell_{like}^j}{\partial p_i} &=-\delta_{i2}\Bigg(\psi\left(\tilde{C}_{t_0 + j} + \phi\right) - \psi(\phi) + \frac{\eta(t_0 + j)C(t_0 + j) - \tilde{C}_{t_0 + j}}{\eta(t_0 + j)C(t_0 + j) + \phi}+\log(\phi) - \\
&\quad \log\left(\eta(t_0 + j)C(t_0 + j)+\phi\right)\Bigg)\phi^2.
\end{align*}
The forward sensitivity analysis is characterised by the following extended system of ODEs
\begin{align*}
\dot{\mathbf{y}}(t)&=\mathbf{f}(t,\mathbf{y},\mathbf{p}), \quad	\mathbf{y}(t_0)=(p_0,p_1,0, \dots, 0, N - (p_0 + p_1), N - p_0)',\\
\dot{\mathbf{s}}_i(t) &=  \frac{\partial \mathbf{f}(t)}{\partial \mathbf{y}}\mathbf{s}_i(t) + \frac{\partial \mathbf{f}(t)}{\partial p_i},\quad\mathbf{s}_i(t_0)=(-\delta_{i0},\delta_{i1},0,\dots,0,-(\delta_{i0} + \delta_{i1}), -\delta_{i0})'.
\end{align*}
\subsection{Jacobian matrices}\label{sec:jacobian_SIR}
 Differentiating the right hand sides of system (\ref{SEMIKR_ODES_2}), one obtains the following Jacobian matrix:
\begin{footnotesize}
\begin{align*}
\left[\frac{\partial \mathbf{f}(t)}{\partial \mathbf{y}}\right]_{1,1} &= -\exp\left(\sum_{i=1}^{m}\beta_iB_{i}(t)\right) \frac{1}{N}\sum_{j=2}^{K + 1}y_j(t),\left[\frac{\partial \mathbf{f}(t)}{\partial \mathbf{y}}\right]_{2,1} =- \left[\frac{\partial \mathbf{f}(t)}{\partial \mathbf{y}}\right]_{1,1}, \left[\frac{\partial \mathbf{f}(t)}{\partial \mathbf{y}}\right]_{K+3,1} =- \left[\frac{\partial \mathbf{f}(t)}{\partial \mathbf{y}}\right]_{1,1},\\
\left[\frac{\partial \mathbf{f}(t)}{\partial \mathbf{y}}\right]_{j,1} &= 0,\quad 2<j< K + 3,\\
\left[\frac{\partial \mathbf{f}(t)}{\partial \mathbf{y}}\right]_{1,2} &= -\exp\left(\sum_{i=1}^{m}\beta_iB_{i}(t)\right) \frac{y_1(t)}{N}, \left[\frac{\partial \mathbf{f}(t)}{\partial \mathbf{y}}\right]_{2,2} = -\left[\frac{\partial \mathbf{f}(t)}{\partial \mathbf{y}}\right]_{1,2} - K\gamma, \left[\frac{\partial \mathbf{f}(t)}{\partial \mathbf{y}}\right]_{K+3,2} = -\left[\frac{\partial \mathbf{f}(t)}{\partial \mathbf{y}}\right]_{1,2},\\
\left[\frac{\partial \mathbf{f}(t)}{\partial \mathbf{y}}\right]_{3,2} &= K\gamma, \left[\frac{\partial \mathbf{f}(t)}{\partial \mathbf{y}}\right]_{j,2} = 0,\quad 3<j<K+3,\\
\left[\frac{\partial \mathbf{f}(t)}{\partial \mathbf{y}}\right]_{1,i} &= -\exp\left(\sum_{i=1}^{m}\beta_iB_{i}(t)\right) \frac{y_1(t)}{N}, \left[\frac{\partial \mathbf{f}(t)}{\partial \mathbf{y}}\right]_{2,i} = -\left[\frac{\partial \mathbf{f}(t)}{\partial \mathbf{y}}\right]_{1,i}, \left[\frac{\partial \mathbf{f}(t)}{\partial \mathbf{y}}\right]_{K+3,i} = -\left[\frac{\partial \mathbf{f}(t)}{\partial \mathbf{y}}\right]_{1,i},\\
\left[\frac{\partial \mathbf{f}(t)}{\partial \mathbf{y}}\right]_{i,i} &= -K\gamma,  \left[\frac{\partial \mathbf{f}(t)}{\partial \mathbf{y}}\right]_{i+1,i} = K\gamma, \left[\frac{\partial \mathbf{f}(t)}{\partial \mathbf{y}}\right]_{j,i} = 0,\quad j\neq 1,2,i,i+1,K+3\quad i = 3, \dots, K+1,\\
\left[\frac{\partial\mathbf{f}(t)}{\partial \mathbf{y}}\right]_{j,i} &= 0,\quad j = 1, \dots,  K+3,\quad i = K+2, K+3.
\end{align*}
\end{footnotesize}
The Jacobian with respect to the parameters is characterised by 
\begin{align*}
\frac{\partial \mathbf{f}}{\partial p_0}(t) &=\cdots=\frac{\partial \mathbf{f}}{\partial p_3}(t) = 0,\\
\frac{\partial \mathbf{f}}{\partial p_i}(t) &= \Bigg(-\exp\left(\sum_{i=1}^{m}\beta_iB_{i}(t)\right) B_{i-4}(t)\frac{y_1(t)\sum_{j=2}^{K + 1}y_j(t)}{N},\\
&\exp\left(\sum_{i=1}^{m}\beta_iB_{i}(t)\right) B_{i-4}(t)\frac{y_1(t)\sum_{j=2}^{K + 1}y_j(t)}{N},0,\dots,0,\\
&\exp\left(\sum_{i=1}^{m}\beta_iB_{i}(t)\right) B_{i-4}(t)\frac{y_1(t)\sum_{j=2}^{K + 1}y_j(t)}{N}\Bigg)'\quad i=4,\dots, m+3.\\
\end{align*}

\section{SIR/SEIR-like models with a diffusion-based transmission rate}\label{SODE_section}
Using the ideas in \cite{Dureau2013}, we provide stochastic ordinary differential equations (SODEs) for SE$_M$I$_K$R and SI$_K$R models, where $\log\beta(t)$ is governed by a diffusion process. An additional parameter $\beta_0$ to model the initial transmission rate and $\sigma$ to model the smoothness of the diffusion process are introduced. 

In this setting, the SE$_M$I$_K$R equations become
\begin{align*}
\label{SODES}
\frac{dS(t)}{dt}&= -\beta_0\beta(t) S(t)\frac{I(t)}{N},\\
\frac{dE_1(t)}{dt}&=\beta_0\beta(t)S(t)\frac{I(t)}{N}-M\alpha E_1(t),\\
\frac{dE_2(t)}{dt}&=M\alpha E_1(t)-M\alpha E_2(t),\quad \dots,\quad\frac{dE_M(t)}{dt}=M\alpha E_{M - 1}(t)-M\alpha E_M(t),\\
\frac{dI_1(t)}{dt}&=M\alpha E_M(t)-K\gamma I_1(t),\numberthis\\
\frac{dI_2(t)}{dt}&=K\gamma I_1(t)-K\gamma I_2(t),\quad\dots,\quad\frac{dI_K(t)}{dt}=K\gamma I_{K-1}(t)-K\gamma I_K(t),\\
\frac{dR(t)}{dt}& = \frac{d}{dt}\left(N - S(t) - E(t) - I(t)\right) =  K\gamma I_K(t),\\
\frac{dC_I(t)}{dt}&=\beta_0\beta(t) S(t)\frac{I(t)}{N},\\
\frac{d\log\beta(t)}{dt}&=\sigma dB(t),\\
\end{align*}
with initial conditions:
\begin{align*}
S(t_0)&= S_0, E_1(t_0) = E_0, I_1(t_0)=I_0, R(t_0) =R_0=N - (S_0+E_0+I_0).\\
E_2(t_0)&=\dots=E_M(t_0), I_2(t_0)=\dots = I_K(t_0)= 0,\\
C_I(t_0) &= N - S_0, \beta(t_0) = 1,\\
\end{align*}
where $B(t)$ denotes the Standard Brownian Motion, $E(t) = \sum_{i = 1}^{M}E_i(t)$, $I(t) = \sum_{j= 1}^{K}I_j(t)$, and $N = S(t)+E(t)+I(t)+R(t)$ is the fixed size of the total population.

The parameters of interest for model (\ref{NegBinModel}) in the main text, based on the dynamics (\ref{SODES}) are
$$\mathbf{p} = (p_0,p_1,p_2,p_3,p_4, p_5, p_6)' = (\alpha, S_0, E_0, I_0, \phi^{-1}, \beta_0, \sigma)'.$$

For completeness we provide the SODEs corresponding to the SI$_K$R model
\begin{align*}
\label{SODES_2}
\frac{dS(t)}{dt}& = -\beta_0\beta(t) S(t)\frac{I(t)}{N},\\
\frac{dI_1(t)}{dt}&=\beta_0\beta(t) S(t)\frac{I(t)}{N}-K\gamma I_1(t),\\
\frac{dI_2(t)}{dt}&=K\gamma I_1(t)-K\gamma I_2(t),\quad\dots,\quad\frac{dI_K(t)}{dt}=K\gamma I_{K-1}(t)-K\gamma I_K(t),\numberthis\\
\frac{dR(t)}{dt}& = \frac{d}{dt}\left(N - S(t) - I(t)\right) =  K\gamma I_K(t),\\
\frac{dC_I(t)}{dt}&=\beta_0\beta(t) S(t)\frac{I(t)}{N},\\
\frac{d\log\beta(t)}{dt}&=\sigma dB(t),\\
\end{align*}
with initial conditions:
\begin{align*}
S(t_0)&= S_0, I(t_0)=I_1(t_0) = I_0, R(t_0)=R_0, C_I(t_0) = N - S_0, \beta(t_0)=1.
\end{align*}
The corresponding parameters to estimate are 
$$\mathbf{p} = (p_0,p_1,p_2,p_3,p_4)' = (S_0, I_0,\phi^{-1}, \beta_0,\sigma)'.$$  

\section{Generalised Hamiltonian Monte Carlo}\label{Hamiltonian_dynamics}

Generalised Hamiltonian Monte Carlo (GHMC) \citep{Kennedy2001} is a Markov Chain Monte Carlo (MCMC) method for sampling from a target probability distribution by generating proposals informed by the gradient of its log-density. It extends the standard Hamiltonian Monte Carlo (HMC) \citep{duane1987hmc, Brooks2011} framework by introducing partial momentum refreshment, which provides a tunable balance between deterministic Hamiltonian dynamics and stochastic momentum randomization, enabling better control over mixing and sampling efficiency.

In GHMC, sampling is performed from the augmented distribution
\begin{equation}\label{eq:InvariantJointDistribution}
\pi(\mathbf{p}, \mathbf{q})
=
\pi(\mathbf{p})\, n(\mathbf{q})
\propto
\exp\{-H(\mathbf{p}, \mathbf{q})\},
\end{equation}
where $\mathbf{p}$ denotes the model parameters of interest,
$\pi(\mathbf{p})$ is their posterior distribution,
$\mathbf{q}\sim n(\mathbf{q})=\mathcal{N}(0,\mathsf{M})$ is an auxiliary
momentum variable, and marginalising over $\mathbf{q}$ recovers
$\pi(\mathbf{p})$. In \eqref{eq:InvariantJointDistribution},
$H(\mathbf{p},\mathbf{q})$ is a separable Hamiltonian function
\begin{equation*}\label{eq:HamiltonianSeparable}
H(\mathbf{p},\mathbf{q})
=
\frac{1}{2}\mathbf{q}'\mathsf{M}^{-1}\mathbf{q}
+
U(\mathbf{p}),
\end{equation*}
where $\mathsf{M}$ is a symmetric positive-definite mass matrix and the
potential energy $U(\mathbf{p})$ encodes the target distribution via
\begin{equation*}
U(\mathbf{p}) = -\ell(\mathbf{p}) + \mathrm{const}.
\end{equation*}

Each GHMC iteration consists of two main steps: a partial momentum update
(PMU) \citep{horowitzGHMC} and Hamiltonian dynamics integration. The PMU reads
\begin{equation}\label{eq:PMU}
\mathbf{q}
\leftarrow
\sqrt{1-\varphi}\,\mathbf{q}
+
\sqrt{\varphi}\,\mathbf{u},
\end{equation}
where $\varphi\in(0,1]$ controls the degree of momentum refreshment, and
$\mathbf{u}\sim\mathcal{N}(0,\mathsf{M})$. Setting $\varphi=1$ recovers
standard HMC.

The updated state is then evolved according to Hamilton's equations
\begin{equation}\label{eq:HamiltonianSystemSeparable}
\frac{d\mathbf{p}}{dt}
=
\mathsf{M}^{-1}\mathbf{q},
\qquad
\frac{d\mathbf{q}}{dt}
=
-\nabla_{\mathbf{p}} U(\mathbf{p}),
\end{equation}
which are integrated numerically using a symplectic scheme
\citep{numerical_hamiltonian_problems} with step size $\Delta t$ for $L$
steps, producing a proposal $(\mathbf{p}',\mathbf{q}')$. Finally, the latter
is accepted with probability
\begin{equation*}
\mathrm{accep\_prob}
=
\min\left\{
1,
\exp\left[-\left(
H(\mathbf{p}',\mathbf{q}')
-
H(\mathbf{p},\mathbf{q})
\right)\right]
\right\},
\end{equation*}
ensuring detailed balance with respect to the augmented target distribution.

\section{Prior on the initial compartment proportions}\label{sup_sec:Initial_Ocuppancy}

This appendix justifies the Dirichlet prior used for the initial proportions
\((S_0,E_0,I_0,R_0)/N\) and describes the fast sampler that perturbs the
MAP estimate within a total-variation (TV) ball during Stage 2 of the workflow.

\subsection{Dirichlet specification}\label{subsec:dirichlet}

Let \(\mathbf d=(d_1,\dots,d_K) \in \Delta^{K-1}\) follow a
\(\operatorname{Dirichlet}(a_1,\dots,a_K)\) distribution with density  
\[
f(\mathbf d;\mathbf a)=
\frac{\Gamma(a_0)}{\prod_{i=1}^K \Gamma(a_i)}
\;\prod_{i=1}^K d_i^{\,a_i-1},\qquad 
a_0=\sum_{i=1}^K a_i.
\]
The mean and variance are
\[
\mathbb E[d_i]=\tilde a_i=\frac{a_i}{a_0},\qquad
\operatorname{Var}[d_i]=\frac{\tilde a_i\,(1-\tilde a_i)}{a_0}.
\]

\paragraph{Early-epidemic prior.}
We choose a prior concentrated around \(S_0 = N - 10\) and \(E_0 = 10\) (\(\mathbb E[E_0] = 10\)), while keeping \(I_0\) and \(R_0\) close to zero, as in an early epidemic.  
When \(a_i < 1\), the Dirichlet marginal is highly skewed towards zero, making the mean a poor descriptor. To obtain a distribution concentrated near the desired value, we require \(a_i \ge 1\), in which case
\[
\mathrm{Mode}[d_i]=\frac{a_i -1}{a_0-K}.
\]

With \(\mathbf a=(a_S,a_E,a_I,a_R)\), we set \(a_I = a_R = 1\) to give minimal nonzero mass to \(I_0\) and \(R_0\). For \(E\), we fix
\[
\tilde a_E = \frac{10}{N},\qquad a_0 = 10^{6},
\]
yielding
\[
\mathbf a
=
\bigl(
a_0 - (10/N)\,a_0 - 1 - 1,\;
(10/N)\,a_0,\;
1,\;
1
\bigr).
\]
For all \(N < 10^7\), \(a_E > 1\). The large \(a_0\) ensures small variances, encoding strong prior knowledge of \(S_0/N\) and \(E_0/N\) while keeping \(I_0/N\) and \(R_0/N\) essentially zero.

\paragraph{Mid-epidemic prior.}
At later times, expected proportions \(E^*/N\), \(I^*/N\), \(R^*/N > 0\) may be preferred.  
Defining \(\tilde a_E=E^*/N\), \(\tilde a_I=I^*/N\), \(\tilde a_R=R^*/N\) and choosing any \(a_0\) with \(a_E,a_I,a_R>1\) gives
\[
a_i = a_0\,\tilde a_i,\qquad
\tilde a_S = 1 - \tilde a_E - \tilde a_I - \tilde a_R.
\]
This produces the desired Dirichlet vector \((a_S,a_E,a_I,a_R)\).

\subsection{Perturbing the MAP within a TV ball}\label{subsec:tv_sampler}

For Stage 2 we draw candidate vectors \(\mathbf d\in\Delta^{3}\) such that  

\[
\operatorname{TV}(\hat p^{\mathrm{prop}},\mathbf d)
=
\frac12\sum_{i=1}^4|\hat p^{\mathrm{prop}}_i-d_i|
< \mathrm{TV}_0.
\]
Exact uniform sampling in this convex set is feasible
(e.g.\ \citealp{smith1996hit}); here we use a simpler \emph{non-uniform}
surrogate. The algorithm samples components sequentially while tracking
(i) the remaining $\ell^1$ budget and (ii) the remaining mass on the
simplex. It runs in $\mathcal{O}(K)$ per \emph{proposal} and may reject if
the final coordinate would violate feasibility. Here \(\hat p=(\hat p_1,\dots,\hat p_4)=\hat p^{\mathrm{prop}}\).

\begin{algorithm}[H]
    \caption{Greedy TV-ball sampler (feasible, non-uniform)}
    \label{alg:tv_sampler}
    \begin{algorithmic}[1]
        \Require $\widehat p\in\Delta^K$, TV radius $\mathrm{TV}_0\ge 0$, maxTrials
        \For{$t=1$ to maxTrials}
          \State $b\gets 2\,\mathrm{TV}_0$ \Comment{remaining $\ell^1$ budget}
          \For{$i=1$ to $K-1$}
            \State $m \gets 1-\sum_{j=1}^{i-1} d_j$ \Comment{remaining mass}
            \State $\ell \gets \max\{0,\ \widehat p_i-b\}$,\quad
                   $u \gets \min\{1,\ \widehat p_i+b,\ m\}$
            \State sample $d_i\sim\operatorname{Unif}([\ell,u])$
            \State $b\gets b-\lvert d_i-\widehat p_i\rvert$
          \EndFor
          \State $d_K\gets 1-\sum_{i=1}^{K-1} d_i$ \Comment{close the simplex}
          \If{$0\le d_K\le 1$ \textbf{and} $\lvert d_K-\widehat p_K\rvert \le b$}
            \State \Return $d=(d_1,\dots,d_K)$
          \EndIf
        \EndFor
        \State \textbf{fail} \Comment{no sample within maxTrials}
    \end{algorithmic}
\end{algorithm}

Although the distribution induced by Algorithm \ref{alg:tv_sampler} is not
uniform on the TV ball, it provides inexpensive, well-scattered seeds
for Stage 2, which is all that is needed for effective MAP-centred
initialisation.

\section{Tuning procedure for GHMC}\label{Tuning_HMC}
The accuracy and performance of an HMC-based simulation strongly depend on a choice of free parameters and settings. The set of tuneable parameters and building blocks of a GHMC simulation includes a numerical scheme for integrating the Hamiltonian dynamics \eqref{eq:HamiltonianSystemSeparable}, an integration step size, $\Delta t$, a number of integration steps per iteration, $L$, and an amount of random noise for the PMU \eqref{eq:PMU}, $\psi$.

In this work, we adopt the ATune method proposed by \cite{AKHMATSKAYA2026116892}, a computationally inexpensive adaptive tuning procedure that automatically identifies a system-specific optimal numerical integrator together with reliable GHMC hyperparameters and their optimal randomisation intervals. 
Here, we summarise the key elements of ATune for detecting optimal GHMC settings. For full details, see \cite{AKHMATSKAYA2026116892}.

\begin{itemize}
    \item \textbf{Numerical integrator} \\
    ATune employs the adaptive integration approach s-AIA3 \citep{nagar2023adaptive}, a three-stage palindromic splitting integrator \citep{bcss_paper, campos_sanz-serna2017} designed to optimise energy conservation for harmonic forces at any chosen simulation step size $\Delta t$.
    The method combines analytical results for multivariate Gaussian models with simulation data collected during the GHMC burn-in stage to select the most suitable numerical integrator and to estimate system-specific stability interval. 
    This analysis is further exploited by ATune to determine optimal randomisation intervals for the GHMC hyperparameters, as discussed below.
    \item \textbf{Step size $\Delta t$} \\
    Following the recommendations by \cite{bcss_paper} and \cite{mazur}, together with the results of \cite{nagar2023adaptive}, ATune selects the optimal randomisation interval for $\Delta t$ around the center of the stability interval estimated by s-AIA3. This choice minimises the energy error (thereby increasing acceptances) while maintaining sufficiently large moves to reduce sample autocorrelation. 
    \item \textbf{Randomisation interval for the momentum refreshment parameter $\Delta\psi$} \\
    ATune selects a randomisation interval for the GHMC momentum-refreshment
    parameter $\psi$ by extending the PMU tuning strategy of
    \cite{akhmatskaya_etal_2017}. Following \cite{AKHMATSKAYA2026116892}, this interval
    is computed from the system dimension, the s-AIA3 integrator coefficients, and
    a target PMU acceptance rate.  
    \item \textbf{Number of integration steps per iteration $L$} \\
    The hyperparameter settings obtained by ATune yield high acceptance rates, enabling the use of short trajectories ($L = 1$) that reproduce the effect of long but flexible trajectories while enhancing sampling efficiency \citep{Fang2014}. 
    However, the underlying s-AIA3 analysis is based on harmonic (Gaussian) assumptions, which may not hold for systems with strong anharmonic behaviour. To mitigate potential inaccuracies in such cases, ATune adapts $L$ accordingly: it uses a fixed $L = 1$ for harmonic-like models and randomises $L$ uniformly in the range $\mathcal{U} \{2, 5, 7\}$ for anharmonic ones.
\end{itemize}

The settings of simulation parameters identified by ATune and used for the numerical experiments are summarized in Table \ref{GHMC_parameters_setting}.

\begin{table}[]
\centering
\begin{tabular}{c|c|c|c|c}
Data & Model & $\Delta t$ & $L$ & $\Delta \psi$ \\
\hline
\multirow{4}{*}{Synthetic} 
& SIR & $(0.02381, 0.03433)$ & \multirow{4}{*}{$\{2, 5 , 7\}$} & $(0.02412, 0.15383)$ \\
& SI$_3$R & $(0.02330, 0.03369)$ & & $(0.02412, 0.15383)$ \\
& SEIR & $(0.02877, 0.04156)$ & & $(0.02306, 0.13875)$  \\
& SEI$_3$R & $(0.02941, 0.04113)$ & & $(0.02306, 0.13875)$ \\
\cline{1-5}
\multirow{4}{*}{Basque C.} 
& SIR & $(0.00309, 0.00447)$ & \multirow{4}{*}{$\{2, 5 , 7\}$} & $(0.01565, 0.09415)$\\
& SI$_3$R & $(0.00110, 0.00158)$ & & $(0.01565, 0.09415)$\\
& SEIR & $(0.00792, 0.01144)$ & & $(0.01460, 0.08787)$\\
& SEI$_3$R & $(0.00913, 0.01318)$ & & $(0.01460, 0.08787)$\\
\end{tabular}
\caption{\label{GHMC_parameters_setting} Tuned values of the GHMC parameters for the experiments of Section \nameref{section_results_and_discussion} in the main text.}
\end{table}

\section{Details on comparison methods}\label{comparison_methods}
Regarding the diffusion-based models presented in \ref{SODE_section}, the parameter estimation for such models is performed using a sophisticated particle filter MCMC sampler, called SMC$^2$ \citep{Chopin2013}. The prior distributions used in this case were
\begin{align}
\label{priors_SMC2}
\alpha &\sim \mathcal{N}(0.5, 0.05), \quad
\frac{(S_0,E_0,I_0,R_0)}{N}\sim\operatorname{Dirichlet}(999{,}993.4246,4.5754,1,1)\nonumber\\
\phi^{-1}&\sim U(0.01,1),\quad \sigma\sim U(0.01,1), \quad \beta_{0}\sim U(0.01,1).
\end{align}
Note that for $\alpha$ and $(S_0,E_0,I_0,R_0)$ we considered the same prior distributions as in the Hamiltonian-based Monte Carlo approach (see \eqref{priors_HMC} in the main text). However, for $\phi^{-1}$ (the dispersion parameter in (\ref{NegBinModel}) in the main text) we chose a uniform prior. The reason for this choice (based on trial and error) comes from the complete failure of SMC$^2$ when initialised with the well grounded exponential prior used for the proposed spline-based dynamic models. The prior for $\beta_0$ and $\sigma$ try to be not very informative but they do fix a reasonable range of variation. In all synthetic-data comparisons reported in the main text and in this appendix, the case-detection fraction is fixed at $\eta(t)=1$ for all $t$. Hence, differences between methods are not driven by assumptions on under-ascertainment.

We briefly describe how SMC$^2$ works. At $t=0$ it samples $N_p$ particles (values) from the prior distribution of the parameters \eqref{priors_SMC2}, i.e., $\{\mathbf{p}^i=(\alpha_i,\gamma_i, E_{0,i}, \phi^{-1}_i,\sigma_i,\allowbreak\beta_{0,i})\}_{i=1}^{N_p}$ for SEIR models and $\{\mathbf{p}^i=(\gamma_i, I_{0,i}, \phi^{-1}_i,\sigma_i,\allowbreak\beta_{0,i})\}_{i=1}^{N_p}$ for SIR models. Also, for each $\mathbf{p}^i$, at each time step $t$ it samples $N_x$ instances of the stochastic $\beta(t)$ and evolves the trajectories of the mechanistic dynamics (the SODEs \eqref{SODES} and \eqref{SODES_2}), which are linked to the observations via (\ref{NegBinModel}) in the main text. Next, the parameters are reweighted based on the conditional probability of the observed value at time $t$, $P(\tilde{C}_t|\tilde{C}_{t-1},\dots,\tilde{C}_1,\mathbf{p}^i)$. If a degeneracy criteria is reached at time $t$ (based on the weights of the proposed parameters) there is a rejuvenation mechanism that proposes, accepts or rejects new candidates. Therefore, SMC$^2$ returns a sample that is in correspondence with the evidence at each time $t$. This means that at time $t=100$ (the last time point of the synthetic data) the posterior sample obtained by SMC$^2$ is comparable to the one obtained by the four-stage GHMC sampling workflow described in \nameref{implementation} in the main text. 5 chains with 1000 production steps and 700 particles were run, where the number of production steps was chosen to get a Gelman-Rubin reduction factor close to 1.

The EpiEstim model is the following. Given past incidence cases $I_0,\dots,I_{t-1}$, a time dependent reproduction number $\mathsf{R_t}$, constant in a time window $\xi$, and a serial interval distribution $w$, one has
\begin{equation*}
    P(I_{t-\xi},\dots,I_{t}|I_0,\dots,I_{t-\xi-1},\mathsf{R_t},w)=\prod_{k=t-\xi}^t\frac{\left(\mathsf{R_t}\Lambda_k(w)\right)^{I_k}e^{-\mathsf{R_t}\Lambda_k(w)}}{I_k!}
\end{equation*}
where $\Lambda_k(w)=\sum_{s=1}^kI_{k-s}w_s$. Under the Bayesian paradigm and suitable assumptions on $w$ one has
\begin{equation*}
    P(\mathsf{R_t}|I_0,\dots,I_{t-\xi-1},I_{t-\xi},\dots, I_t)\propto P(I_{t-\xi},\dots,I_{t}|I_0,\dots,I_{t-\xi-1},\mathsf{R_t}, w)P(\mathsf{R_t})
\end{equation*}
and estimates of $\mathsf{R_t}$ can be obtained. Details can be found in \cite{thompson2019improved} and references therein. We follow the standard approach for applying EpiEstim where $w$ is a discretised 1-day offset gamma distribution parameterised by a mean $\mu$ and a variance $\sigma^2$. To take into account the uncertainty in $w$, one takes $\mu$ and $\sigma$ as parameters and puts generators on them (this corresponds with \texttt{method = "parametric\_si"} in the function \texttt{estimate\_R} in the \texttt{EpiEstim} R-package \citep{EpiEstim}). To make use of the available prior information and make a fair comparison, we place the average of the infectious profile around 10 days which is the true average time spent being infectious, i.e.,  $\mu\sim \mathcal{N}(10, 0.5^2)$. The generator on the variance, $\sigma\sim N_{[5, 15]}(10, 4^2)$,  was chosen by trial and error, seeking to obtain a posterior sample that captured well the true generator (which is unavailable in real situations). The window parameter was chosen to obtain good performance while being as small as possible, hence we chose it to be $\xi = 7$ days.

\section{Monte Carlo Detailed Results}\label{monte_carlo_details}

In this section further details and results of the experiments carried out in Section \nameref{section_results_and_discussion} of the main text are presented. 

\subsection{Case study 1: Synthetic data}
\begin{table}[]
\centering
\begin{tabular}{lllll}
            & \multicolumn{4}{c}{$\hat R$}                            \\
            & SEI$_3$R & SEIR  & SI$_3$R                      & SIR   \\
$\alpha$    & 1.000    & 1.001 &                              &       \\
$S_0$       & 1.010    & 1.015 & {1.094} & 1.049 \\
$E_0$       & 1.005    & 1.009 &                              &       \\
$I_0$       & 1.048    & 1.046 & {1.058} & 1.047 \\
$\phi^{-1}$ & 1.001    & 1.003 & 1.001                        & 1.001 \\
$\tau^2$    & 1.039    & 1.038 & {1.060} & 1.022
\end{tabular}
\caption{Convergence analysis for the synthetic data on 10 chains of 100000 production steps.}
\label{sup-table:r-hat_synthetic}
\end{table}
\begin{figure*}[h!]
\begin{center}
    \includegraphics[scale=0.27]{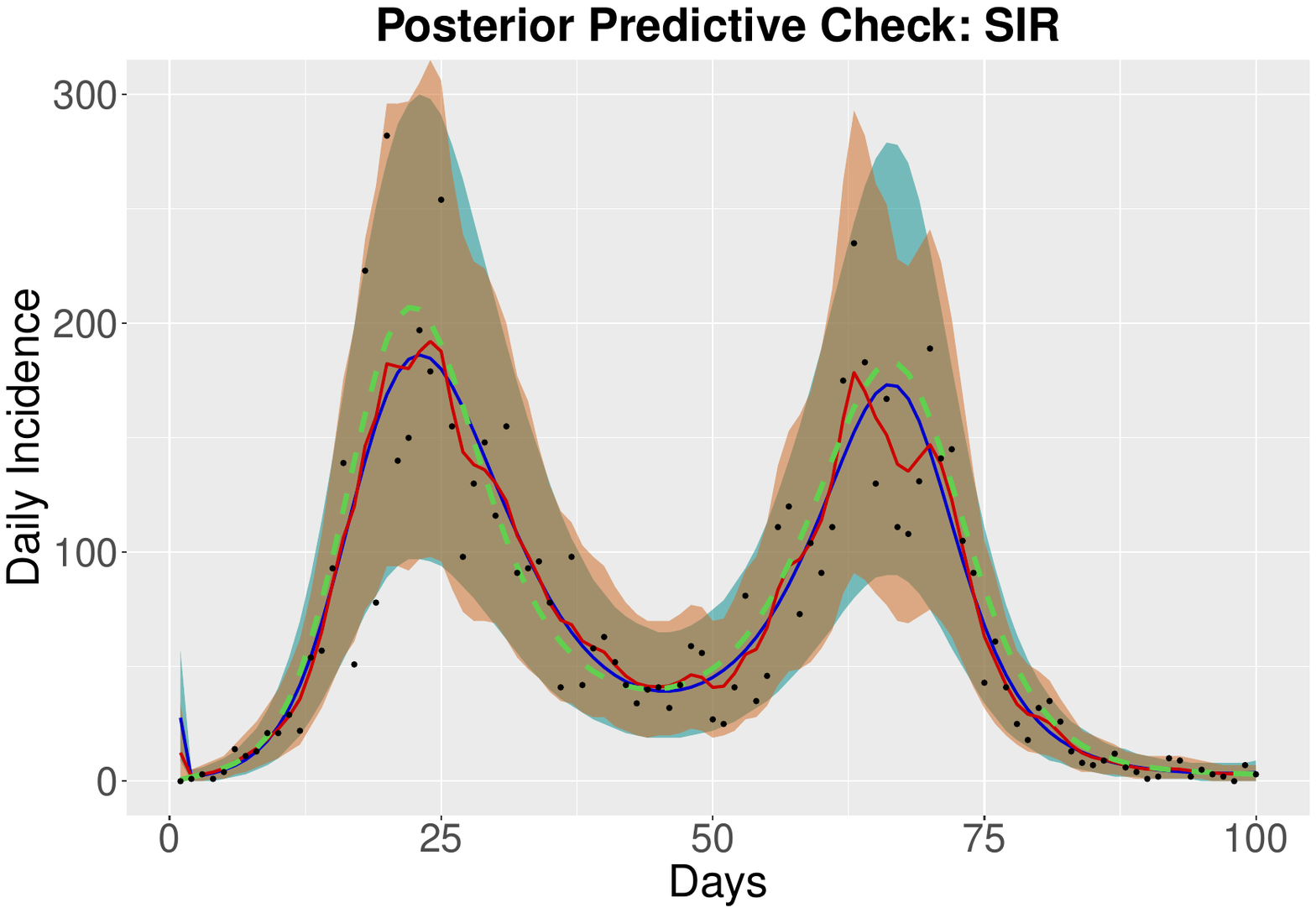}\includegraphics[scale=0.27]{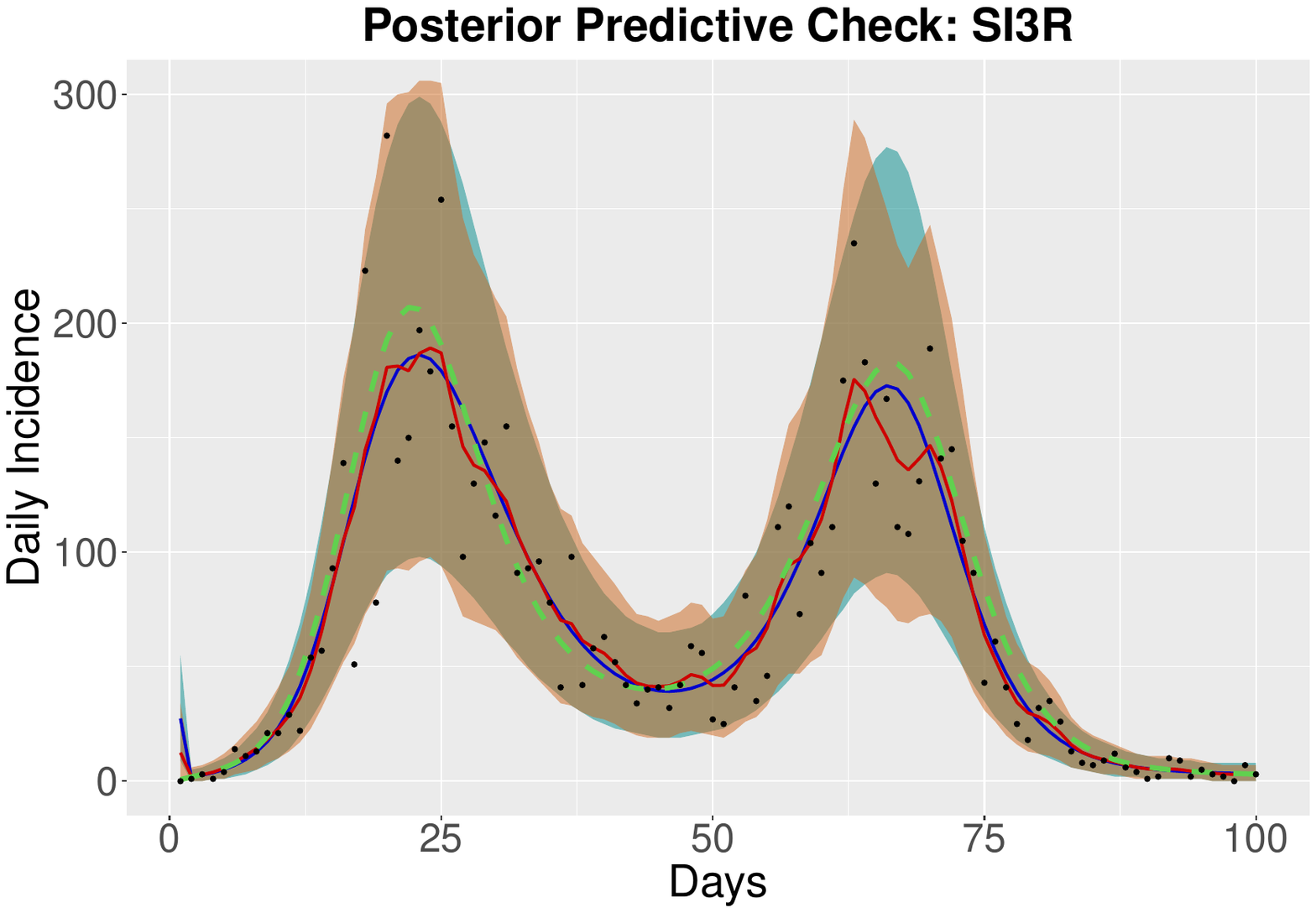}
\includegraphics[scale=0.27]{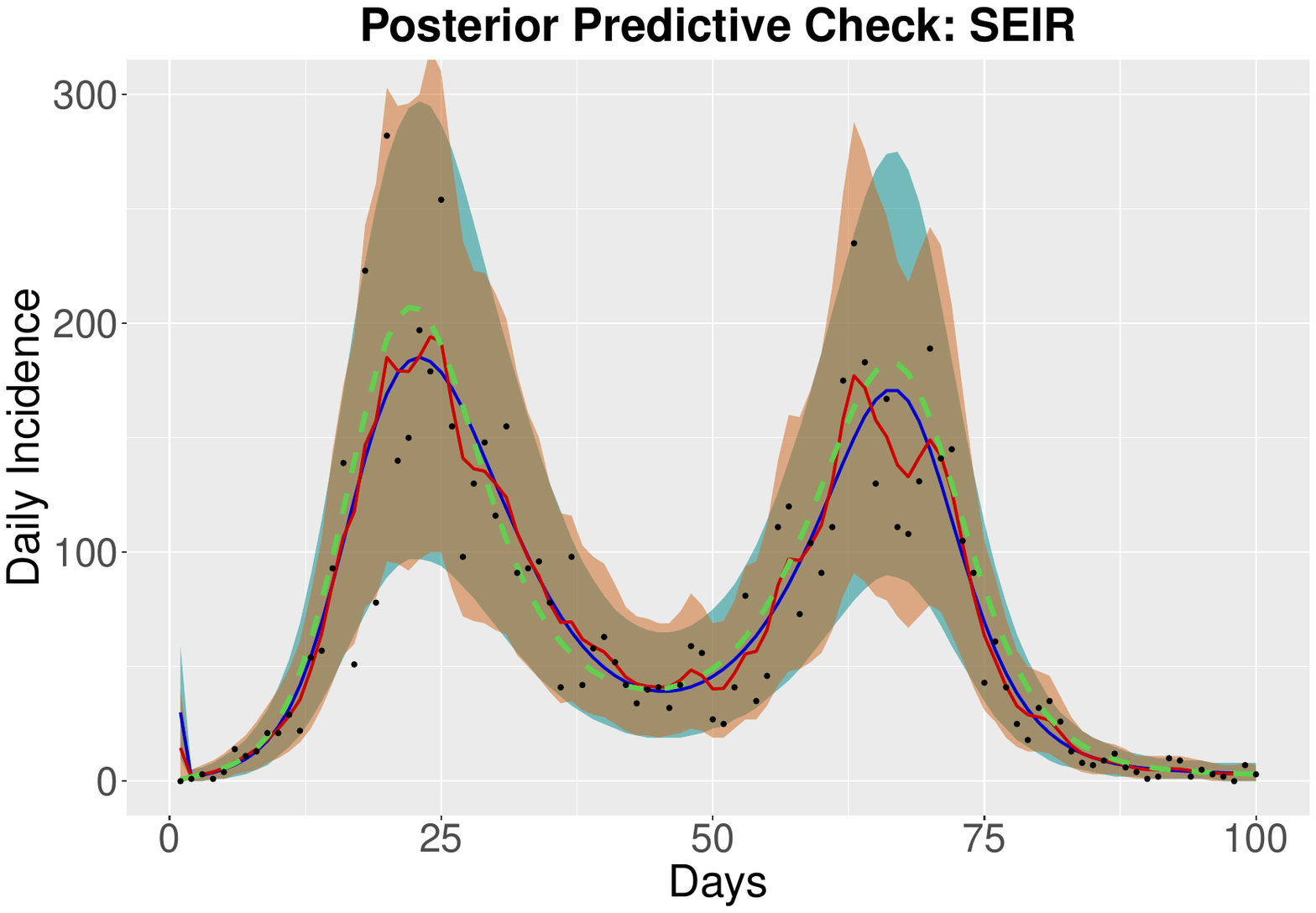}\includegraphics[scale=0.27]{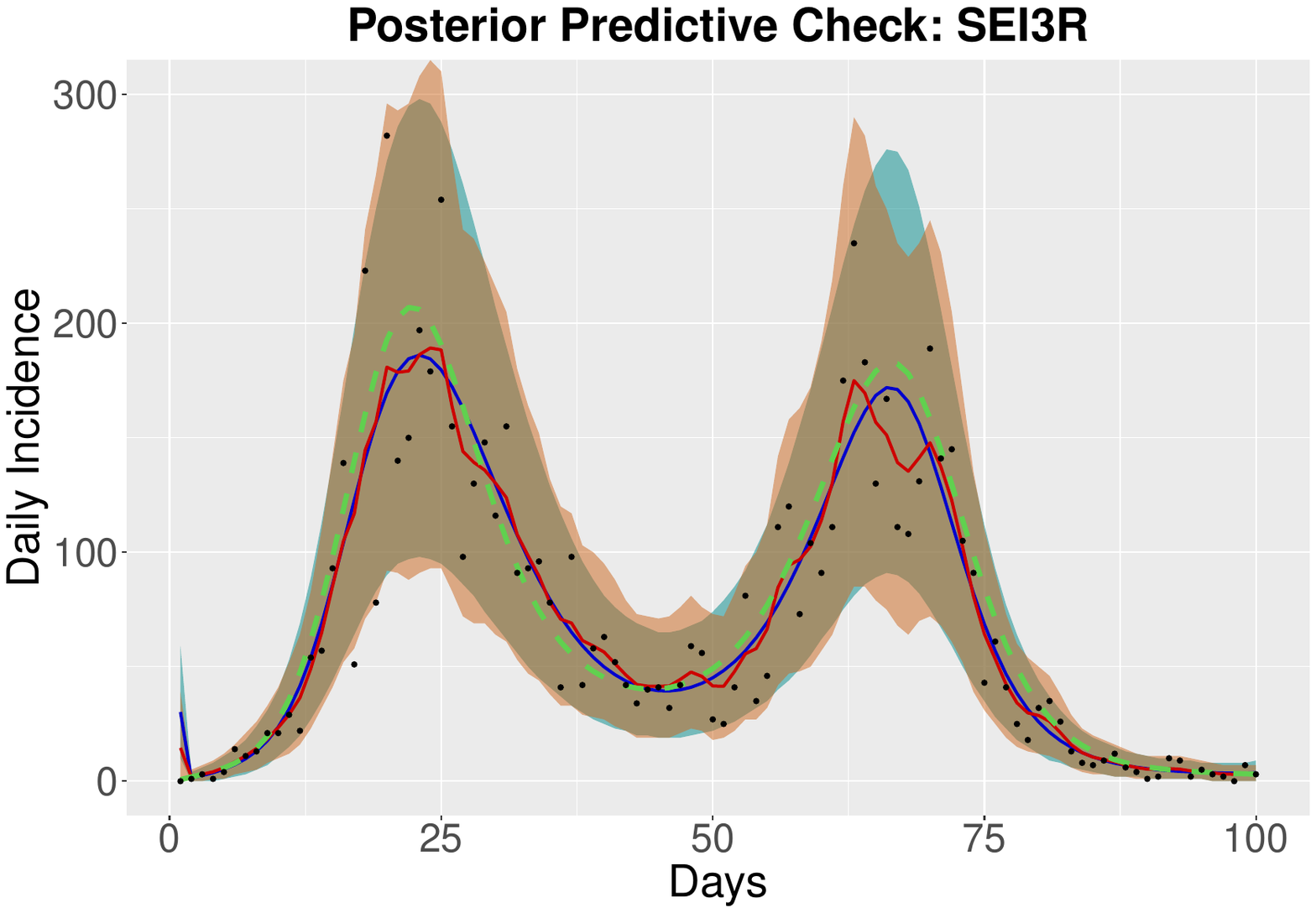}
\includegraphics[scale=0.67]{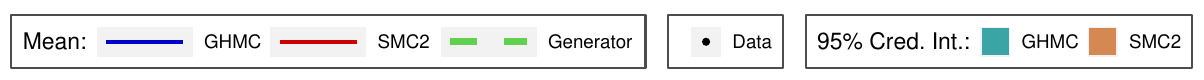}
\end{center}
\caption{For the synthetic data: posterior predictive checks on daily incidence for a spline-based dynamics sampled with GHMC (combination of 10 chains with 100000 production steps) and a diffusion-based dynamics sampled with SMC$^2$(combination of 5 chains with 1000 particles and 1000 production steps) for four different compartmental models SIR, SI$_3$R, SEIR and SEI$_3$R. Dashed green lines represent the corresponding values of the true generator $\mathbf{p}^{syn}$ (see (\ref{synt_params}) in the main text).}
\label{synt_inc_fig}    
\end{figure*}
Web Figure \ref{synt_inc_fig} depicts the posterior predictive checks on daily incidence for four different models -- SIR, SI$_3$R, SEIR and SEI$_3$R -- and their spline-based and diffusion-based versions, sampled with the GHMC and SMC$^2$ methodologies, respectively. We stress that EpiEstim is a direct estimate of the time-dependent reproduction number and does not provide a straightforward incidence posterior predictive check. As follows from the figure, the posterior predictive distribution samples for all models, regardless of the sampler, can generate data similar to the sample at hand (in black dots). Furthermore, the posterior predictive means, in solid lines, are very close to the true average of the generator (in dashed green). This means that our approach is comparable in its description of the data to the alternative diffusion-based mechanistic model (the SODEs \eqref{SODES} and \eqref{SODES_2} in \ref{SODE_section}) with the sophisticated SMC$^2$ sampler.
\subsubsection{Additional plots}\label{sec_additional_plots}
\begin{figure*}[h!]
\begin{center}
\includegraphics[scale=0.27]{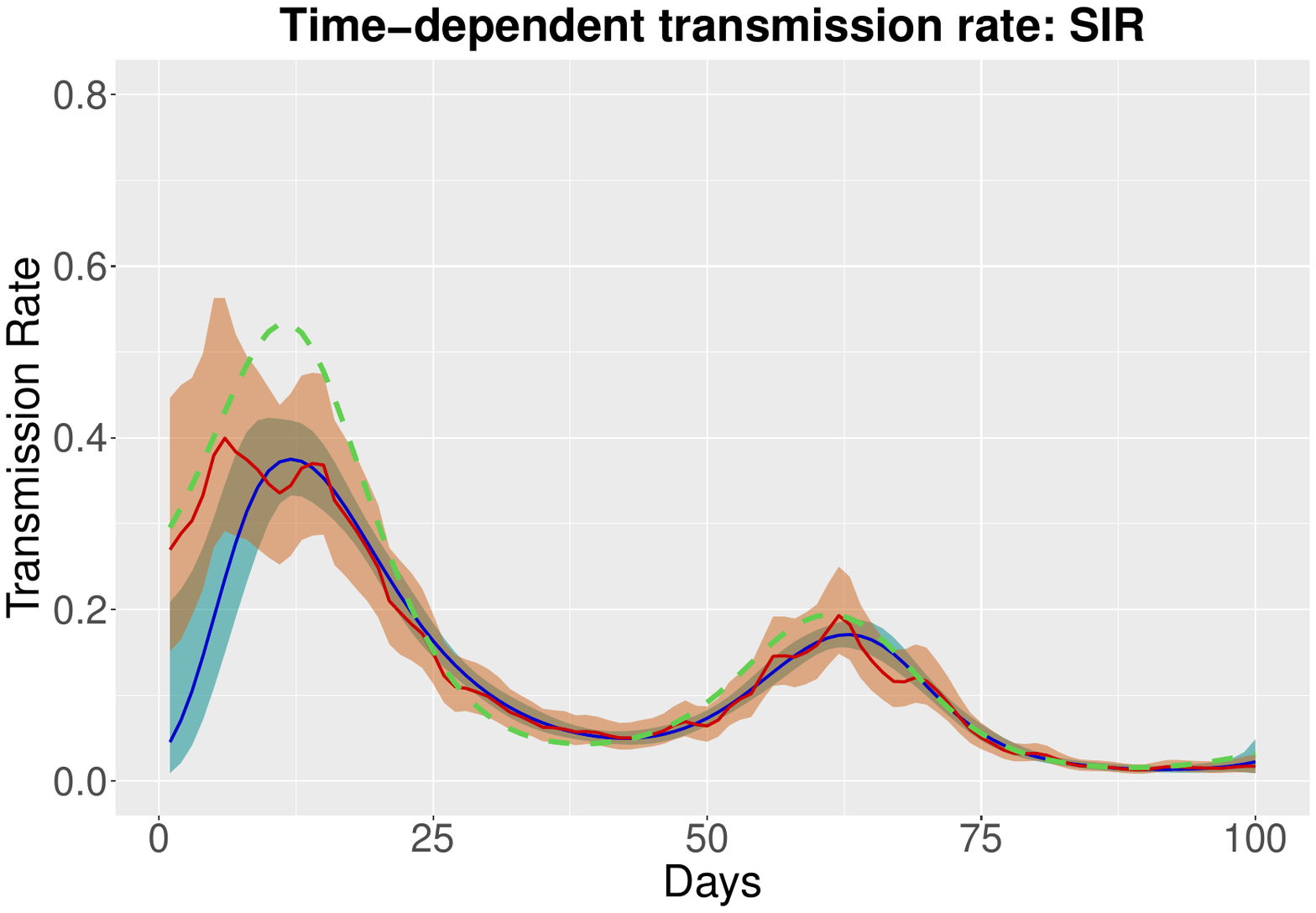}\includegraphics[scale=0.27]{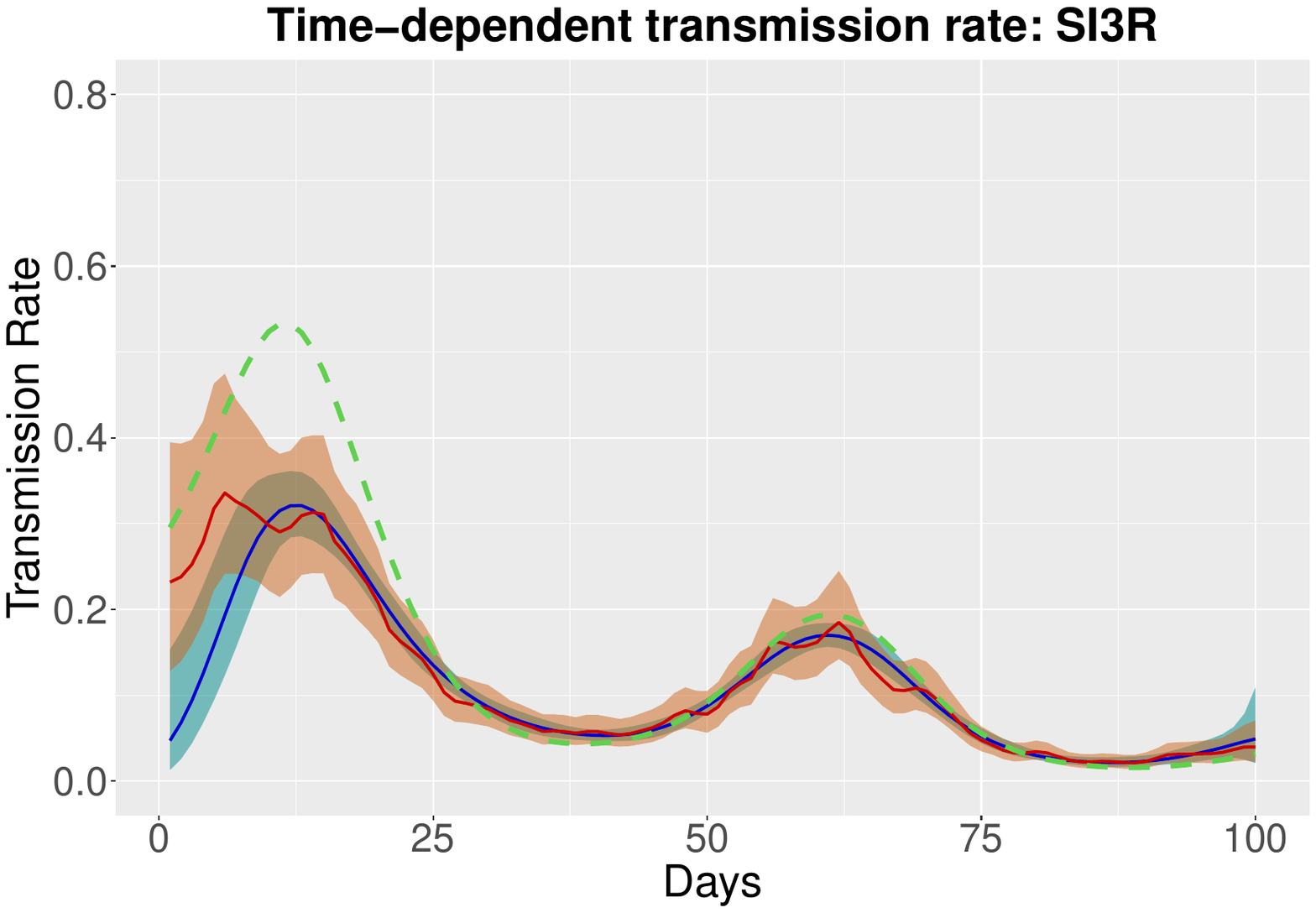}
\includegraphics[scale=0.27]{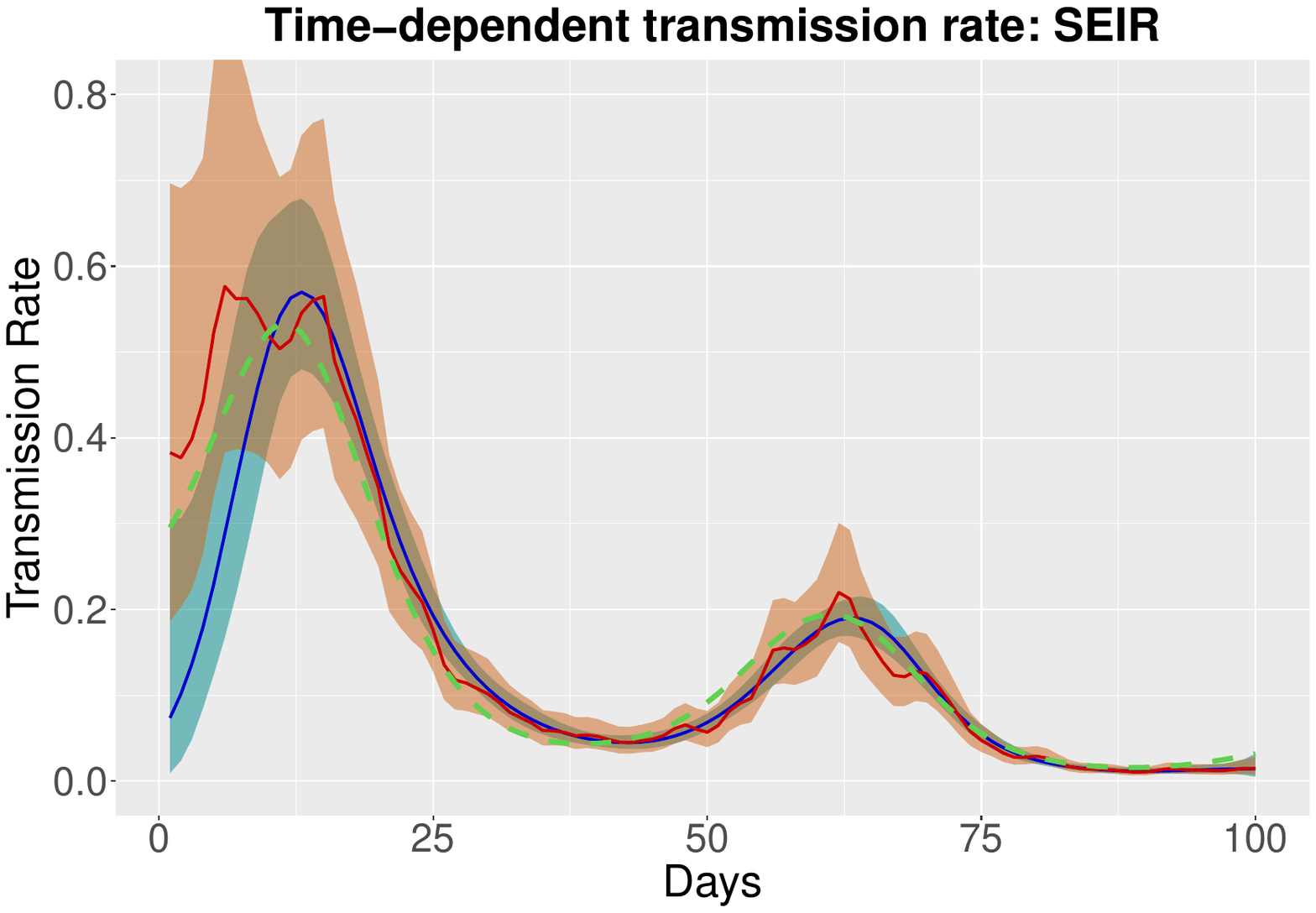}\includegraphics[scale=0.27]{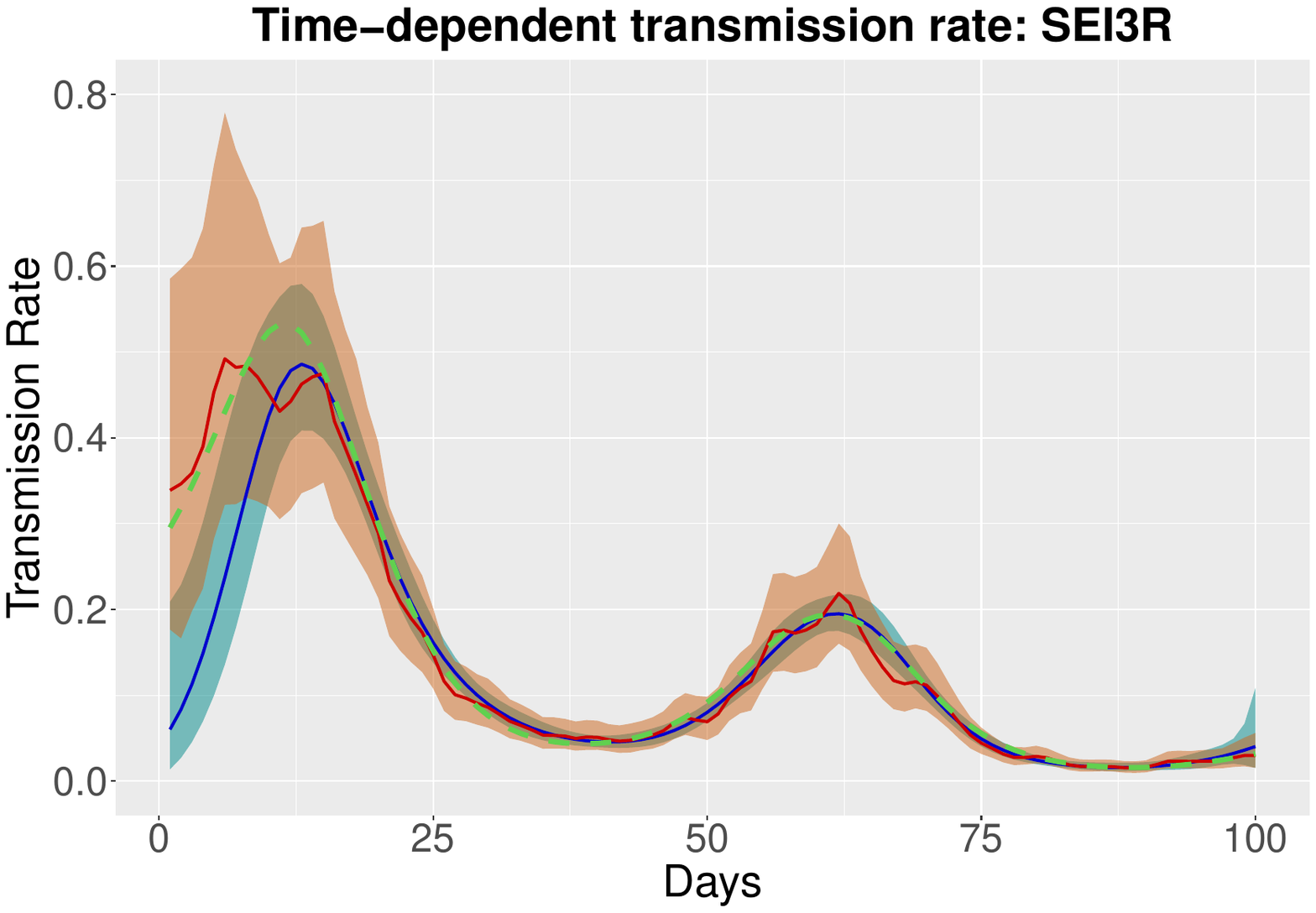}
\includegraphics[scale=0.67]{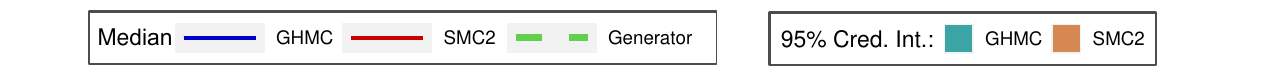}
\end{center}
\caption{For the synthetic data: posterior medians (solid lines) and 95\% credible intervals (shaded areas) of the time-dependent transmission rate, $\beta(t)$, for a spline-based dynamics sampled with GHMC (combination of 10 chains with 100000 production steps) and a diffusion-based dynamics sampled with SMC$^2$(combination of 5 chains with 1000 particles and 1000 production steps) for four different compartmental models SIR, SI$_3$R, SEIR and SEI$_3$R. Dashed green shows the corresponding values of the true generator $\mathbf{p}^{syn}$ (see (\ref{synt_params}) in the main text).} 
\label{synt_beta_fig}    
\end{figure*}

\begin{figure*}[h!]
\centering
\includegraphics[scale=0.15]{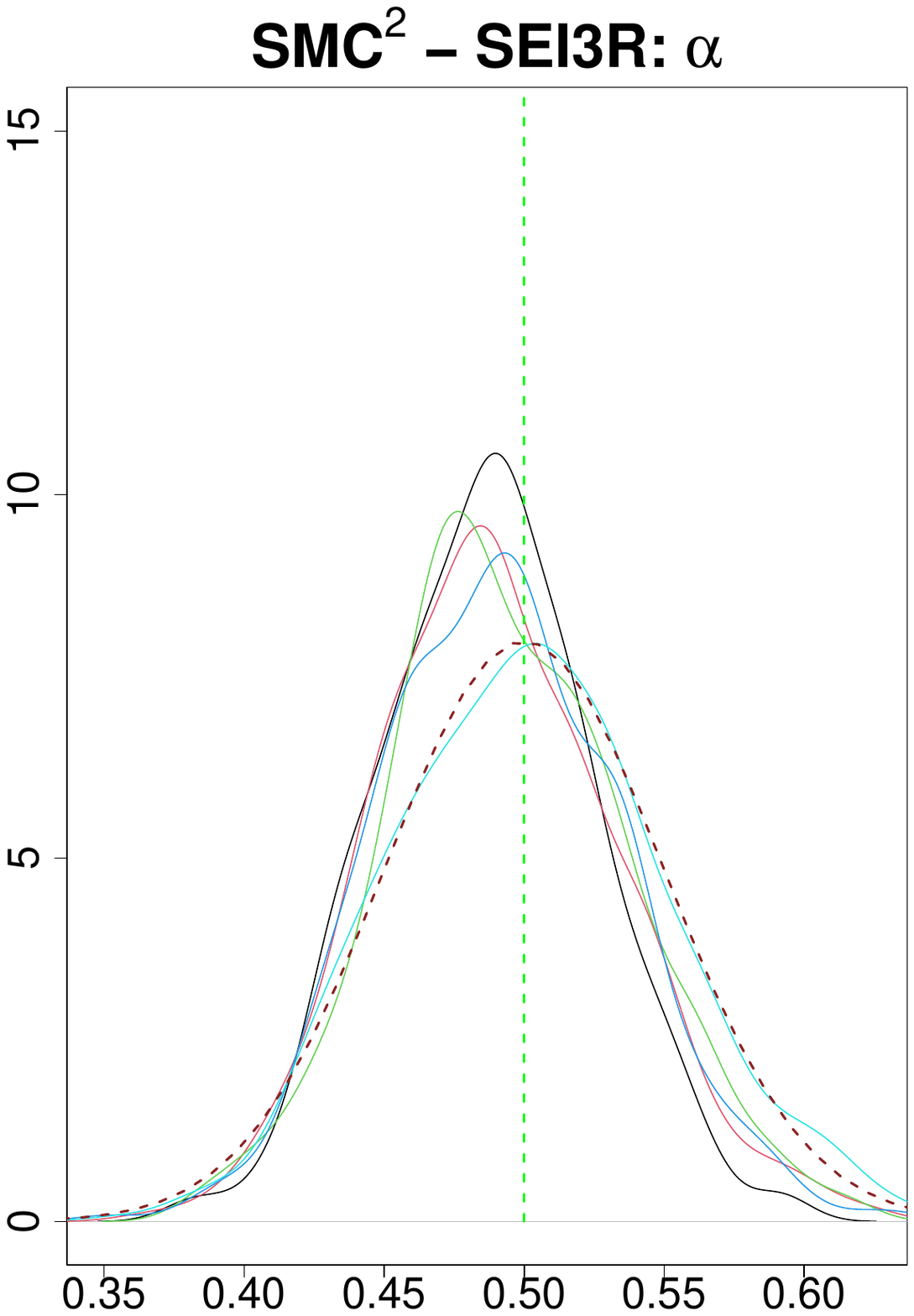}\includegraphics[scale=0.15]{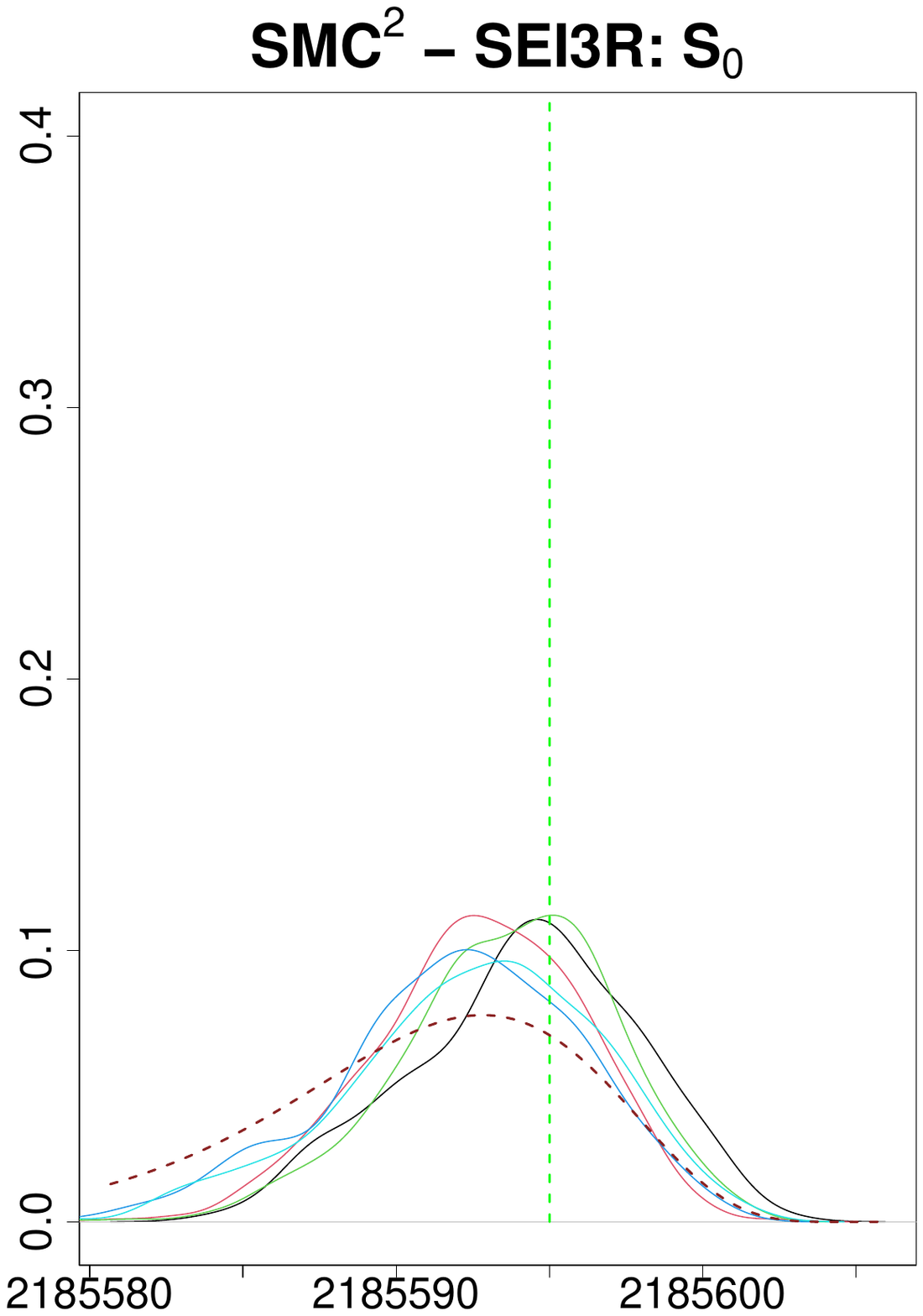}\includegraphics[scale=0.15]{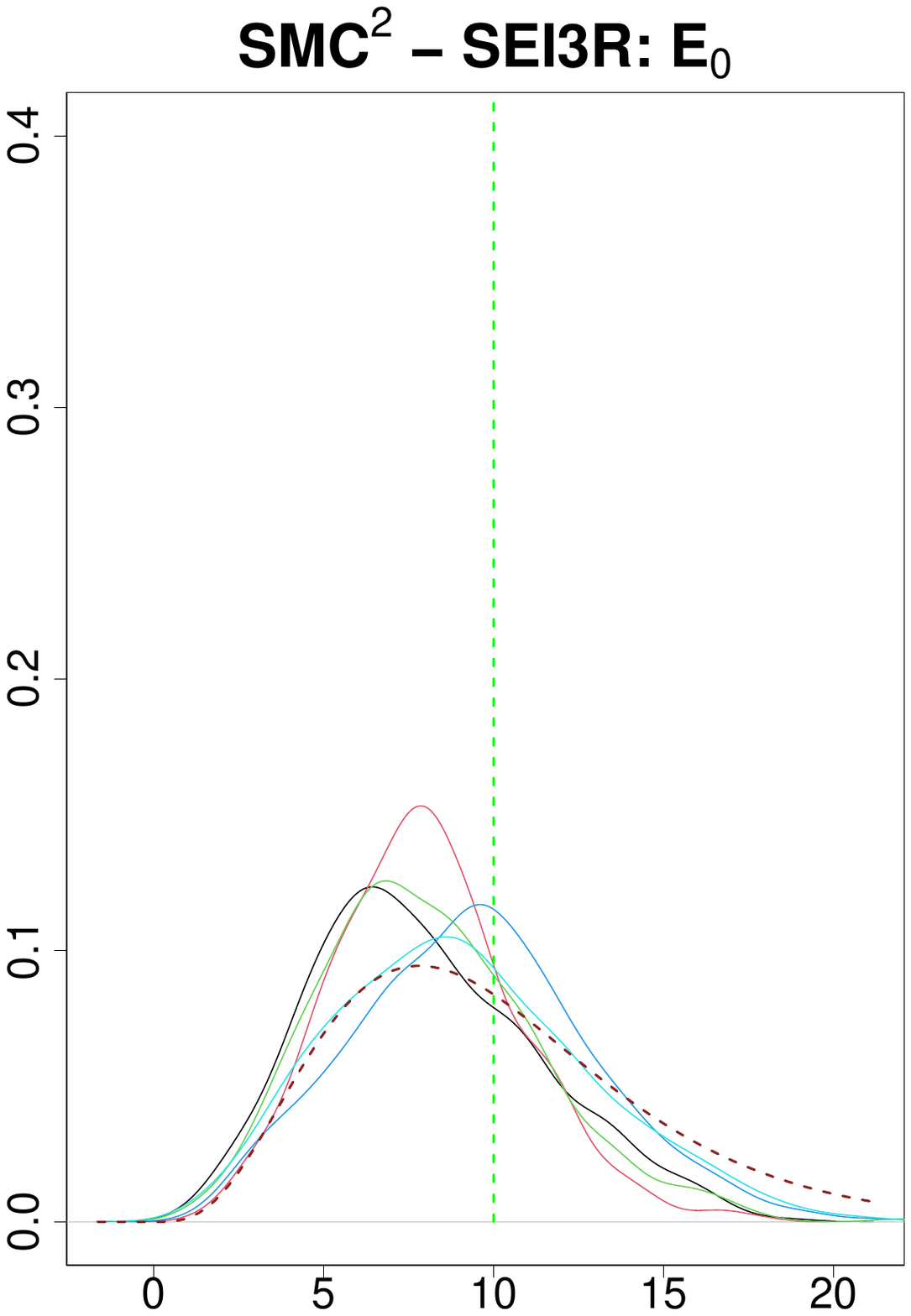}\includegraphics[scale=0.15]{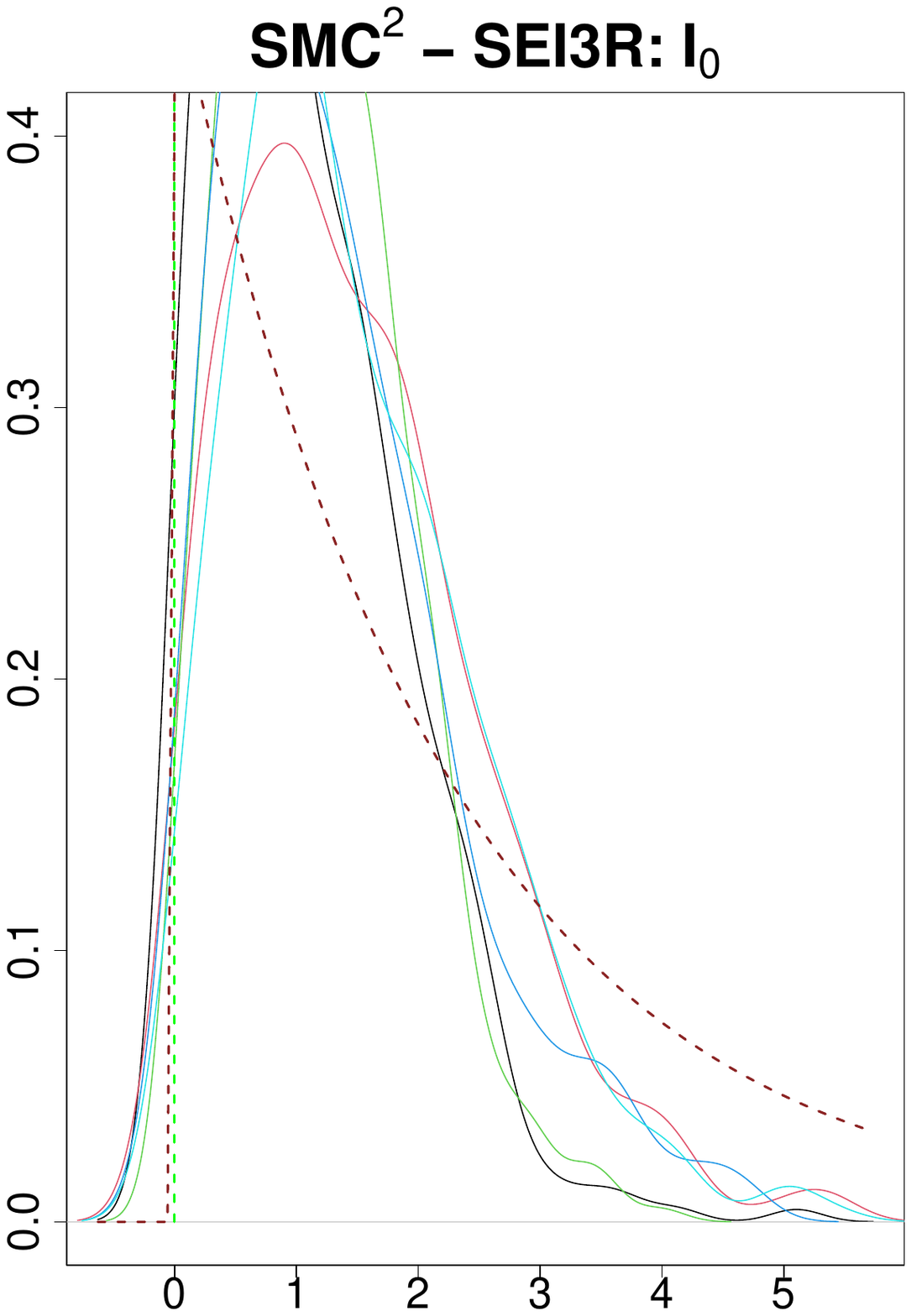}\includegraphics[scale=0.15]{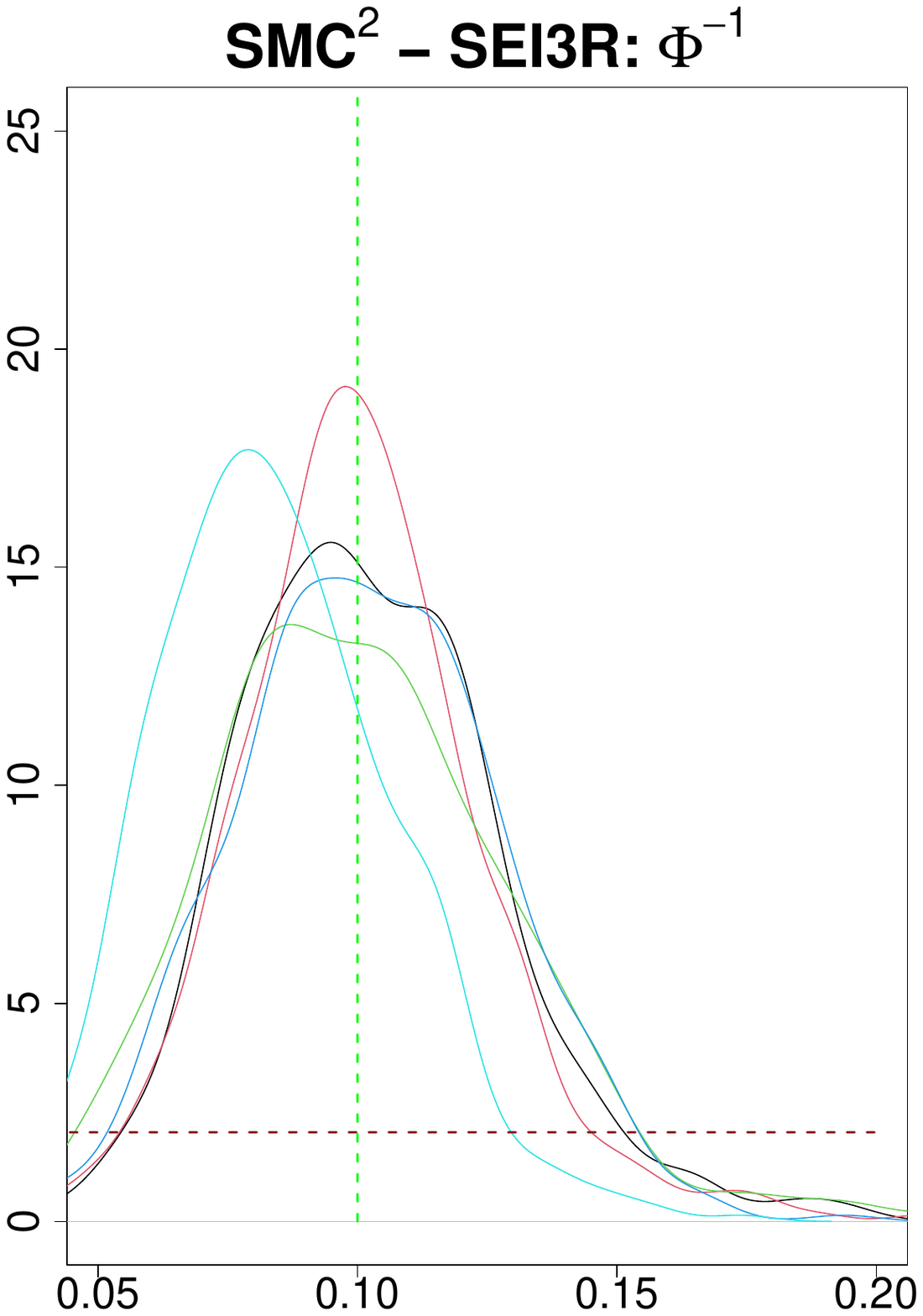}

\includegraphics[scale=0.15]{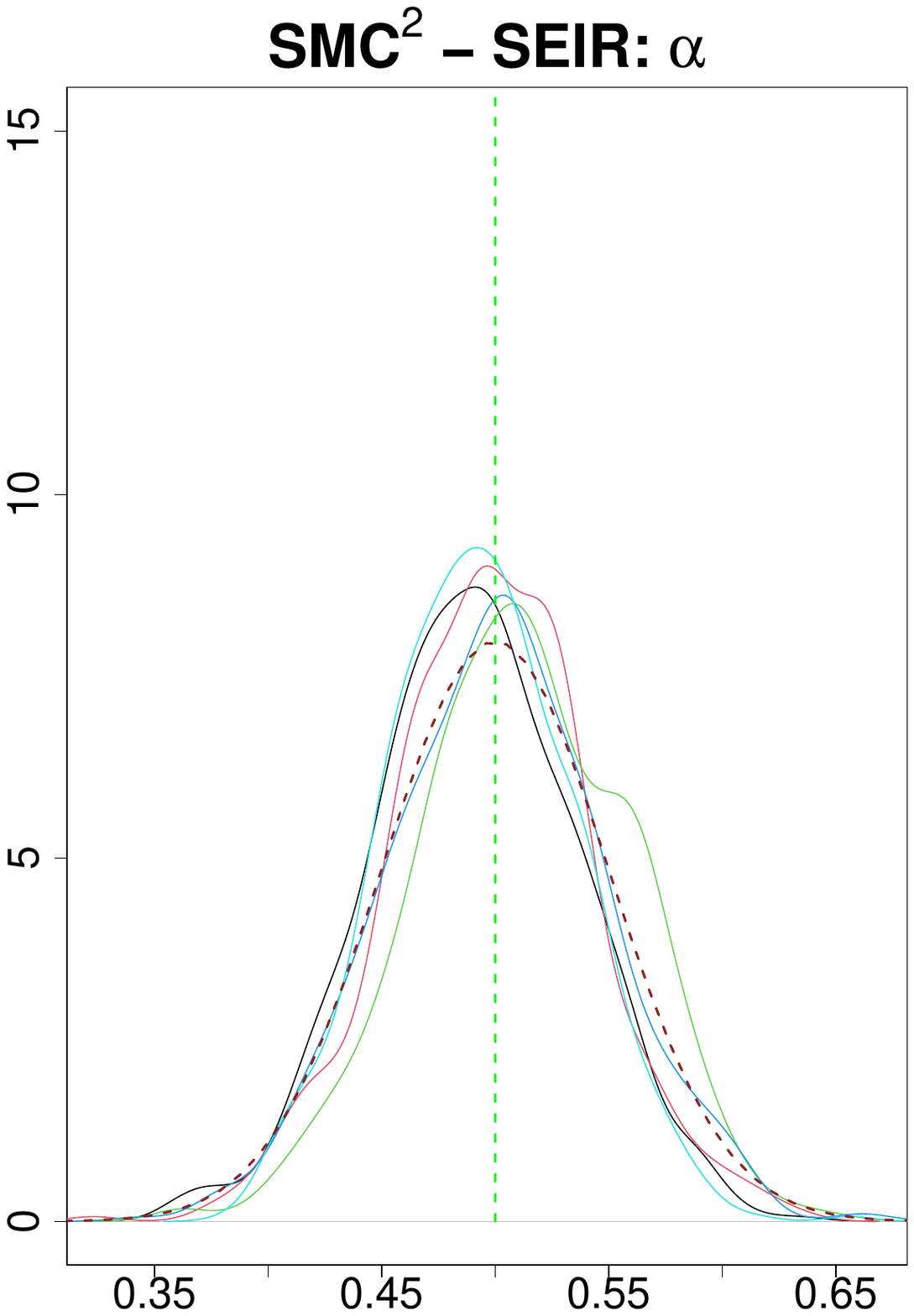}\includegraphics[scale=0.15]{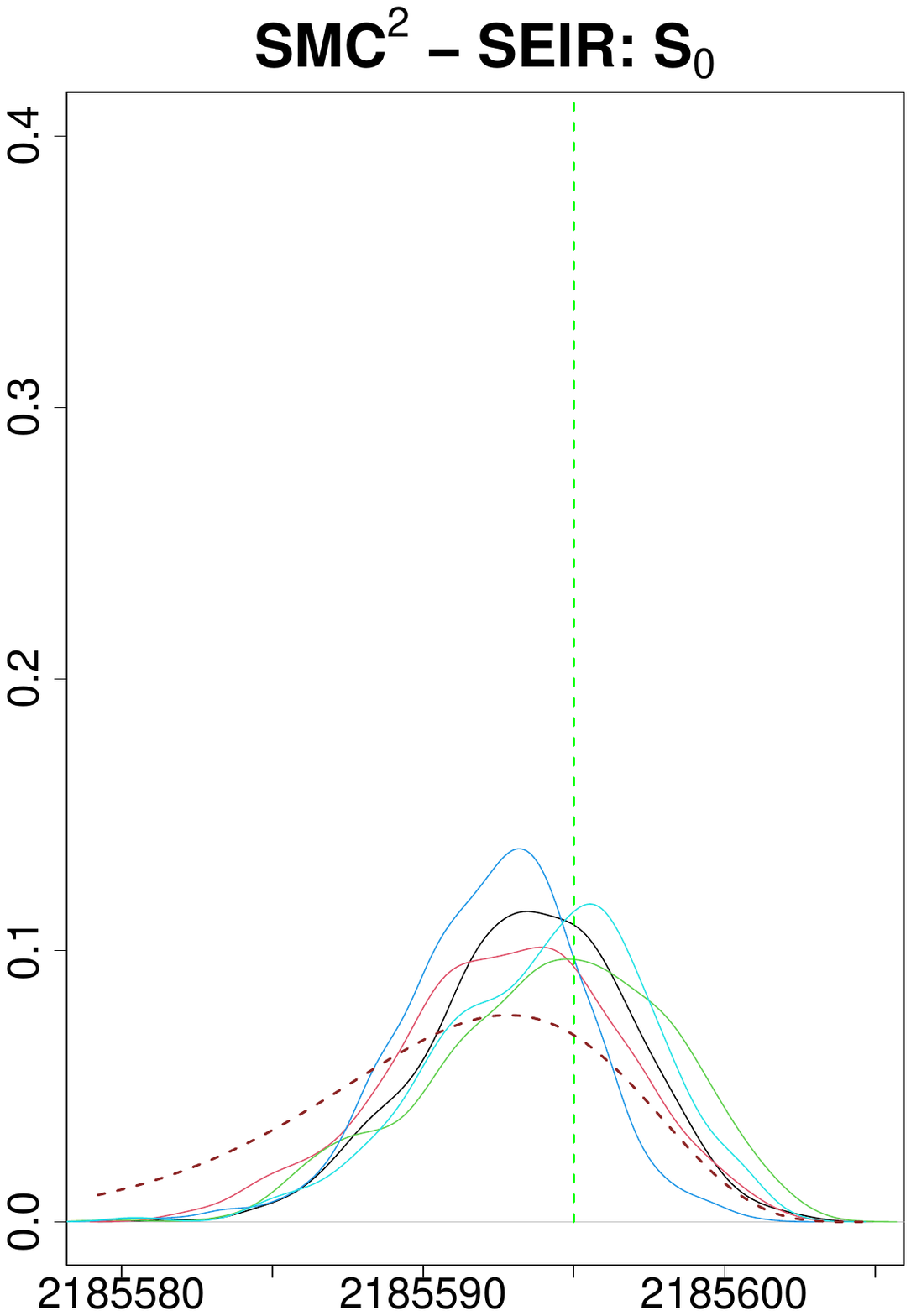}\includegraphics[scale=0.15]{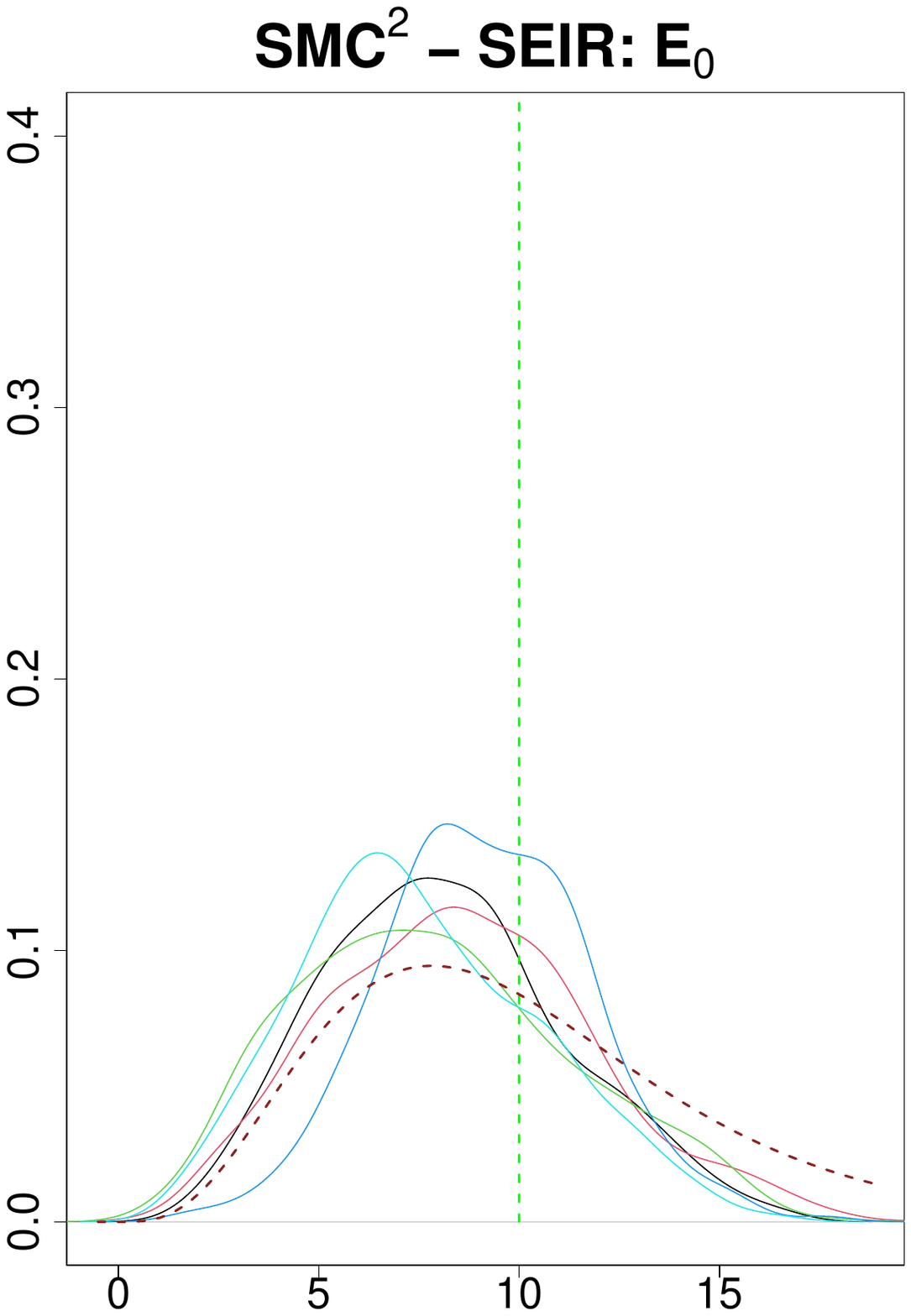}\includegraphics[scale=0.15]{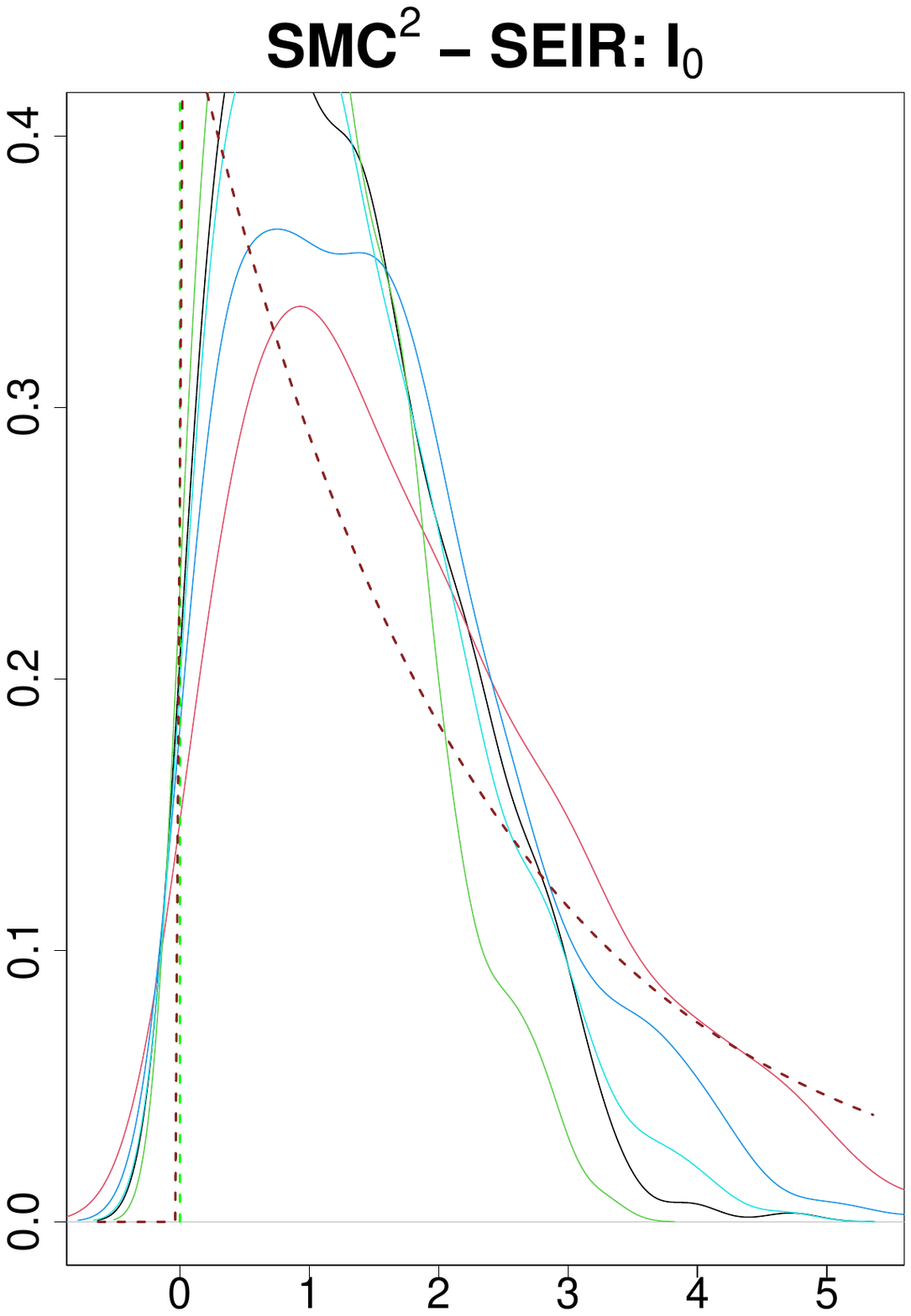}\includegraphics[scale=0.15]{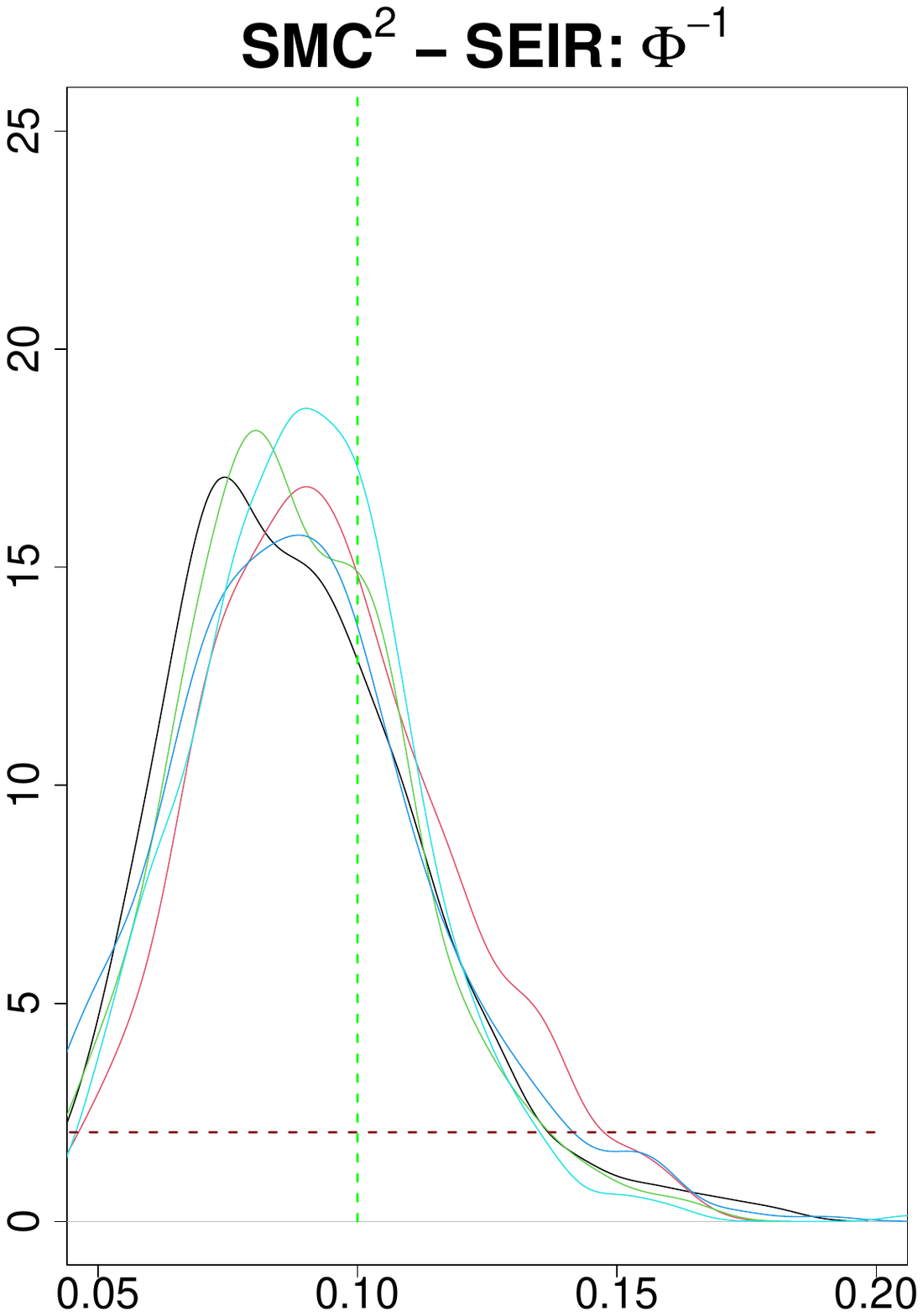}

\includegraphics[scale=0.15]{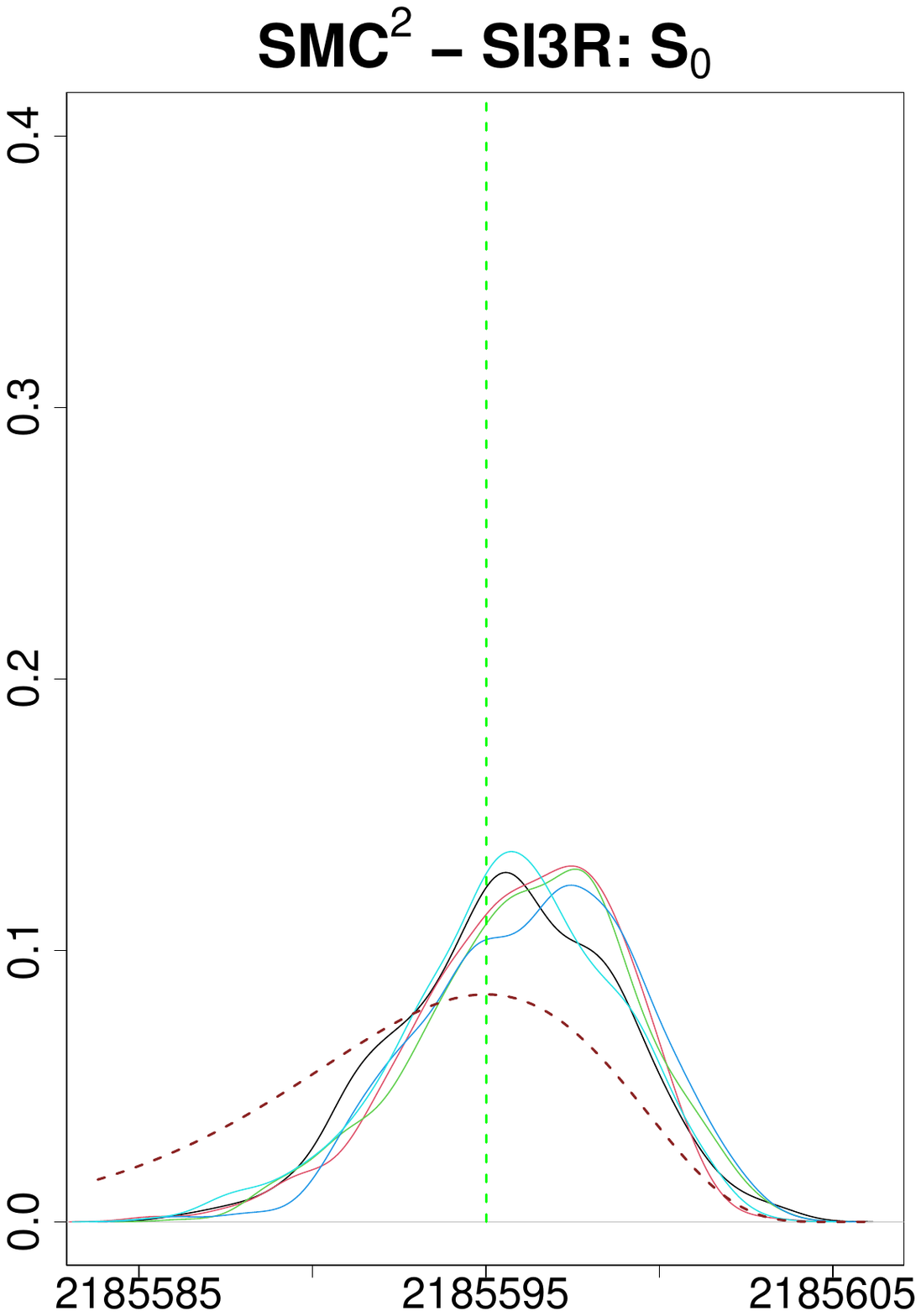}\includegraphics[scale=0.15]{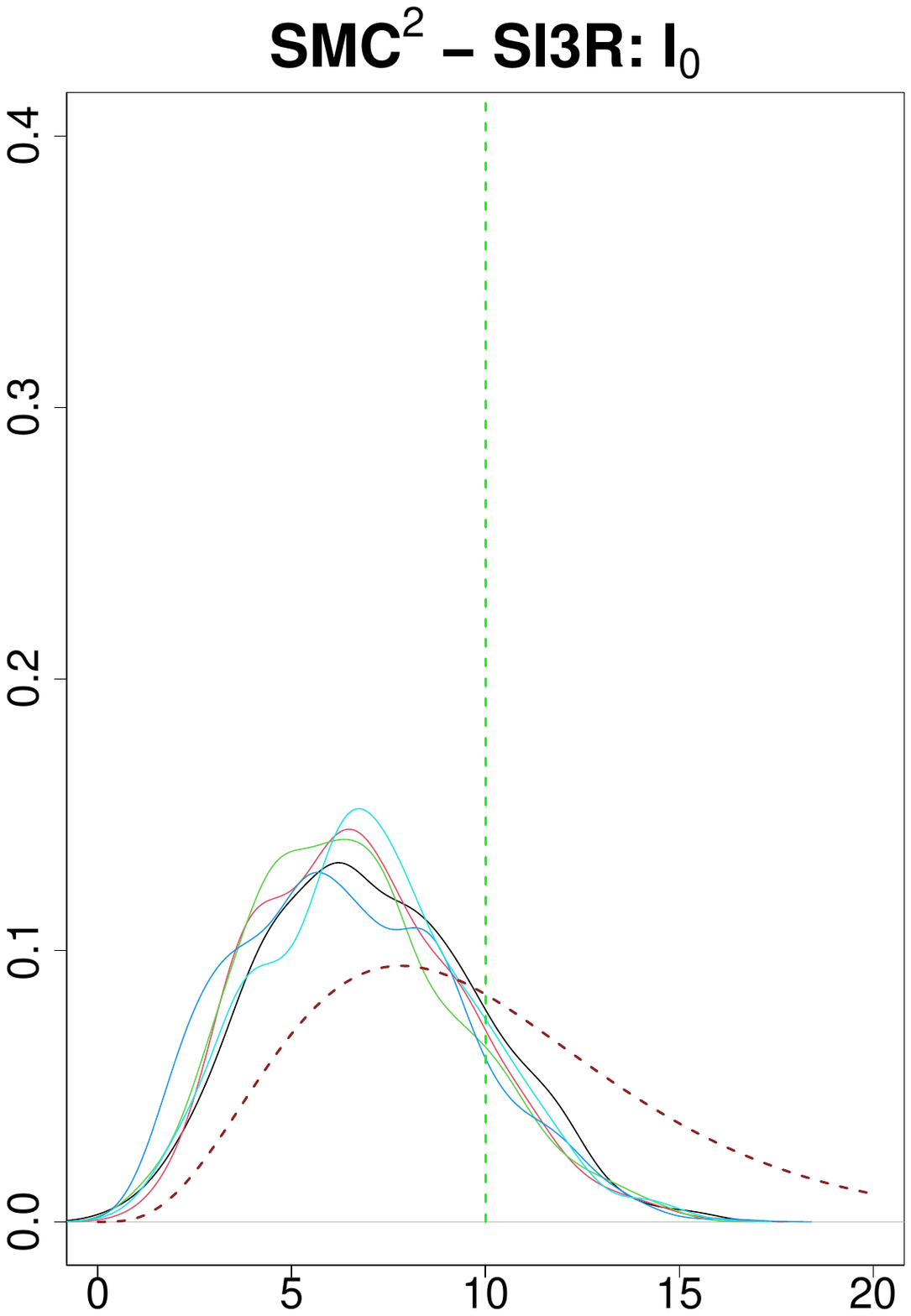}\includegraphics[scale=0.15]{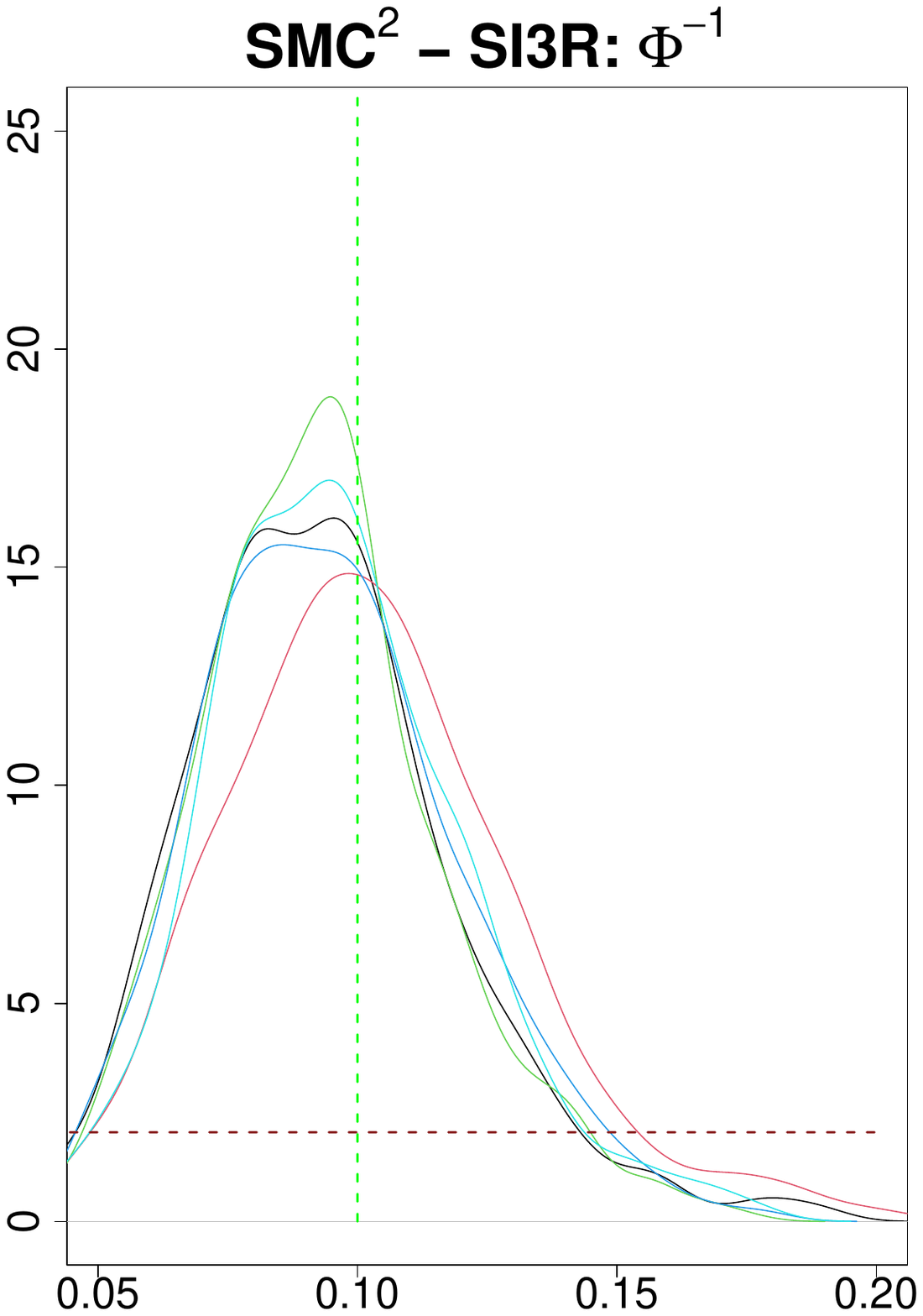}

\includegraphics[scale=0.15]{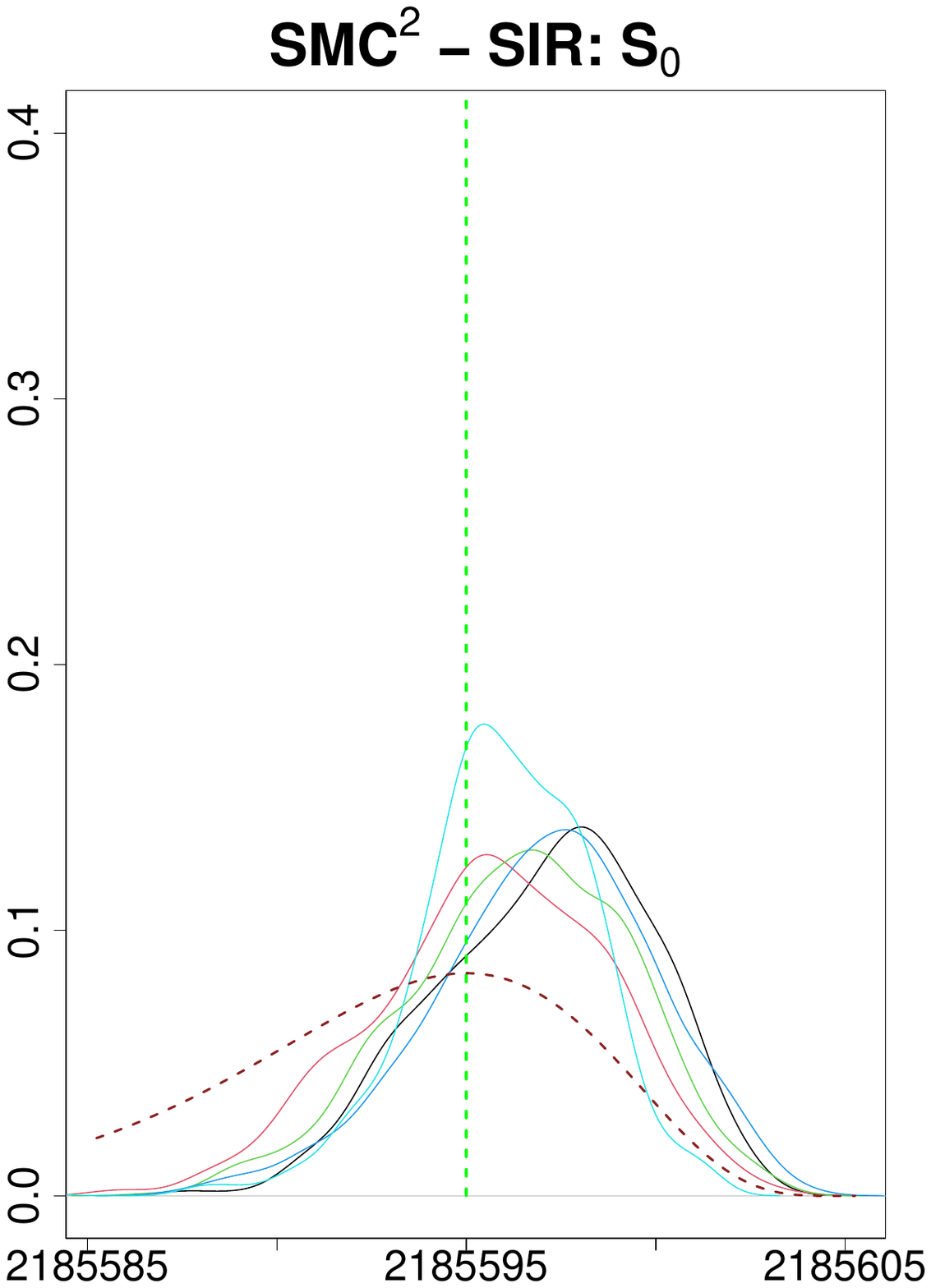}\includegraphics[scale=0.15]{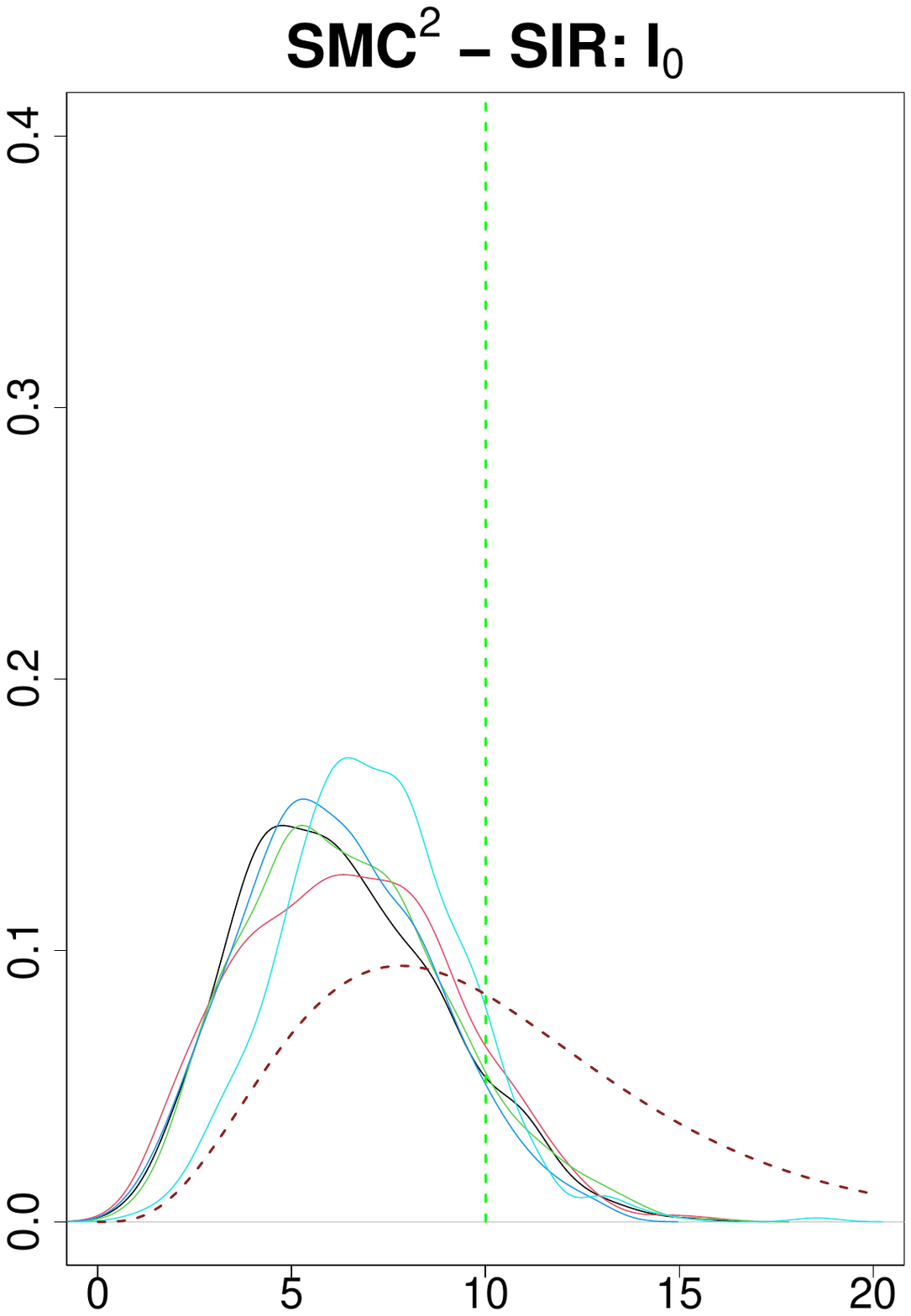}\includegraphics[scale=0.15]{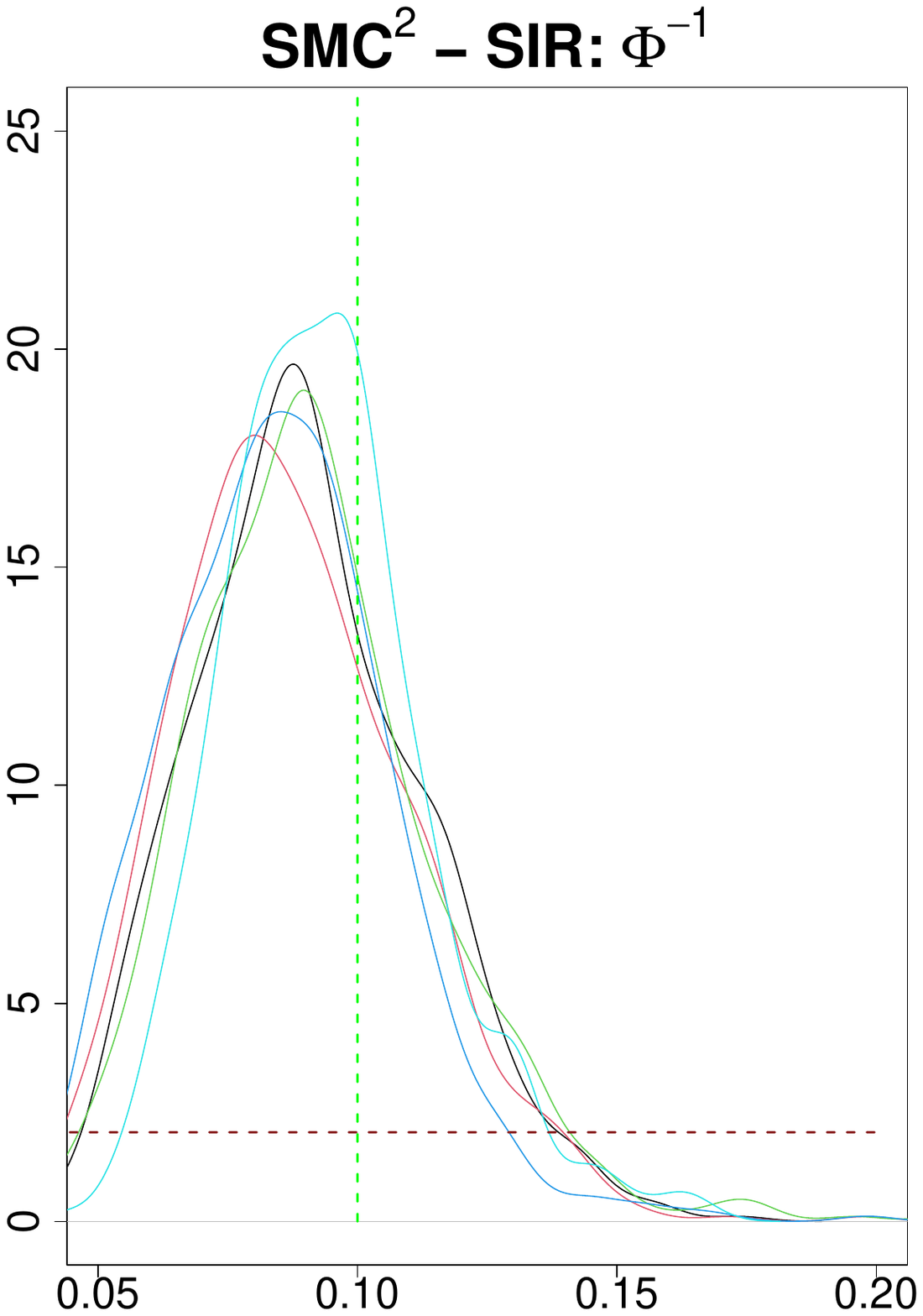}
\includegraphics[scale=0.5]{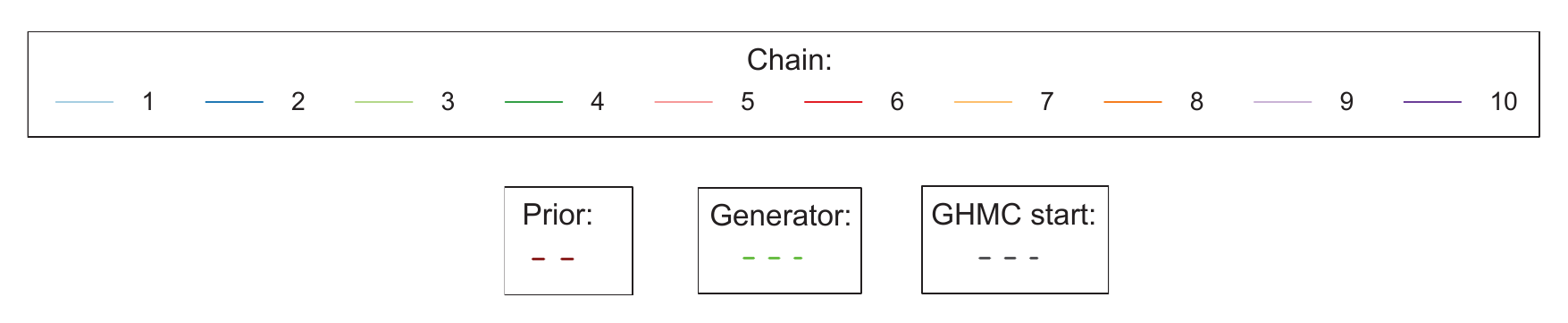}
\caption{For the synthetic data: parameters' posterior densities for  SMC$^2$ corresponding to the four different compartmental models. Dashed green vertical lines correspond to the true values.}
\label{plot_params_4}
\end{figure*}

\clearpage
\subsubsection{Sensitivity analysis}\label{sensitivity_analysis}
\begin{table}[h!]
\centering
\begin{tabular}{llllllllll}
            & \multicolumn{9}{c}{$\hat R$}                                                                                                                     \\
            & Set 01 & Set 02 & Set 03 & Set 04                       & Set 05 & Set 06                       & Set 07 & Set 08 & Set 09                       \\
$\alpha$    & 1.000  & 1.000  & 1.000  & 1.006                        & 1.000  & 1.000                        & 1.000  & 1.001  & 1.000                        \\
$S_0$       & 1.009  & 1.017  & 1.008  & 1.012                        & 1.011  & 1.018                        & 1.038  & 1.018  & 1.016                        \\
$E_0$       & 1.007  & 1.008  & 1.007  & 1.006                        & 1.009  & 1.013                        & 1.035  & 1.009  & {1.407} \\
$I_0$       & 1.019  & 1.039  & 1.028  & {1.060} & 1.014  & 1.033                        & 1.024  & 1.091  & 1.008                        \\
$\phi^{-1}$ & 1.000  & 1.001  & 1.001  & {2.069} & 1.001  & 1.001                        & 1.001  & 1.001  & 1.000                        \\
$\tau^2$    & 1.036  & 1.034  & 1.039  & {1.080} & 1.033  & {1.053} & 1.019  & 1.037  & 1.001                       
\end{tabular}
\caption{Convergence analysis for the sensitivity study of the SEI$_3$R model on 10 chains of 100000 production steps for each parameters set.}
\label{sup-table:r-hat_sensitivity}
\end{table}

\subsection{Case study 2: Basque Country data}
\begin{figure*}[h!]
\begin{center}
\includegraphics[scale=0.55]{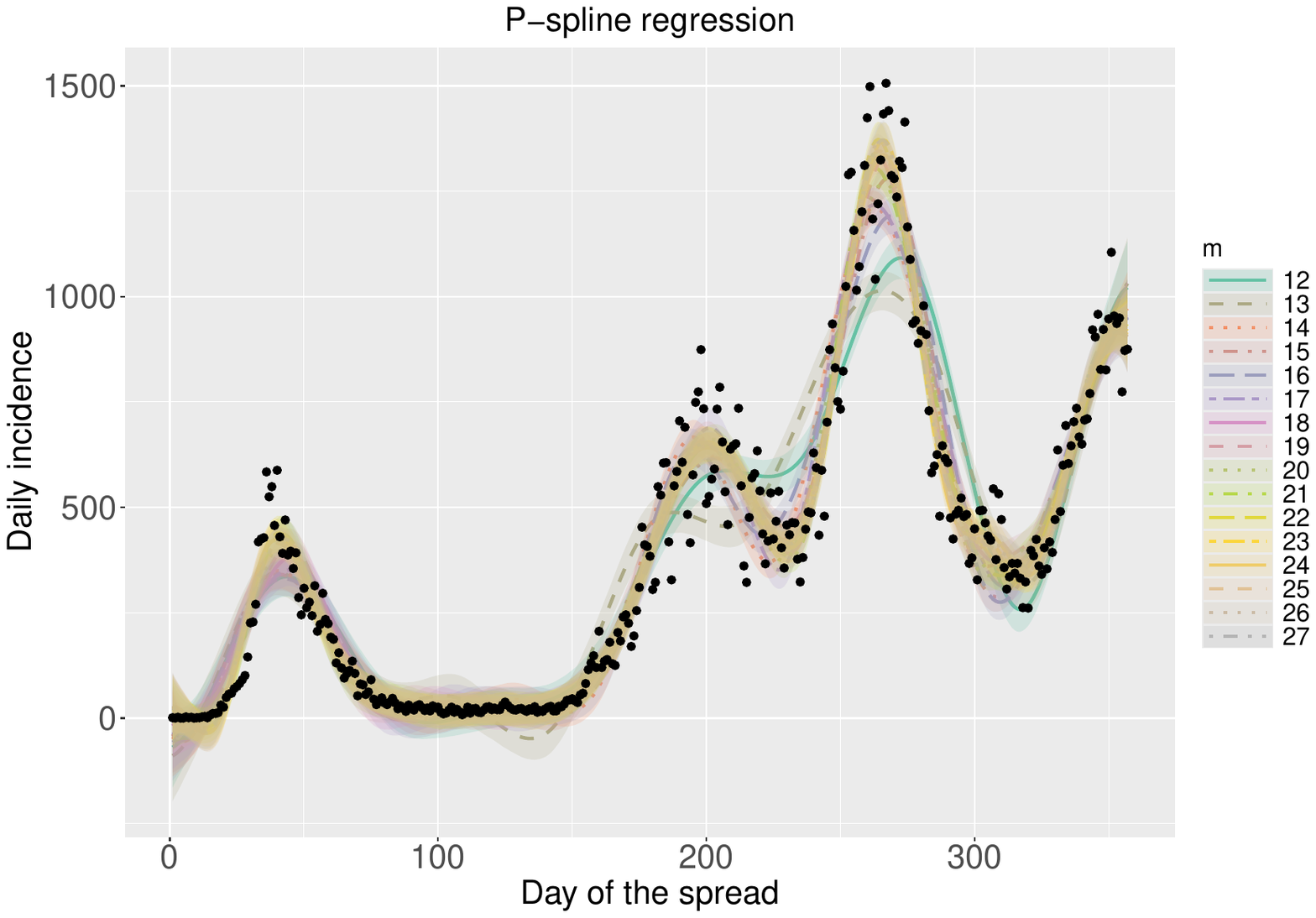}
\end{center}
\caption{Estimated P-spline regression (posterior mean) on Basque Country daily incidence data (black dots) uncorrected for under-reporting for a different number of internal knots $Q = 10, \dots, 25$. The priors were chosen following the suggestions in \cite{Lang2004}. 5000 MCMC production steps with a 1000 burn-in were taken. 
}
\label{regression_fig}    
\end{figure*}

\begin{table}[h!]
\centering
\begin{tabular}{lllllllll}
m    & 12     & 13     & 14         & 15     & 16     & 17     & 18     & 19     \\
WAIC & 296.70 & 365.82 & 112.34 & 71.94  & 164.62 & 90.97  & -19.86 & -18.59 \\
m    & 20     & 21     & 22         & 23     & 24     & 25     & 26     & 27     \\
WAIC & 6.51   & -22.85 & -66.69     & -85.27 & -85.50 & -70.81 & -62.68 & -68.49
\end{tabular}
\caption{WAIC for the P-spline fits with different number of spline basis $m$.}
\label{sup-table:waic}
\end{table}
In Web Figure \ref{regression_fig} one can see the posterior mean of P-spline regressions, with different internal knots, of daily incidence data in the Basque Country. As specified in the main text we use the WAIC criterion to select an appropriate number of spline basis for our procedures in the Basque Country data. From Web Table \ref{sup-table:waic}, the best trade off between small WAIC and small $m$ is clearly $m=23$ which is the one that we selected.  

\begin{table}[]
\centering
\begin{tabular}{lllll}
            & \multicolumn{4}{c}{$\hat R$}                                                                       \\
            & SEI$_3$R                     & SEIR  & SI$_3$R                      & SIR                          \\
$\alpha$    & {1.234} & 1.003 &                              &                              \\
$S_0$       & {1.602} & 1.013 & {1.393} & {1.105} \\
$E_0$       & {1.435} & 1.006 & {}      & {}      \\
$I_0$       & {1.330} & 1.016 & {1.822} & {1.710} \\
$\phi^{-1}$ & 1.008                        & 1.000 & {2.995} & {1.386} \\
$\tau^2$    & {1.132} & 1.047 & {2.184} & {1.228}
\end{tabular}
\caption{Convergence analysis for the Basque Country data on 10 chains of 100000 production steps.}
\label{sup-table:r-hat_basque}
\end{table}

\clearpage

\bibliographystyle{hapalike}
\bibliography{bibliography_splines}
\end{document}